\documentclass{article}
\usepackage{amsmath, amsthm, amssymb}
\usepackage{bbold}
\usepackage{array}
\usepackage{dsfont,esint}
\usepackage[utf8]{inputenc}
\usepackage{graphicx}

\newcommand{\n}{\mathfrak{n}}
\newcommand{\x}{z}
\newcommand{\dd}{\mathrm{d}}
\newcommand{\Res}{\mathop{\,\rm Res\,}}
\newcommand{\q}{q}
\newcommand{\qp}{ q'}

\begin{document}

\title{Purity distribution for generalized random Bures mixed states}

\author{Ga\"etan Borot$^1$ $\,\,$ and $\,\,$ C\'eline Nadal$^2$}
\makeatletter{\renewcommand*{\@makefnmark}{}
\footnotetext{$^1$ Department of Mathematics, University of Geneva (on leave from Institut de Physique Th\'eorique, CEA Saclay), gaetan.borot@unige.ch }\makeatother}
\makeatletter{\renewcommand*{\@makefnmark}{}
\footnotetext{$^2$ Rudolf Peierls Centre for Theoretical Physics, University of Oxford, and All Souls College (Oxford),
celine.nadal@physics.ox.ac.uk }\makeatother}

\maketitle

\vspace{2cm}

\begin{abstract}

We compute the distribution of the purity for random density matrices
(i.e.random mixed states) in a large quantum system, distributed according to the Bures
measure.
% The Bures measure is a probability measure based on a
% distance for density matrices.
The full distribution of the purity is computed using a mapping to
random matrix theory and then a Coulomb gas method. We find three
regimes that correspond to two phase transitions in the associated Coulomb
gas. The first transition is characterized by an explosion of the third derivative
on the left of the transition point.
The second transition is of first order, it is characterized by
the detachement of a single charge of the Coulomb gas.
A key remark in this paper is that the random Bures states are
closely related to the $O(n)$ model for
$n=1$. This actually led us to study ``generalized Bures states'' by keeping $n$ general instead of
specializing to $n=1$.

\end{abstract}

\newpage

\tableofcontents

\section{Introduction}
\label{sec:intro}

\subsection{Why random states ?}

In the last decades, there has been an increasing interest in
 random states in quantum physics.
Taking a state at random corresponds in some sense to assume minimal prior
knowledge about the quantum system~\cite{Hall}. Random states are thus very general, they can be seen as typical states.
They play the role of reference states to which the evolution in time of
a given physical state can be compared. 
The idea behind random states is close to the one of Wigner when he introduced random matrices to describe the Hamiltonian
of a large nucleus -a very complex system with many unknown details. 
In a similar way, in a large quantum system when the Hamiltonian is not known precisely
one can model the state of the system by a random state.

Let us consider a quantum system described by a Hilbert space $\mathcal{H}$ of dimension $N$.
A pure state is a vector $|\psi\rangle \in \mathcal{H}$ of unit norm, it describes the state of the quantum system. 
There is a natural way of choosing pure states at random: it is to pick them according to the uniform distribution.
The uniform distribution -also called Haar measure- is indeed
the unique probability measure that is invariant under a change of basis
(unitary transformation). Random pure states have been studied extensively~\cite{Cappellini2006,Facchi2008,Giraud2007,Lubkin1978,Majumdar2008,Page1993,Zyczkowski2001}. In particular there has been a lot of studies about entanglement properties of such random states in a bipartite system~\cite{Facchi2008,Facchi2010,Giraud2007,Lubkin1978,BE,BElong,Page1993},
 i.e. when the system is made of two subsystems $\mathcal{H} = \mathcal{H}_A\otimes \mathcal{H}_B$.
 % $\mathbb{C}^N = \mathbb{C}^{N_1}\otimes\mathbb{C}^{N_2}$.

The notion of quantum state can be generalized by introducing some statistical uncertainty.
A generalized quantum state or ``mixed state'' is described by a density matrix $\sigma$,
that is a $N\times N$ matrix on $\mathcal{H}$ with the following properties:
\\
-it is Hermitian, i.e. $\sigma^{\dagger}=\sigma$
\\
-it is semi-definite positive, i.e. $\langle v|\sigma|v\rangle \geq 0$ for all $|v\rangle \in\mathcal{H}$,
\\
-it has trace $1$, i.e. ${\rm Tr}\,\sigma=1$.
\\
The matrix $\sigma$ can thus be diagonalized in an orthonormal basis
$\{|v_i \rangle\}_{1\leq i\leq N}$, it has $N$ nonnegative eigenvalues
$\lambda_i$ that sum up to one. The spectral decomposition of $\sigma$ is thus of the form:
\begin{equation}\label{eq:melstatdec}
\boxed{\sigma=\sum_{i=1}^N \lambda_i |v_i \rangle \langle v_i|\;\;\textrm{with }\;\; \lambda_i\geq 0, \forall i \;\;\textrm{and}
\;\; \sum_{i=1}^N \lambda_i=1}
\end{equation}
The eigenvalue $\lambda_i$ can be interpreted as the probability
to find the system in pure state  $|v_i\rangle$.
$\sigma$ is a statistical superposition of pure states. If only one eigenvalue, say
 $\lambda_{i_0}$,
is non-zero, then $\lambda_{i_0}=1$ and
$\sigma=|v_{i_0} \rangle \langle v_{i_0}|$
is the projector on $|v_{i_0}\rangle$, $\sigma$ is thus simply
the density matrix representation of pure state
 $|v_{i_0}\rangle$ -we will call $\sigma$ a pure state in this case.
If several $\lambda_i$ are non-zero then there is an uncertainty about the state of the quantum system.
In general, such a mixed state can describe a system that is not isolated but in contact with an environment.
The interaction with the environment, more precisely the entanglement between the system and the environment,
  is responsible for a loss of information
for the observer that can do measurements only on the system.
The more mixed the state of the system is, the more entangled the system is with its environment.

One can now be interested in taking at random mixed states.
There is no single natural probability measure for mixed states (unlike pure states), as the 
unitary invariance only implies that the measure depends on the eigenvalue spectrum alone
(thus not on the eigenbasis of the density matrix).
Several probability measures have been proposed for mixed states and
motivated on physical grounds.
One possibility
is to construct a  measure
 from a unitarily-invariant distance
between mixed states. Such a distance induces
a measure on the space of density matrices
(volume element associated with the distance).
By construction, this measure is invariant under unitary transformations,
it is thus of the form
$\mathcal{P}(\lambda_1,\ldots,\lambda_N)\mathrm{d}\lambda_1\cdots\mathrm{d}\lambda_N \times \mathrm{d}\nu$
where $d\nu$ is the Haar measure on the group of unitary matrices $U(N)$.
The measure
is thus characterized by the marginal distribution of the eigenvalues
 $\mathcal{P}(\lambda_1,\ldots,\lambda_N)$.

A natural choice of distance is
the Hilbert-Schmidt distance \cite{Hall}
defined by:
\begin{equation}\label{eq:dHS}
D_{HS}\left(\rho,\sigma\right)^2={\rm Tr}\left\{ (\rho-\sigma)^2\right\}
\end{equation}
When both density matrices are pure states, i.e.
$\rho=|\psi\rangle\langle \psi|$ and $\sigma=|\phi\rangle\langle \phi|$,
then $\rho^2=\rho$, $\sigma^2=\sigma$ and the distance
 reduces to
$D_{HS}\left(\rho,\sigma\right)^2=2-2\left|\langle \psi |\phi \rangle\right|^2$.
The Hilbert-Schmidt distance induces a probability measure for mixed states.
We thus obtain random states that we will call ``random Hilbert-Schmidt states''.
The eigenvalues of the density matrix
of such random mixed states are distributed according to the law:
\begin{equation}\label{distriEVhaar}
\mathcal{P}_{HS}(\lambda_1,...,\lambda_N)=A_N\, \delta\left(\sum_i \lambda_i -1\right)
\prod_{j<k} \left(\lambda_j -\lambda_k \right)^2
\end{equation}
where $A_N=\Gamma(N^2)/\left[\prod_{j=0}^{N-1}\Gamma(N-j)\Gamma(N-j+1)\right]$.
This distribution can actually also be obtained in another natural way.
If we consider an auxiliary space $\mathcal{H}_E$ of same dimension $N$
(the ``environment'') and choose a pure state $|\psi\rangle$ at random (uniform Haar measure) in
the bipartite space $\mathcal{H} \otimes \mathcal{H}_E$,
the eigenvalues of the reduced density matrix of our system
$\sigma = {\rm Tr}_E\left\{|\psi\rangle \langle \psi|\right\}$
will be distributed exactly according to the law in Eqn~\eqref{distriEVhaar}.
${\rm Tr}_E$ is the partial trace over the environment $E$\footnote{
If $\rho$ is a density matrix on $\mathcal{H}\otimes \mathcal{H}_E$,
then $\sigma={\rm Tr}_E\rho=\sum_{k=1}^N \langle m_k |\rho| m_k\rangle$
where $\{|m_k\rangle\}_k$ is a basis of $\mathcal{H}_E$, and $\sigma$ is a density matrix on $\mathcal{H}$.}.
The mapping to random matrix theory for Hilbert-Schmidt states
can be guessed from the presence of the
 Vandermonde determinant $\det(\lambda_i^{j-1})=\prod_{i<j}(\lambda_j-\lambda_i)$ in $\mathcal{P}_{HS}$.
 The distribution $\mathcal{P}_{HS}(\lambda_1,...,\lambda_N)$ is actually the same as the distribution
 of the eigenvalues of a so-called Wishart matrix but with additional constraint $\sum_i\lambda_i=1$
 \cite{BElong}.

A second physically relevant choice of distance is the Bures distance, and this is the topic of this article. It is defined by
\begin{equation}
\label{eq:bures}
\boxed{D_B\left(\sigma,\rho\right)^2=2-2 \sqrt{F\left(\sigma,\rho\right)}}
\end{equation}
where $F\left(\sigma,\rho\right)$ is the fidelity:
\begin{equation}
\label{eq:fidelity}
\sqrt{F\left(\sigma,\rho\right)}={\rm Tr}\left\{\left[\sqrt{\rho}\sigma
\sqrt{\rho}\right]^{1/2} \right\}
\end{equation}
The distance $D_B(\sigma,\rho)=D_B(\rho,\sigma)$
is nonnegative, it is zero if and only if
$\rho=\sigma$.
When both density matrices are pure states, i.e.
$\rho=|\psi\rangle\langle \psi|$ and $\sigma=|\phi\rangle\langle \phi|$
and thus $\sqrt{\rho}=\rho$,
it reduces to $\sqrt{F\left(\sigma,\rho\right)}=
\left|\langle \psi |\phi \rangle\right|$ and
$D_B\left(\sigma,\rho\right)^2=2-2 \left|\langle \psi |\phi \rangle\right|$.It is the topic of this article. The Bures metric is a metric in the mathematical sense
for density matrices. This choice has been motivated
on both measurement and statistical grounds \cite{Braunstein1994,Braunstein1996,Hall,Josza1994}. These random mixed states will be called ``random Bures states''.
The distribution of the
eigenvalues of the density matrix for random Bures states is given by~\cite{SommersBures,Hall}:
\begin{equation}\label{distriEVbures}
\boxed{\mathcal{P}_B(\lambda_1,...\lambda_N)=C_N\, \frac{\delta( \sum_i \lambda_i-1)}{
\prod_i \lambda_i^{1/2}} \, \prod_{j<k}
 \frac{(\lambda_{j}-\lambda_{k})^2}{\lambda_{j}+\lambda_{k}}}
\end{equation}
where
the delta function enforces the condition
$\sum_i\lambda_i={\rm Tr}\,\sigma=1$
and where $C_N$ is a normalisation constant:
$ C_N=2^{N^2-N} \;\, \Gamma(N^2/2)/\left[ \pi^{N/2}\, \prod_{j=1}^N \Gamma(j+1)\right]$.
Note that the same Vandermonde determinant as in Eqn. \eqref{distriEVhaar} appears in \eqref{distriEVbures},
but there is an additionnal term in the denominator of the form $\prod_{i<j}(\lambda_j+\lambda_i)$.
Osipov, Sommers and \.Zyczkowski explain in \cite{SommersPur} how to generate numerically Bures density matrices.

%The distribution $\mathcal{P}_B(\lambda_1,...\lambda_N)$ is actually also related to random matrix theory
%as it corresponds to the so-called $O(n)$ matrix model for $n=1$ as we will see below. Let us mention that They also show how to generate numerically at random Bures density matrices.
%\newline

\subsection{Statistical properties of random mixed states: purity
  distribution and entanglement properties}
 
One can ask several questions about statistical properties of random
mixed states. The most natural one is to ask how mixed (or pure) such
states are.
For a random mixed state described by a density matrix $\sigma$,
the purity defined as $\Sigma_2={\rm Tr}\,\sigma^2=\sum_i\lambda_i^2$ provides an
indicator to answer this question.
The higher
the purity, the purer the state. As we have seen above, a purer state
corresponds to a state that is less entangled with its environment. 

For Hilbert-Schmidt random mixed states, the distribution of the purity
has been computed recently by one of the authors in collaboration with
Majumdar and Vergassola for a large system \cite{BE,BElong}, it features two interesting phase transitions. 
For random Bures states, the average of the purity has been computed
by Sommers and and \.{Z}yczkowski \cite{SommersDens}, and the moments were computed by the same authors and Osipov \cite{SommersPur}. However the
distribution of the purity for Bures states was up to now not known
analytically, and the goal of this paper is to compute it for large
systems ($\mathcal{H} \simeq \mathbb{C}^N$, $N \rightarrow
\infty$). 
%We use a mapping to random matrix theory -more precisely to the $O(n)$
%matrix model as explained below- and a Coulomb gas method
%to compute this distribution.
We find phase transitions that are very similar to the ones
appearing in the Hilbert-Schmidt case, but with important changes in
the analytical expressions. The order of the first transition is
different for Bures states and Hilbert-Schmidt states, but the second
transition is identical in the two cases:  it is a first order
transition that corresponds to a sudden jump of the maximal
eigenvalue.
Another very interesting question is to consider random states on a
bipartite (or even multipartite) system, ie
$\mathcal{H}=\mathcal{H}_A\otimes\mathcal{H}_B$ (of dimension $N=N_A
N_B$),
and to study entanglement between the subsystems $A$ and $B$.
A mixed state described by a density matrix $\sigma$ on $\mathcal{H}$
is said to be separable if it can be written as a convex sum of tensor
products, ie as $\sigma=\sum_k p_k
\sigma_A^k \otimes \sigma_B^k$ with $p_k\geq 0$ and $\sum_k p_k=1$.
Otherwise it is entangled.
For mixed states, unlike pure states, one cannot use Von Neumann or Renyi
entropies as a measure of entanglement, but several
measures of entanglement have been introduced and studied (such as the entanglement of formation or a
distance to separable states) \cite{NadalThesis,Vedral98} but they are all quite difficult to
compute, especially for random states. 
An easier question is to ask what is the probability that a random
mixed state is separable,
or what is the probability that its partial transpose is positive
(which is implied by separability)  \cite{Peres}.
 Those probabilities have been studied by Aubrun et Szarek
 \cite{Aubrun,Szarek} for the Hilbert-Schmidt case, and by Ye
 \cite{Ye} for the Bures measure. To summarize the conclusion of
 \cite{Ye}, 
 random Bures states are almost surely non separable when $N \rightarrow \infty$, even when conditioned to have a positive partial transform. 

%Let us mention that other indicators of entanglement have been
%introduced and studied.
% In particular, when one considers $\mathbb{C}^N$ as
% $\mathbb{C}^{N_1}\otimes\cdots\otimes\mathbb{C}^{N_r}$,
% one can wonder what is the probability for a random state to be
% separable (i.e. to be a convex sum of tensor product states), 
% or what is the probability that its partial transpose is positive
% (which is implied by separability)
% \cite{Peres}. Those probabilities have been studied by Aubrun et
% Szarek \cite{Aubrun,Szarek} for the Hilbert-Schmidt case,
% and by Ye \cite{Ye} for the Bures measure. To summarize the
% conclusion of \cite{Ye}, 
%Bures random states are almost surely non separable when $N \rightarrow \infty$, even when conditioned to have a positive partial transform. 

\subsection{Relation with the $O(n)$ matrix model}

A key remark in this article is that the distribution in Eqn~\eqref{distriEVbures}
is also the same as the distribution of the eigenvalues for the $O(n)$ matrix model, for the special case $n=1$, except for the additional constraint $\sum_i \lambda_i=1$. And, at least for large $N$ results, this constraint can be handled with help of Lagrange multipliers. The $O(n)$ model was originally introduced by Kostov \cite{Kostov} in a very different context, namely quantum gravity and enumeration of random discrete surfaces with self-avoiding loops (see \cite{2Dgravity} for a review of quantum gravity techniques). The partition function of the $O(n)$ model is:
\begin{equation}
Z=\int dM dA_1 ... dA_n \, e^{-\frac{N}{g} {\rm Tr} \left[V(M)+
M\left(A_1^2+...+A_n^2 \right)\right]}
\end{equation}
where the matrices $M$ and $A_i$ are Hermitian $N\times N$ matrices
and $V(M)$ is for example a polynomial in $M$. The Gaussian integral over matrices $A_i$ can be performed. Then $M$ can be parametrized by a unitary transformation and its eigenvalues
and the integral over the unitary transformation can also be done.
Finally, $Z$ becomes an integral over the eigenvalues $\lambda_i$:
\begin{equation}
Z=\int \prod_i \mathrm{d}\lambda_i \,
 \frac{ \prod_{j<k}  \left(\lambda_{j}-\lambda_{k}\right)^2}{
\prod_{j,k} \left(\lambda_{j}+\lambda_{k}\right)^{n/2}}\,
e^{-\frac{N}{g}\sum_i V(\lambda_i)}
\end{equation}
In this form, the model can be extended to real values of $n$
(not necessarily integer).
For the special case of a linear potential $V(\lambda)=\lambda$
and for $n=1$ (case of a two-matrix model),
if we add the constraint $\sum_i \lambda_i=1$, we recover
the joint distribution of Bures eigenvalues in Eqn~\eqref{distriEVbures}.
We can also remark that this provides a realization of the Bures measure as the probability measure induced on eigenvalues in a two Hermitian matrices model.

In this paper, we actually study ``generalized Bures states'', for which the eigenvalues of the density matrix $\sigma$ is distributed according to the probability density:
\begin{equation}\label{eq:genBures}
\boxed{\mathcal{P}_{B,n}(\lambda_1,...,\lambda_N) = D_{N,n}\,\delta\Big(\sum_{i = 1}^{N}\lambda_i - 1\Big)\,\frac{\prod_{j < k} (\lambda_j - \lambda_k)^2}{\prod_{j,k} (\lambda_j + \lambda_k)^{n/2}}}
\end{equation}
for $n \in ]0,2[$. Other values of $n$ could be handled in a similar way, but we focus on the range $]0,2[$,
 which contains $n = 1$ (``physical'' Bures case) and for which the
 results can be expressed without cases discussions. We shall
 specialize at the end our results to $n = 1$, i.e. for the
 random Bures states.
 
It is not clear to us if Eqn~\eqref{eq:genBures} can be derived from a distance on the set of density matrices, which could be interesting on its own.
 For the Bures metric, cf. Eqn~\eqref{eq:bures}, one can show that the infinitesimal distance is given by
 $(ds_B)^2=D_B\left(\sigma,\sigma+\delta \sigma \right)^2
=\frac{1}{2}\sum_{i,j}\frac{\left|
\langle v_i|\delta \sigma | v_j \rangle\right|^2}{\lambda_{i}+\lambda_{j}}$
where the $\lambda_{i}$ are the eigenvalues of $\sigma$ and  $|v_i \rangle$
its eigenvectors~\cite{NadalThesis}. If there exists a distance $D_{B,n}$ corresponding to the generalized Bures measure, then
the associated infinitesimal distance must be given by:
 \begin{equation}\label{eq:dsGenBures}
(ds_{B,n})^2=D_{B,n}\left(\sigma,\sigma+\delta \sigma \right)^2
=\frac{1}{2}\sum_{i,j}\frac{\left|
\langle v_i|\delta \sigma | v_j \rangle\right|^2}{(\lambda_{i}+\lambda_{j})^n}
\end{equation}

\subsection{Outline}

Our results on the statistical properties of random Bures states (the case $n = 1$) are summarized in section \ref{sec:statprop}. In section
\ref{sec:Coulombgas} we explain the Coulomb gas and saddle point method we use to compute the purity distribution when $N \rightarrow \infty$. With this method we derive an integral equation for the eigenvalue density. The distribution of the
purity for random Hilbert-Schmidt states has been computed with the same
method \cite{BElong}. However, in the Hilbert-Schmidt case, this leads to integral equations for the eigenvalue density that can be solved using a theorem by Tricomi. This method cannot be used in the Bures case as the equations are more involved.

In the distribution of the purity, we find three regimes. In section \ref{sec:regII}, we derive the explicit solution for regime {\bf II}, i.e. in the case where the density has a support of the form $[0,b]$. One can use for this purpose a theorem of B\"uckner which is the appropriate generalization of the result of Tricomi \cite{Bueckner}. In section \ref{sec:regIII} we compute the solution for regime {\bf III} again using B\"uckner method. This regime is characterized by an isolated eigenvalue (the maximal one) and a continuous density for the other eigenvalues. A form of sudden condensation happens at the transition between {\bf II} and {\bf III}:
the maximal eigenvalue jumps from a small value to a much larger
value, thus detaching from the sea of the other eigenvalues. In section \ref{sec:regI} we derive the explicit solution for regime {\bf I}, characterized by a density with support $[a,b]$ away from $0$ (the most involved case). For this purpose, we use a technique developed in the context of the $O(n)$ model by Eynard and Kristjansen \cite{EK95}, and reformulated in a more algebraic geometric way recently by one of the authors and Eynard \cite{GaetanB,GBthese}. This solution involves elliptic functions and Jacobi theta functions. We also analyze the limit of low purity (completely mixed) states, and the transition between regimes {\bf I} and {\bf II}. Some intermediate steps of computations are included in Appendix.

\section{Statistical properties of random Bures states}
\label{sec:statprop}

We recall that, for random Bures states, the eigenvalues of the density matrix $\sigma$
are distributed according to the law~\cite{Hall,SommersBures}:
\begin{equation}\label{distriEVbures2}\begin{aligned}
\mathcal{P}_B(\lambda_1,...\lambda_N)&= C_N\, \frac{\delta( \sum_i \lambda_i-1)}{
\prod_i \lambda_i^{1/2}} \, \prod_{j<k}
 \frac{(\lambda_{j}-\lambda_{k})^2}{\lambda_{j}+\lambda_{k}}\\
 &=D_N\, \delta( \sum_i \lambda_i-1) \,
 \frac{\prod_{j<k} (\lambda_{j}-\lambda_{k})^2}{\prod_{j,k} (\lambda_{j}+\lambda_{k})^{\frac{1}{2}}}
 \end{aligned}
\end{equation}
where $C_N$ and $D_N$ are normalisation constants and where
the delta function enforces the condition
$\sum_i\lambda_i={\rm Tr}\,\sigma=1$.

\subsection{Average density and purity}

Sommers and \.{Z}yczkowski
have computed in \cite{SommersDens} the average density of eigenvalues
$\rho(\lambda,N)=\frac{1}{N}\sum_{i=1}^N\langle \delta\left(
\lambda-\lambda_i\right)\rangle$.
The quantity $\rho(\lambda,N)\mathrm{d}\lambda$ is the probability
to find an eigenvalue between $\lambda$ and $\lambda+\mathrm{d}\lambda$.
For a random Bures state, i.e. when the eigenvalues
$\lambda_i$ are distributed according to the joint distribution
in Eqn~\eqref{distriEVbures2}, the density of states
is given for large $N$ by
$\rho(\lambda,N)=N \rho^*(\lambda \, N)$
where the rescaled density  $\rho^*$  has a finite support over
$]0,b]$ with $b=3 \sqrt{3}$ and is given by
\begin{equation}
\rho^*(x)=\frac{1}{4 \pi \sqrt{3}}\left\{
\left(\frac{b}{x}+\sqrt{\frac{b^2}{x^2}-1}\right)^{2/3}
-\left(\frac{b}{x}-\sqrt{\frac{b^2}{x^2}-1}\right)^{2/3}
\right\}
\end{equation}
The scaling of the density was expected.
 A typical eigenvalue indeed
scales as $\lambda_{\rm typ}\sim 1/N$ for large $N$
because of the constraint $\sum_{i=1}^N \lambda_i=1$.

%Another quantity of interest is the purity:
As we have seen in the introduction, another quantity of interest (and the one we focus on in this paper)
is the purity defined as:
\begin{equation}\label{eq:defpur}
\boxed{\Sigma_2={\rm Tr}\left[\sigma^2\right]=\sum_{i=1}^N \lambda_i^2}
\end{equation}
$\Sigma_2$ measures how mixed (or how pure) a  state described by a density matrix $\sigma= \sum_{i=1}^N \lambda_i |v_i\rangle \langle v_i|$
is.
Let us consider two limiting cases:\\
$\bullet$ When only one eigenvalue,
say $\lambda_{i_0}$ is nonzero and thus equal to one,
the purity is maximal $\Sigma_2=1$.  In this case, the quantum state is pure, $\sigma$
is the projector on state $|v_{i_0}\rangle$: $\sigma=|v_{i_0}\rangle \langle v_{i_0}|$.
This means that there is no interaction (thus no entanglement) with the environment.\\
 %(see Eqn~\eqref{densmatrix}).\\
$\bullet$  When all eigenvalues
are equal, i.e. $\lambda_j=1/N$ for all $j$,
the purity is minimal $\Sigma_2=1/N$. The quantum state is
then completely mixed, it is a mixture of all
possible pure states $|v_j \rangle$ with equal probability $1/N$.
In this case,  entanglement with the environment is maximal.
%In reference \cite{SommersDens}, Sommers and \.{Z}yczkowski
%compute
%the average purity

The authors of \cite{SommersDens} compute, for random Bures states, the average of the purity $\Sigma_2={\rm Tr}\left[\sigma^2\right]=\sum_{i=1}^N \lambda_i^2$ at finite $N$, and the average of the generalized purity $\Sigma_q=\sum_{i=1}^N \lambda_i^q$ for large $N$, and compare their results to the Hilbert-Schmidt case. They find $\langle \Sigma_2\rangle \sim \frac{5}{2 N}$ for large $N$ in the Bures case whereas $\langle \Sigma_2\rangle \sim \frac{2}{N}$ for large $N$ in the Hilbert-Schmidt case. Later, Osipov, Sommers and \.{Z}yczkowski \cite{SommersPur} have shown how to compute recursively the moments of the purity at finite $N$. Though effective, this method does not allow to write an analytical answer for the distribution of purity for random Bures states.

For Hilbert-Schmidt random states,  the
 distribution of $\Sigma_q$ (and thus of the Renyi entropy $S_q=\frac{1}{1-q} \ln \Sigma_q$) was computed for large $N$ and all $q > 1$ in \cite{BE,BElong}. It shows interesting phase transitions, that we find again in the distribution of purity for Bures random states. Our method could also be applied on to compute the distribution of $\Sigma_q$ at large $N$ for Bures random states (it amounts to take a potential of the form $V(x) = t_0 + t_1 x + t_q x^q$ below), and we expect similar phase transitions although the analytical expression of the thresholds will be different. 
 
%It has been found that the distribution
%of $\Sigma_q$ shows unusual phase transitions.
%In this paper, using the same Coulomb gas method as we used for the Hilbert-Schmidt case, we compute analytically the distribution of the purity
%$\Sigma_2$ for random Bures states
%in the large $N$ limit. We find similar unusual phase transitions but of different order.
%However, because of the form of the eigenvalue distribution, the Coulomb gas method leads to
%integral equations that are much more involved.

\subsection{Our results: purity distribution for random Bures states}

\begin{figure}\begin{center}
\includegraphics[width=0.75\textwidth]{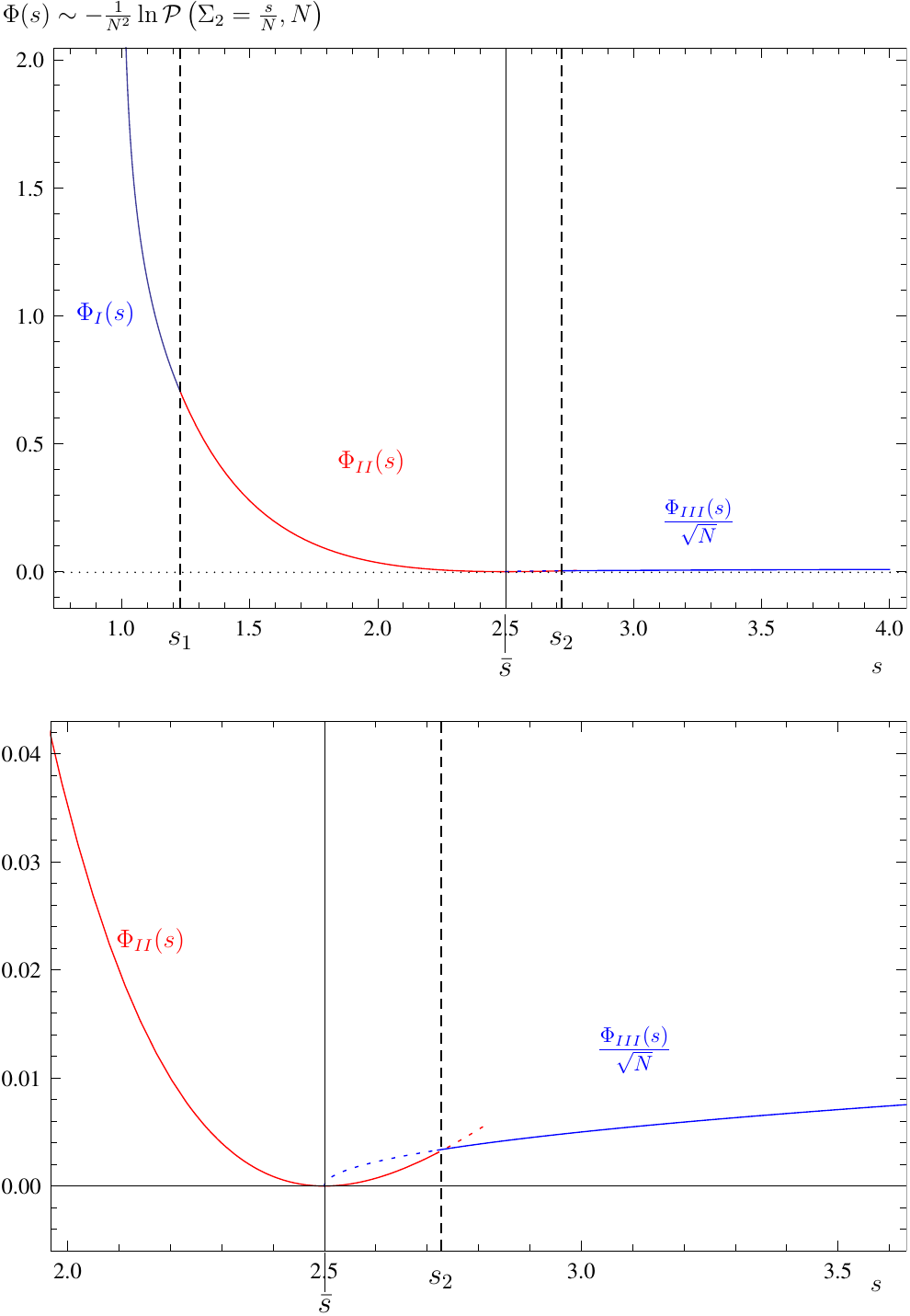}
\caption{Distribution of the purity $\Sigma_2$ for random Bures states ($n=1$, $\mu=2/3$)
for $N=5000$:
(a) Upper figure: plot of the rate function $\Phi(s)\sim -\frac{1}{N^2} \ln \mathcal{P}\left(\Sigma_2=\frac{s}{N},N\right)$.
The three regimes are shown. Regime {\bf I} goes from $s=1$ to $s=s_1=\frac{315}{256}\approx 1.23$.
Regime {\bf II} goes from $s=s_1$ to $s=s_2\approx 2.70$ for $N=5000$. It includes the mean value $s=\bar{s}=5/2$.
Regime {\bf III} is valid for $s>s_2$. The scaling with $N$ (for large $N$) is different for regime {\bf III}:
$\Phi(s)=\Phi_{III}(s)/\sqrt{N}$. This is the reason why $\Phi$ seems very close to zero within regime {\bf III}.
(b) lower figure: zoom of the previous figure around $s=s_2$.
 At the transition {\bf I}-{\bf II}, ie at $s=s_2$, we clearly see that the first derivative of $\Phi(s)$
is discontinuous.
This figure also shows that there is a small range where the solutions of regime {\bf II} and {\bf III} coexist
 (solid and dashed lines).
The one which dominates (solid line) at a given value of $s$ is the one with lowest rate function (lowest energy). 
}\label{fig:plotphi}\end{center}
\end{figure}

%As $\sum_{i=1}^N\lambda_i=1$, we typically expect for large $N$ that $\lambda_i\asymp \frac{1}{N}$
%(i.e. $\lambda_i$ proportional to $1/N$ for large $N$) and thus
%$\Sigma_2=\sum_i\lambda_i^2 \asymp \frac{1}{N}$.
%This scaling is in agreement with the result of Sommers and \.{Z}yczkowski \cite{SommersPur}
%who have shown that the mean value of the purity is given for large $N$ by
%$\langle \Sigma_2 \rangle\sim \bar{s}/N$ where $\bar{s}=5/2$.
Since a typical eigenvalue is of order $1/N$
(because $\sum_{i=1}^N\lambda_i=1$), we will be interested in the behaviour of $\mathcal{P}\left(\Sigma_2= \frac{s}{N},N\right)$ for $s$ fixed and $N\to\infty$. To compute this distribution, we use a mapping to random matrix theory
and a Coulomb gas method as explained in section \ref{sec:Coulombgas}. Similarly to the Hilbert-Schmidt case~\cite{BE},
we find three regimes for the distribution of the purity
for random Bures states.
These regimes follow from two phase transitions in the associated Coulomb gas (cf. section \ref{sec:Coulombgas}).
We show that the distribution of the purity is given for large $N$ by:
\begin{equation}\label{distriPur}
\boxed{\mathcal{P}\left(\Sigma_2=
\frac{s}{N},N\right)\approx
\left\{\begin{array}{ll}
\exp\left\{-N^2 \,\Phi_I(s)\right\}&{\rm for}\;\; 1<s\leq s_1\\
\\
\exp\left\{-N^2 \, \Phi_{II}(s)\right\}&{\rm for}\;\;
s_1<s\leq s_2\\
\\
\exp\left\{-N^{3/2} \, \Phi_{III}(s)\right\}
&{\rm for}\;\; s> s_2\\
\end{array}\right.}
\end{equation}
 where the critical points $s_1$ and $s_2$ are given by:
 \begin{equation}
 s_1=\frac{315}{256}\;, \;\;\;\; s_2=\frac{5}{2}\left[1+\left(\frac{5}{2 N}\right)^{\frac{1}{3}}+...\right]
 \end{equation} 
 thus  $s_2\sim \frac{5}{2}$ when $N\to \infty$.
The exact mathematical meaning of the symbol $\approx$ is a logarithmic equivalent
whereas $\sim$ denotes an equivalent: for example
$\mathcal{P}\left(\Sigma_2=
\frac{s}{N}\right)\approx
\exp\left\{-N^2 \Phi_I(s)\right\}$ for  $1<s\leq s_1$ means
$\ln\mathcal{P}\left(\Sigma_2=
\frac{s}{N}\right)\sim
-N^2 \Phi_I(s)$ for large $N$.
Let us define
the rate function $\Phi$  by:
\begin{equation}
\ln\mathcal{P}\left(\Sigma_2=
\frac{s}{N}\right)\sim
-N^2 \Phi(s)\;\;\textrm{as $N\to \infty$}
\end{equation}
$\Phi$ describes the large deviations of the purity.
We show in this paper that
$\Phi(s)=\Phi_{I}(s)$ if $s<s_1$,
$\Phi(s)=\Phi_{II}(s)$ if $s_1<s<s_2$ and $\Phi(s)=\Phi_{III}(s)/\sqrt{N}$ if $s>s_2$,
where $\Phi_{I,II,III}$ do not depend on $N$ when $N\to\infty$. Fig. \ref{fig:plotphi} shows a plot of the rate function $\Phi(s)$ (we
plot here our analytical results), in
particular one can see the three regimes. In the three subsections below, we give a summary of our results for each of the three regimes,
in particular we give the expression that we obtain respectively for
$\Phi_{II}$, $\Phi_{III}$ and $\Phi_{I}$.

\subsubsection{Regime {\bf II}}

Regime {\bf II} corresponds to $s_1<s<s_2$ with $s_1=\frac{315}{256}$ and
$s_2=\frac{5}{2}\left[1+\left(\frac{5}{2 N}\right)^{\frac{1}{3}}+...\right]$ for large $N$.
It describes the central region, around the mean value of the purity $s=\bar{s}=\frac{5}{2}$.
In this regime, the density of eigenvalues is a continuous density with support $]0,b]$
that vanishes at $b$ but diverges at the origin, see Eqn~\eqref{eq:reg2optDens1}.
We compute explicitly the rate function (cf. Eqn~\eqref{PhiII}):
\begin{equation}\label{eq:PhiII}
\boxed{
\begin{aligned}
\Phi_{II}(s)=-\frac{\ln b}{2}+\frac{81}{8 b^2}-\frac{15 \sqrt{3}}{4 b}
+\frac{7}{8}+\frac{1}{2}\ln\left(3 \sqrt{3}\right)\\
\textrm{with}\quad b=b(s)=\frac{9}{2} \sqrt{3} \left(1-\sqrt{1-\frac{16 s}{45}}\right)
\end{aligned}}
\end{equation}
$b=b(s)$ is the upper bound of the density support.

Regime {\bf II}
 is well defined  only for $\frac{315}{256}<s<\frac{45}{16}$.
Indeed $b(s)$ is real (and positive) only for $s<\frac{45}{16}$,
and
the density $\rho_c$ becomes negative for $s<\frac{315}{256}$, which is non-physical.

For $s<s_1$ with $s_1=\frac{315}{256}$, the density has a support
over $[a,b]$ with $a>0$, this is regime {\bf I}.
For $s>s_2$  ($s_2=5/2$ when $N\to\infty$), we find that
 there is actually another solution that becomes more stable:
this is a solution with one eigenvalue much larger than the other,
described by one single eigenvalue plus a continuous density, this is regime {\bf III}.
Therefore regime {\bf II} is only valid for $s_1<s<s_2$.
\\

As the distribution of the purity is highly peaked for large $N$, the mean value
of $\Sigma_2=s/N$ is also the most probable value. Thus
$\langle \Sigma_2 \rangle \sim \frac{\bar{s}}{N}$ for large $N$ where
$\bar{s}$ minimizes $\Phi_{II}(s)$, thus
$\frac{d\Phi_{II}}{ds}\Big|_{\bar{s}}=0$, cf Fig. \ref{fig:plotphi}.
We recover  the result of Sommers and \.{Z}yczkowski~\cite{SommersDens}
\footnote{They indeed showed that
$\langle \Sigma_2 \rangle=\frac{5 N^2+1}{2 N (N^2+2)}\sim
\frac{5}{2 N}$ for large $N$.}:
\begin{equation}
\boxed{\langle \Sigma_2 \rangle\sim\frac{\bar{s}}{N}\sim \frac{5}{2 N}\;\;\textrm{ as $N\to\infty$,}
\;\; \textrm{ie}\;\;  \bar{s}=\frac{5}{2}}
\end{equation}
Around its minimum
$\bar{s}=5/2$,
the rate function $\Phi_{II}$ has a quadratic behaviour $ \Phi_{II}(s)\sim
\frac{2}{25}\left(s-\frac{5}{2}\right)^2$ as $s\to \frac{5}{2}$, which means that the distribution of the purity
has a Gaussian behaviour
around its average:
\begin{equation}
\mathcal{P}\left(\Sigma_2=
\frac{s}{N}\right)\approx
\exp\left\{- \frac{2\, N^2}{25}\left(s-\frac{5}{2}\right)^2
\right\}\;\;\textrm{for $s$ close to $\frac{5}{2}$}
\end{equation}
Again, as the distribution is highly peaked for large $N$,
the variance of $\Sigma_2$ is given to leading order in $N$
by the variance of the Gaussian  describing the neighbourhood
 of the mean value:
\begin{equation}
\boxed{{\rm Var}\; \Sigma_2=
\langle \Sigma_2^2\rangle -\langle \Sigma_2\rangle^2
\sim \frac{25}{4\, N^4}\;\;\textrm{ as $N\to\infty$}}
\end{equation}
Our result is in agreement with \cite{SommersPur}, where the authors find for the first moments\footnote{
In this paper the general formula for $\mu_k$ is correct but
in the application to $k = 2$, there is a small mistake, one should read $\frac{5 (5 N^4 +47 N^2+32)}{
4 (N^2+2)(N^2+4)(N^2+6)}$ instead of $\frac{5 (5 N^4 +47 N^2+32)}{
2 (N^2+2)(N^2+4)(N^2+6)}$.}
$\mu_1 \equiv \langle \Sigma_2 \rangle=
\frac{5 N^2+1}{2 N (N^2+2)}$ and
$\mu_2\equiv \langle \Sigma_2^2 \rangle=\frac{5 (5 N^4 +47 N^2+32)}{
4 (N^2+2)(N^2+4)(N^2+6)}$.
With their results, one obtains for large $N$:
\begin{equation}
{\rm Var}\,\Sigma_2 = \mu_2 -\mu_1^2=
\frac{25 N^6-71 N^4+70 N^2-24}{
4 N^2 (N^2+2)^2 (N^2+4)(N^2+6)}
\sim \frac{25}{4\, N^4}
\end{equation}
Since the eigenvalues are typically of order
$1/N$ for large $N$, 
the density of eigenvalues of $\sigma$ (for fixed purity $s$)
 is expected to be of the form $\rho_N(\lambda)=N\rho_c(\lambda N)$ where $\rho_c$ is the rescaled density
independent of $N$ for large $N$.
In regime {\bf II} the rescaled density $\rho_c$ 
has a finite support $]0,b]$.
It is explicitly given by:
\begin{equation}\label{eq:reg2optDens1}
\begin{aligned}
\rho_c(x)=&\frac{\alpha}{\pi} \left\{\left(\frac{b}{x}-\sqrt{\frac{b^2}{x^2}-1}\right)^{\frac{2}{3}}-
\left(\frac{b}{x}+\sqrt{\frac{b^2}{x^2}-1}\right)^{\frac{2}{3}}\right\}\\
& +\frac{\beta}{\pi} \;\frac{x}{b}\;  \left\{
\left(\frac{b}{x}-\sqrt{\frac{b^2}{x^2}-1}\right)^{\frac{5}{3}}
-\left(\frac{b}{x}+\sqrt{\frac{b^2}{x^2}-1}\right)^{\frac{5}{3}}\right\}
\end{aligned}
\end{equation}
with
$\alpha
=\frac{3}{4 b^2}(45 \sqrt{3}-16 b)$ and
$ \beta
 =\frac{27}{5 b^2}(b-3 \sqrt{3})$
where the upper bound of the density support
is given by $b=b(s)=\frac{9}{2} \sqrt{3} \left(1-\sqrt{1-\frac{16 s}{45}}\right)$.
The density vanishes at $b$ but diverges at the origin:
\begin{equation}\begin{aligned}
\rho_c(x) \sim\frac{c_1}{x^{2/3}}\quad\textrm{as $x\to 0$}, \qquad
\rho_c(x) \sim c_2\, \sqrt{b- x} \quad\textrm{as $x\to b$}
\end{aligned}
\end{equation}
where $c_1=\frac{3 \left(-9 \sqrt{3}+8 b\right)}{10\ 2^{1/3} b^{4/3} \pi  }$ and
$c_2=\frac{\left(-9 \sqrt{3}+2 b\right) \sqrt{2}}{b^2 \pi  \sqrt{b}}$. A plot of the density for $s=2$ (thus within regime {\bf II}) is shown
in Fig. \ref{fig:dens} (blue solid line).

 \subsubsection{Regime {\bf III}}

 Regime {\bf III} corresponds to higher values of the purity, $s> s_2$
 where $s_2=\frac{5}{2}\left[1+\left(\frac{5}{2 N}\right)^{\frac{1}{3}}+...\right]$ for large $N$.
 Larger values of $s$ correspond to purer states.

 In regime {\bf III} (for $s>s_2$), there is one eigenvalue, the maximal one $\lambda_{\rm max}$
 that is much larger than the other eigenvalues.
 Typically  one has in this regime $\lambda_{\rm max}\asymp \frac{1}{\sqrt{N}}$
 whereas $\lambda_{i}\asymp\frac{1}{N}$ for $i\neq {\rm max}$.
 The density is
 thus made of two parts: a continuous density with finite support $]0,b]$
 describing the $(N-1)$ smallest eigenvalues plus a delta peak at $\lambda=\lambda_{\rm max}$.
 
Very large values of $s$ correspond to purer states. In particular the limit $S=s/N\to 1$ is reached when
 the system is in a pure state,  i.e. when there is no interaction (thus no entanglement) with
 the environment.
This limit corresponds to the case where only one eigenvalue is non-zero, i.e.
 $\lambda_{i_0}=1$ and $\lambda_j =0$ for $j\neq i_0$.
Therefore the maximal eigenvalue is expected
 to become larger and larger when $s$ increases, but what is more surprising is that there is one point, $s=s_2$
 where the maximal eigenvalue suddenly jumps from a value of order $1/N$ to a much larger value of order
  $1/\sqrt{N}$.
  This transition is very similar to the one observed in the Hilbert-Schmidt case between
  the second and third regime,
  cf. ref. \cite{BElong}.
\\

In regime {\bf III}, we show that the rate function has a different scaling with $N$,
i.e. $\Phi(s)=\frac{\Phi_{III}(s)}{\sqrt{N}}$ with
\begin{equation}
\boxed{\Phi_{III}(s)=\frac{\sqrt{s-5/2}}{2}=\frac{\sqrt{s-\bar{s}}}{2}}
\end{equation}
  The scaling $N^{3/2}$ and the expression
of $\Phi_{III}$ are actually the same as in the Hilbert-Schmidt case~\cite{BElong}
(except for the fact that $\bar{s}=2$ instead of $5/2$).

We have already seen that the solution of regime {\bf II}
is well defined for $s_1<s<\frac{45}{16}$ with $s_1=\frac{315}{256} < \bar{s}$
($\bar{s}=5/2$). Regime {\bf III} is well defined for $s>\bar{s}$.
On the range $\bar{s}<s<\frac{45}{16}$ the two solutions thus
coexist. For large $N$, the stable solution is the one with lowest energy, i.e. lowest rate function $\Phi$,
cf. Coulomb gas interpretation section \ref{sec:Coulombgas}, see also
Fig. \ref{fig:plotphi}.
The transition between {\bf II} and {\bf III} thus occurs
at $s=s_2$ such that $N^2 \Phi_{II}(s_2)=N^{3/2} \Phi_{III}(s_2)$.
We find:
\begin{equation}
 s_2= \frac{5}{2}\left[1+\left(\frac{5}{2 N}\right)^{\frac{1}{3}}+...\right]\;\;\;\;\textrm{as $N\to\infty$}
\end{equation}
And at this value, $N^2\Phi_{II} = 2^{-5/3}5^{2/3}\,N^{4/3}$. For very large $N$, the transition occurs sharply at $s_2=\bar{s}=5/2$

At the transition point, the rate function $\Phi$ is continuous but its first derivative is discontinuous.
We have indeed: $\frac{d\Phi}{ds}\Big|_{s_2^+}=
\frac{d \Phi_{II}}{d s}\Big|_{s_2}\sim \left(\frac{2}{5}\right)^{2/3} \left(\frac{1}{N}\right)^{1/3}$
and $\frac{d\Phi}{ds}\Big|_{s_2^-}=\frac{1}{\sqrt{N}}\frac{d\Phi_I}{d s}\Big|_{s_2}\sim \frac{1}{ 2^{4/3} 5^{2/3}}\, \left(\frac{1}{N}\right)^{1/3}$
as $N\to\infty$.

%In this regime, 
%the maximal eigenvalue is much larger than the other eigenvalues.
%The density is thus made of a continuous part $\rho(\lambda)=N\rho_c(\lambda N)$ where the rescaled density
%$\rho_c$ has a finite support $[0,b]$, plus a delta peak describing the isolated maximal eigenvalue
%$\lambda_{\rm max}$.
%We show that in this regime the maximal
%eigenvalue is of order $\lambda_{\rm max}\asymp 1/\sqrt{N}$ whereas the other eigenvalues are
%still of order $1/N$
%(as in regimes {\bf I} and {\bf II}).
%%%%%%%%%%%%%%%%%%%%%%%

 \subsubsection{Regime {\bf I}}

\begin{figure}\begin{center}
\includegraphics[width=0.8\textwidth]{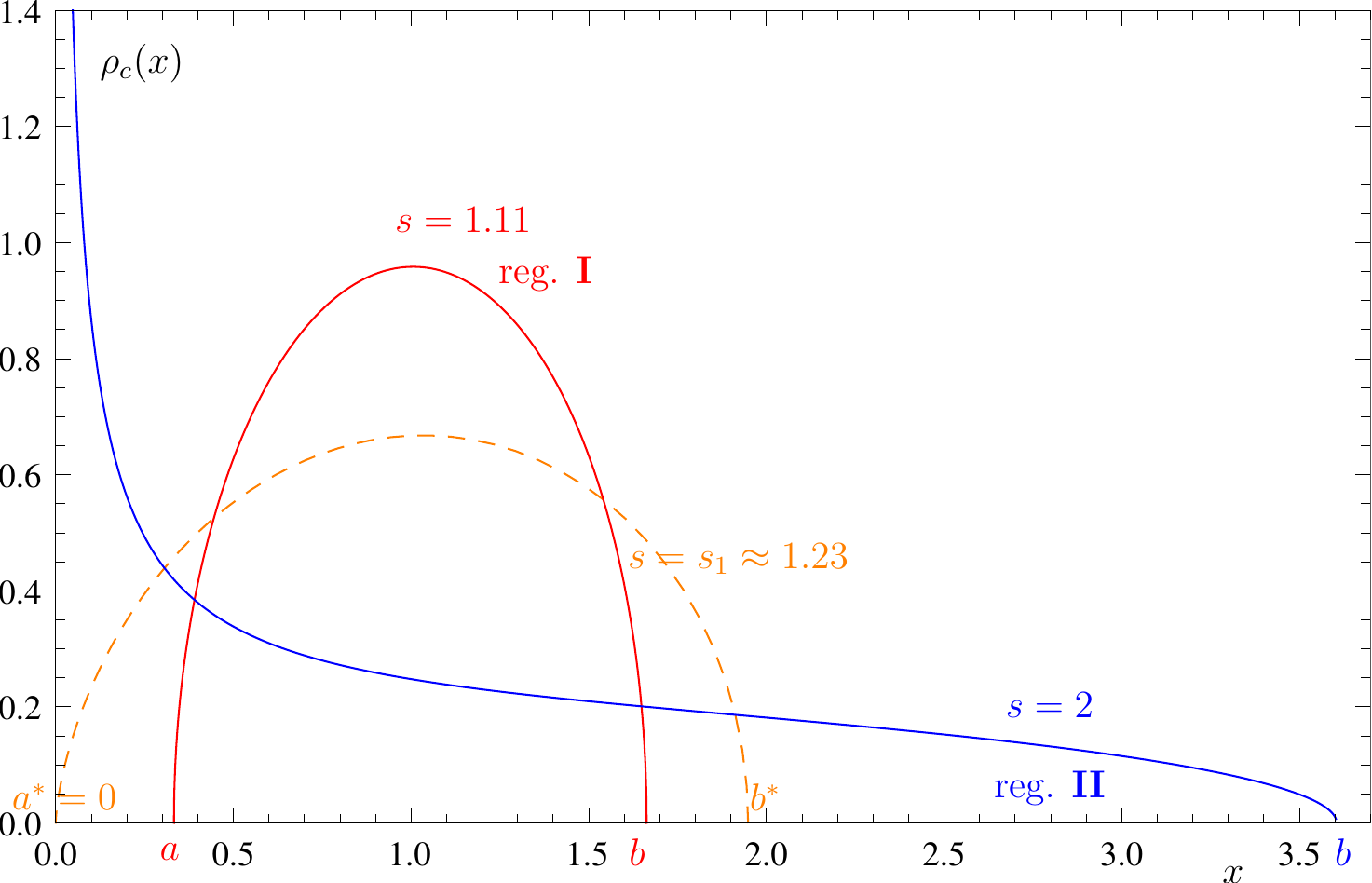}
\caption{Plot of the (rescaled) density $\rho_c(x)$
at fixed purity $\Sigma_2=s/N$ for a random Bures state ($n=1$, $\mu=2/3$), for $s=2$ 
$s=1.23$ and $s=1.11$. The blue solid line corresponds to $s=2$ within regime {\bf II}: the density
has a support $]0,b]$  and diverges at the origin.
The red solid line corresponds to $s=1.11$ (or $k=0.2$) within regime {\bf I}: the density has a finite support
over $[a,b]$ and vanishes at $a$ and $b$.
The orange dashed line is the density at the transition point $s=s_1=\frac{315}{256}\approx 1.23$
between {\bf I} and {\bf II}.
Here the exact expressions of the density are used to plot this figure.
}\label{fig:dens}\end{center}
\end{figure}

 Regime {\bf I} corresponds to small values of the purity, $1<s\leq s_1$
 where $s_1=\frac{315}{256}$.
 The limit $s\to 1$, i.e. $\Sigma_2\to 1/N$ describes a completely mixed state, thus a state
 that is maximally entangled with the environment.
 In regime {\bf I}, the density is a continuous density with finite support $[a,b]$
 with $0<a<b$ that vanishes at $a$ and $b$, cf Fig. \ref{fig:dens}
 (red solid line).

 In this regime, the computations are more involved.
 However we are able to compute
 explicitly the Stieltjes transform of the density $W(z)=\int_{a}^b \mathrm{d}x\, \frac{\rho_c(x)}{z-x}$
for $z\in\mathbb{C}\setminus [a,b]$,
  (and then we can come back to the density): we get a quite complicated expression
  given by a sum of elliptic functions, see below Eqn~\eqref{eq:Wresume}.
We compute the solution parametrically: all the parameters of the problem including
the rescaled purity $s$ are expressed in terms of $k=a/b$, cf. Eqn~\eqref{eq:t12bs}
with $\mu=2/3$, $n=1$.
The expression for the rescaled purity $s$ as a function of $k$ can in principle be inverted
to get the parameter $k$ as a function of $s$. Thus all the parameters are implicitly functions
of $s$.

   We obtain
the following expression for the rate function $\Phi_{I}(s)$:
\begin{equation}\begin{aligned}
&\Phi_{I}(s)=-\frac{t_0}{2}-\frac{t_1}{2}-\frac{t_2\, s}{4}-\frac{1}{4}-\frac{\ln 2}{2}\quad
\textrm{where}
\;\; t_0=-
t_1 b
-t_2 \frac{b^2}{2}+\ln b+2
I_1-I_2\\
&
\textrm{with}\;\;
 I_1=\int_b^{\infty} \mathrm{d}z \left(\frac{1}{z}-W(z)\right), \;\;
I_2
=\int_{-\infty}^{-b} \mathrm{d}z \left(W(z)-\frac{1}{z}\right)
\end{aligned}
\end{equation}
and where $t_1$, $t_2$, $b$  are expressed as functions of $k$ in Eqn~\eqref{eq:t12bs},
and thus implicitly functions of $s$.
The Stieltjes transform of the density is given by
\begin{equation}\label{eq:Wresume}
W(z)=\frac{t_1}{3}+t_2 z + A\;\chi_0(z) + B\; \chi_1(z)
\end{equation}
where
$A = -t_2$ , $B = -t_1 + t_2\,\gamma\,\sqrt{3} $ with $\gamma$ given as a function of $k$ in Eqn~\eqref{eq:clambdagamma}
with $\mu=2/3$, $n=1$,
and the special functions $\chi_0$ and $\chi_1$ are given in Eqn~\eqref{eq:chi01}
with $\mu=2/3$, $n=1$, they are expressed
in terms of elliptic theta functions.
\\

At the critical point $s=s_1=\frac{315}{256}$, the rate function $\Phi(s)$
has  a weak non-analyticity. It is continuous and twice differentiable with
continuous first and second derivatives, but its third derivative does not exist.
We show that:
%\begin{equation}\begin{aligned}
%E_s[\rho_c]&=\frac{11}{24}+2 \ln 2-\frac{\ln 3}{2}+\frac{75 q^{\frac{2}{3}}}{2}-246 q
%+\frac{7101 q^{\frac{4}{3}}}{4}-\frac{62352 q^{\frac{5}{3}}}{5}+O(q^2)\\
%&= \frac{11}{24}+2 \ln 2-\frac{\ln 3}{2}+\frac{64\,  \delta }{27}+\frac{8192 \, \delta^2}{2025}
%+\frac{16777216 \sqrt{2}\,  \delta^{5/2}}{11390625}+O(\delta^3)\\
%&\qquad\qquad\;\; \textrm{where}\;\; \delta=s_1-s=\frac{315}{256}-s\to 0^+
%\end{aligned}
%\end{equation}
%Compare with regime {\bf II}, cf. Eqn~\eqref{eq:EsrhocReg2}:
%$E_s[\rho_c]=
%-\frac{\ln b}{2}+\frac{81}{8 b^2}-\frac{15 \sqrt{3}}{4 b}+\frac{9}{8}+3 \frac{\ln 3}{4}+\frac{\ln 2}{2}$
%with $b=b(s)=\frac{9}{2} \sqrt{3} \left(1-\sqrt{1-\frac{16 s}{45}}\right)$.
%Thus:
%\begin{equation}\begin{aligned}
%E_s[\rho_c]&= \frac{11}{24}+2 \ln 2
%-\frac{\ln 3}{2}+\frac{64\,  \delta }{27}+\frac{8192 \, \delta^2}{2025}+\frac{4194304 \, \delta^3}{820125}+O(\delta^4)\\
%& \qquad \;\;
%\textrm{as $\delta=s_1-s=\frac{315}{256}-s\to 0^-$}
%\end{aligned}
%\end{equation}
\begin{equation}
\boxed{\Phi_{I}(s)-\Phi_{II}(s)\sim \frac{16777216 \sqrt{2}\, }{11390625}\, (s_1-s)^{5/2} \;\;\; \textrm{as $s\to s_1^-$}}
\end{equation}
More precisely $\Phi_{II}(s_1)=\Phi_I(s_1)$ and for $\delta=s_1-s=\frac{315}{256}-s$
we get:
\begin{equation}\begin{aligned}
\Phi_{II}(s)-\Phi_{II}(s_1)&=\frac{64\,  \delta }{27}+\frac{8192 \, \delta^2}{2025}+\frac{4194304 \, \delta^3}{820125}+O(\delta^4)
\;\;\textrm{as $\delta\to 0^-$}\\
\Phi_{I}(s)-\Phi_I(s_1) &=\frac{64\,  \delta }{27}+\frac{8192 \, \delta^2}{2025}
+\frac{16777216 \sqrt{2}\,  \delta^{5/2}}{11390625}+O(\delta^3)
\;\, \textrm{as}\; \delta\to 0^+
\end{aligned}
\end{equation}

In the limit $s\to 1^+$, which corresponds (for $\sigma$) to completely mixed states, we show:
\begin{equation}
\Phi_I(s)=-\frac{1}{2}\ln (s-1)+O(1)
\end{equation}
The distribution of the purity $\Sigma_2$ vanishes when $s\to 1$ as:
\begin{equation}\label{eq:lims1}
\mathcal{P}\left(\Sigma_2=\frac{s}{N},N\right)\propto \left(s-1\right)^{\frac{N^2}{2}} \;\; \textrm{as $s\to 1^+$}
\end{equation}
The probability that $\sigma$ is almost completely mixed is very small.
Note that this $s \rightarrow 1$ behaviour of the distribution of $\Sigma_2$ for random Bures states is exactly the same as the one of the distribution of $\Sigma_2$ for random Hilbert-Schmidt states \cite{BElong}.

   %%%%%%%%%%%%

\section{Coulomb gas method}
\label{sec:Coulombgas}

The goal is to compute the probability density function of the purity $\Sigma_2=\sum_i\lambda_i^2$
for generalized random Bures states.
By definition it is given by:
\begin{equation}\label{eq:pdfS2def}
\boxed{\mathcal{P}(\Sigma_2=S,N)=\int_{0}^{\infty}\mathrm{d}\lambda_1...\int_{0}^{\infty}\mathrm{d}\lambda_N\; \delta\left(\sum_i\lambda_i^2-S\right)\;
\mathcal{P}_{B,n}(\lambda_1,...,\lambda_N)}
\end{equation}
where the distribution $\mathcal{P}_{B,n}(\lambda_1,...,\lambda_N)$ of the generalized Bures eigenvalues
(the usual Bures case corresponds to $n=1$)
is given in Eqn~\eqref{eq:genBures}:
\begin{equation}\label{eq:distriEVbures1}\begin{aligned}
\mathcal{P}_{B,n}(\lambda_1,...\lambda_N)
 = D_N\, \delta\left( \sum_i \lambda_i-1\right) \,
 \frac{\prod_{j<k} (\lambda_{j}-\lambda_{k})^2}{\prod_{j,k} (\lambda_{j}+\lambda_{k})^{\frac{n}{2}}}
 \end{aligned}
\end{equation}

This computation can be done for large $N$ with a Coulomb gas method, that we explain in this section. This technique is well-known in of random matrix theory and has been used in various contexts such as the distribution of the maximal eigenvalue
of Gaussian or Wishart random matrices \cite{Dean2006,MajumdarVer2009,Vivo2007}, non-intersecting Brownian
interfaces~\cite{Nadal2009}, quantum transport in chaotic cavities~\cite{Vivo2008}, phase transition in the restricted trace ensemble \cite{Akemann1999} or the index distribution for Gaussian random fields \cite{Bray2007,Fyodorov2007} and Gaussian random matrices \cite{Majumdar2009}.

\subsection{Coulomb gas}

The idea is to see the joint distribution $\mathcal{P}_{B,n}(\lambda_1,...\lambda_N)$ in Eqn~\eqref{eq:distriEVbures1}
as a Boltzmann weight:
\begin{equation}\label{eq:boltz}\begin{aligned}
&\mathcal{P}_{B,n}(\lambda_1,...\lambda_N)
 %= D_N\, \delta( \sum_i \lambda_i-1) \,
 %\frac{\prod_{j<k} (\lambda_{j}-\lambda_{k})^2}{\prod_{j,k} (\lambda_{j}+\lambda_{k})^{\frac{1}{2}}}
 =\frac{1}{Z_N} e^{-E_N[\{\lambda_i\}]} \qquad \textrm{where}  \;\;\\
 &  \qquad E_N[\{\lambda_i\}]=-2 \sum_{j<k}\ln|\lambda_j-\lambda_k|+\frac{n}{2}\sum_{j,k}\ln|\lambda_j+\lambda_k|\quad
 \textrm{with $\sum_i\lambda_i=1$}
 \end{aligned}
\end{equation}

The $\lambda_i$ can thus be seen as the positions of $N$ particles on a line ($\lambda_i\in\mathbb{R}_+$)
with effective energy $E_N[\{\lambda_i\}]$ (and partition function $Z_N$).
Then
$\mathcal{P}_{B,n}(\lambda_1,...\lambda_N)$  represents the probability to find these particles at positions
$\lambda_i$ in the canonical ensemble at finite temperature (temperature $1$ with $k_B=1$).
The energy $E_N$ consists in a two-body logarithmic interaction potential.
The part $-2 \sum_{j<k}\ln|\lambda_j-\lambda_k|$ is the Coulomb interaction in two dimensions.
It appears very often in random matrix theory (RMT).
The mapping from random matrix eigenvalues to a Coulomb gas problem is
well-known in random matrix theory~\cite{For} and has been recently used in a variety of
contexts.
% that include the distribution of the extreme eigenvalues of
%Gaussian and Wishart matrices~\cite{DM,vivo1,MV,KC}, purity partition
%function in bipartite systems~\cite{parisi}, nonintersecting Brownian
%interfaces~\cite{nadal1}, quantum transport in chaotic
%cavities~\cite{vivo2}, information and communication
%systems~\cite{kaz},  the index distribution for Gaussian random
%fields~\cite{BD,FW} and Gaussian matrices~\cite{nadal2},
%and the distribution of entropy for a bipartite random pure state~\cite{BE,BElong}.
\\

For large $N$, because of the constraint $\sum_{i=1}^N \lambda_i=1$ we typically expect
$\lambda_i \asymp \frac{1}{N}$ where $\asymp$ means that $\lambda_i$  is proportional to $1/N$
when $N\to\infty$.
Let us
 introduce the rescaled eigenvalues $x_i=\lambda_i N$ that are expected to be typically independent of $N$
for large $N$.
Therefore the typical scaling for the purity $\Sigma_2=\sum_i\lambda_i^2$
is expected to be $\Sigma_2\asymp \frac{1}{N}$ for large $N$.
Let us thus write $S=\frac{s}{N}$ in Eqn~\eqref{eq:pdfS2def}.
The equation \eqref{eq:boltz} can be rewritten as:
\begin{equation}\label{eq:boltz1}\begin{aligned}
&\mathcal{P}_{B,n}(\lambda_1,...\lambda_N)
 =\frac{1}{Z_N} e^{-E_N[\{\lambda_i\}]}=\frac{1}{\mathcal{Z}_N} e^{-\mathcal{E}_N[\{x_i\}]}\\
 &\textrm{where}  \;\;  \mathcal{E}_N[\{x_i\}]=-2 \sum_{j<k}\ln|x_j-x_k|+\frac{n}{2}\sum_{j,k}\ln|x_j+x_k|\quad
 \textrm{with $\sum_i x_i=N$}
 \end{aligned}
\end{equation}
and with $\mathcal{Z}_N= N^{\frac{N^2}{2}-N}\; Z_N$ the new partition function.
\\

Let us introduce the rescaled density $\rho(x)=\frac{1}{N}\sum_i \delta(x-x_i)$.
It is normalized to one by definition ($\int_0^{\infty}\mathrm{d}x\,\rho(x)=1$). When $N\to\infty$, we expect
the density
$\rho(x)$ to have a finite and continuous limit.
We will thus first assume that it is the case here.
It turns out a posteriori that this assumption can hold for a large range of values of the purity $S=s/N$.
However we will see that
 it cannot be true for large values of the purity. We will indeed see
that for $s>s_2$  for some $s_2$ (which will be shown to be $5/2$), there is one eigenvalue that becomes much larger than the other: the density
thus tends to a continuous part plus an isolated delta peak.
We will call regime {\bf III} the regime $s>s_2$ (see section \ref{sec:regIII}).

Let us first assume that $\rho(x)$ tends for large $N$ to a continuous limit.
This will actually  give the first two regimes (corresponding to $s<s_2$).
In this case,
 we expect from Eqn~\eqref{eq:boltz1} that $\mathcal{E}_N\asymp N^2$ for large $N$
 and more precisely\footnote{
Let us introduce the rescaled two-points correlation function
$\rho^{(2)}(x,x')=\frac{1}{N(N-1)}\sum_{i\neq j} \delta(x-x_i)\delta(x'-x_j)$.
$\rho^{(2)}(x,x')$ (exactly as $\rho(x)$) is expected to have a finite and continuous limit when $N\to\infty$.
Therefore we get
$\mathcal{E}_N =N(N-1)\int \mathrm{d}x\int \mathrm{d}x' \rho^{(2)}(x,x') (\ln|x+x'|-2\ln|x-x'|)+\frac{N}{2}\int \mathrm{d}x \rho(x)\ln|2x|$
thus
$\mathcal{E}_N\asymp N^2$ as $N\to\infty$.
Also for large $N$ by definition of $\rho^{(2)}$ one has:
$\rho^{(2)}(x,x')\sim \rho(x)\rho(x')$.}:
\begin{equation}\label{eq:EffEn}\begin{aligned}
&\mathcal{E}_N[\{x_i\}]\sim N^2
\int_0^{\infty}\mathrm{d}x \int_0^{\infty}\mathrm{d}x' \rho(x)\rho(x')\left(-\ln|x-x'|+\frac{n}{2}\ln|x+x'|\right)
\end{aligned}
\end{equation}
where $\sim$ means ``equivalent'' for large $N$.
The integrals run from $0$ to $+\infty$ as the eigenvalues of a density matrix are nonnegative, i.e.
$\lambda_i\geq 0$ for all $i$.
\\

The effective energy $\mathcal{E}_N\asymp N^2$ is large for large $N$, thus the multiple integral in
Eqn~\eqref{eq:pdfS2def} can be computed by a saddle point method. To leading order for large $N$,
only the minimal energy contributes to the integral. We get 
\begin{equation}\label{eq:saddleprop}
\mathcal{P}(\Sigma_2=\frac{s}{N},N)\approx \frac{1}{\mathcal{Z}_N} \; e^{-N^2 E_s[\rho_c]} \;\;\;\textrm{as $N\to\infty$}
\end{equation}
 ($\approx$ means logarithmic equivalent)
where $\rho_c$ minimizes the effective energy $E_s$ that is given by:
\begin{equation}\label{eq:EffEnS}\boxed{\begin{aligned}
E_s[\rho]&=\int_0^{\infty}\mathrm{d}x \int_0^{\infty}\mathrm{d}x' \rho(x)\rho(x')\left(-\ln|x-x'|+\frac{n}{2}\ln|x+x'|\right)\\
+&t_0 \left[\int_{0}^{\infty} \mathrm{d}x \rho(x) -1 \right]
+t_1 \left[\int_{0}^{\infty} \mathrm{d}x x\, \rho(x) -1 \right]
+\frac{t_2}{2} \left[\int_{0}^{\infty} \mathrm{d}x x^2\, \rho(x) -s \right]
\end{aligned}}
\end{equation}
where $t_{0,1,2}$ are Lagrange multipliers that have been added to take
into account three constraints in the minimization\footnote{The choice to write $t_2/2$ instead of $t_2$
is arbitrary.}, respectively
the normalization of the density $\int \rho=1$, the
unit sum of the eigenvalues $1=\sum_i\lambda_i=\int \mathrm{d}x x \rho(x)$
and the fixed value of the purity $\Sigma_2=s/N$, i.e. $\int \mathrm{d}x x^2\, \rho(x) = s$
(it replaces the delta function in Eqn~\eqref{eq:pdfS2def}).

\begin{figure}\begin{center}
\includegraphics[width=0.9\textwidth]{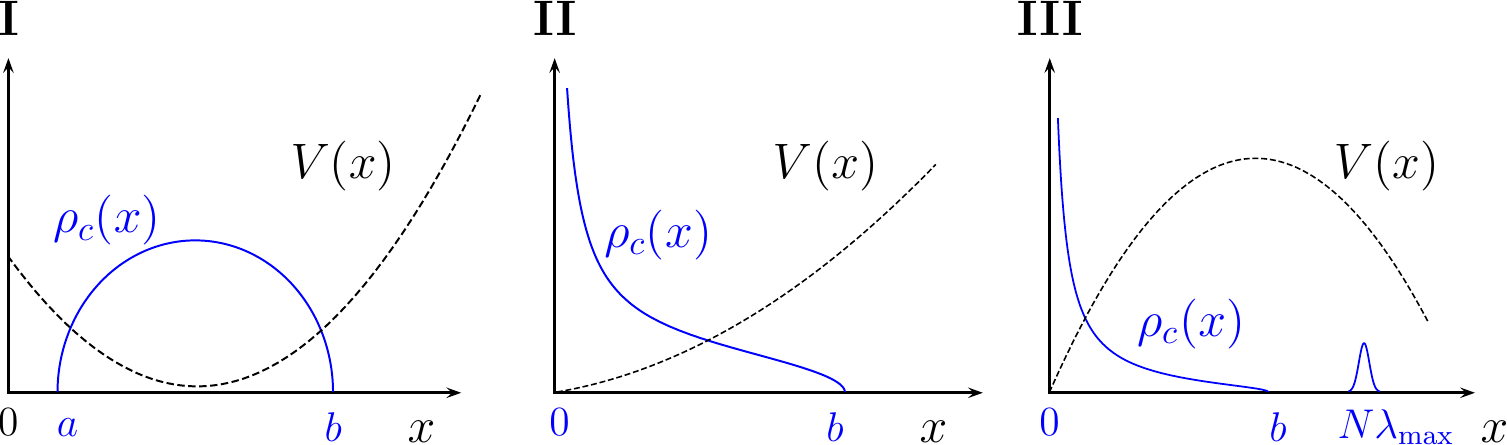}
\caption{Figure showing the (rescaled) density $\rho_c(x)$ and the effective potential $V(x)=t_0+t_1+t_2 x^2/2$
at fixed purity $\Sigma_2=s/N$ in the three regimes.
In regime {\bf I} (left), ie for $1<s<s_1$, the potential has an absolute minimum at a positive $x$,
and the density has a finite support over $[a,b]$ with $0<a<b$ (the Coulomb charges accumulate close to the minimum
of the potential).
In regime {\bf II} (middle), ie for $s_1<s<s_2$, the potential is monotonically increasing for $x>0$. The density has a support
on $]0,b]$ and diverges at the origin (the charges accumulate close to the origin).
In regime {\bf III} (right), ie for $s>s_2$, the potential is not anymore bounded from below.
The maximal eigenvalue becomes much larger than the other eigenvalues (one charge detaches) whereas the other eigenvalues
remain close to each other (described by a continuous density).
}\label{fig:potdens}\end{center}
\end{figure}

The effective energy can be formally written as
\begin{align}
&E_s[\rho]=\int_0^{\infty}\mathrm{d}x \int_0^{\infty}\mathrm{d}x' \rho(x)\rho(x') U_{\rm int}(x,x')
+\int_0^{\infty}\mathrm{d}x \rho(x) V(x)+e_0\\
& \textrm{ with} \quad U_{\rm int}(x,x')=-\ln|x-x'|+\frac{n}{2}\ln|x+x'|\;\;, \quad
V(x)=t_0+t_1 x+t_2 \frac{x^2}{2}\nonumber
\end{align}
and $e_0=-t_0-t_1-\frac{t_2 s}{2}$.
The charges of the Coulomb gas (that interact with the two body-interaction potential $U_{\rm int}$ as in Eqn~ \eqref{eq:boltz}) thus live now in an effective external potential
$V(x)=t_0+t_1 x+t_2 \frac{x^2}{2}$. The shape of this potential depends on the Lagrange multipliers that
are not fixed yet, and that will be functions of the rescaled purity $s$.
 Depending on the value of the rescaled purity $s$, $V$ will have different shapes
which will lead to phase transitions for the Coulomb gas and thus to
different regimes for the purity distribution.
These phase transitions are shown in Fig. \ref{fig:potdens}.
\\

In Eqn~\eqref{eq:saddleprop}, 
 the partition function $\mathcal{Z}_N$
  can be computed by using the same Coulomb gas method as above but
without the constraint on the purity.
Thus $\mathcal{Z}_N \approx e^{-N^2 E[\rho^*]}$ where
 $\rho^*$  minimizes the effective energy without the constraint on the purity, $E[\rho]$ given by:
 \begin{equation}\label{eq:EffEnsansS}\begin{aligned}
E[\rho]=&\int_0^{\infty}\mathrm{d}x \int_0^{\infty}\mathrm{d}x' \rho(x)\rho(x')\left(-\ln|x-x'|+\frac{n}{2}\ln|x+x'|\right)\\
&+t_0 \left[\int_{0}^{\infty} \mathrm{d}x\,\rho(x) -1 \right]
+t_1 \left[\int_{0}^{\infty} \mathrm{d}x\,x\, \rho(x) -1 \right]
\end{aligned}
\end{equation}
To leading order for large $N$, it is not very difficult to see that  $E[\rho^*]$ is actually equal
to $E_{s}[\rho_c]$ evaluated at $s=\bar{s}$ where $\bar{s}$
 is the mean value of the rescaled purity\footnote{We will show later that $\bar{s}=5/2$.}
(or also its most probable value as the distribution is highly peaked for large $N$), thus $E[\rho^*]=E_{\bar{s}}[\rho_c]$ and:
\begin{equation}\label{eq:pdfSigma2Norm}
\boxed{\mathcal{P}\left(\Sigma_2=\frac{s}{N},N\right)\approx e^{-N^2 (E_s[\rho_c]-E_{\bar{s}}[\rho_c])}\quad\textrm{as $N\to\infty$}
}
\end{equation}
where $\approx$ means ``logarithmic equivalent''.%\footnote{The precise mathematical statement is
%$\ln\mathcal{P}(\Sigma_2=\frac{s}{N},N)\sim -N^2 (E_s[\rho_c]-E[\rho^*])$ where $\sim$ means equivalent for large $N$.}.
\\

We need to determine the optimal density
$\rho_c$ that minimizes  $E_s[\rho]$ in Eqn~\eqref{eq:EffEnS}. $\rho_c$ must satisfy
 $\left.\frac{\delta E_s[\rho]}{\delta \rho(x)}\right|_{\rho=\rho_c}=0$:
\begin{equation}\label{eq:saddle}\begin{aligned}
2\int_0^{\infty}\mathrm{d}x' \rho_c(x')\left(\ln|x-x'|-\frac{n}{2}\ln|x+x'|\right)=
t_0
+t_1 x
+t_2 \frac{x^2}{2}\;\;, \quad x\in {\rm Supp}[\rho_c]
\end{aligned}
\end{equation}
where $ {\rm Supp}[\rho_c]$ is the support of the density $\rho_c$, $ {\rm Supp}[\rho_c] \subset \mathbb{R}_+$.
Differentiating with respect to $x$ we get:
\begin{equation}\label{eq:saddle1}\boxed{\begin{aligned}
\fint_0^{\infty}\mathrm{d}x' \frac{\rho_c(x')}{x-x'}-\frac{n}{2}\int_0^{\infty}\mathrm{d}x' \frac{\rho_c(x')}{x+x'}=
%\frac{\mu_1}{2}+\mu_2 x
\frac{t_1+t_2 x}{2}\;\;, \quad x\in {\rm Supp}[\rho_c]
\end{aligned}}
\end{equation}
where $\fint$ is the principal value of the integral\footnote{Principal value:
$\fint_0^{\infty}\mathrm{d}x' \frac{\rho_c(x')}{x-x'}=\lim_{\epsilon\to 0}\left[
\int_0^{x-\epsilon}\mathrm{d}x' \frac{\rho_c(x')}{x-x'}+\int_{x+\epsilon}^{\infty}\mathrm{d}x' \frac{\rho_c(x')}{x-x'}
\right]$}.

We need to find the solution $\rho_c(x)$ of Eqn~\eqref{eq:saddle1}
that is positive on its support and satisfies the three constraints
$\int_{0}^{\infty} \mathrm{d}x \rho(x) =1$,
$\int_{0}^{\infty} \mathrm{d}x\,x\, \rho(x) =1$ and
$\int_{0}^{\infty} \mathrm{d}x\,x^2\, \rho(x) =s$ for a given $s>1$.

The support of the density $ {\rm Supp}[\rho_c] \subset \mathbb{R}_+$ must be finite.
Otherwise, by taking the limit $x\to \infty$ in  Eqn~\eqref{eq:saddle1} we get a contradiction.
The most natural finite support we can imagine is a finite segment $[a,b]$.
Let us thus assume that the density support is of the form $ {\rm Supp}[\rho_c]=[a,b]\subset \mathbb{R}_+$.
If $a>0$,
 we expect by continuity that the density vanishes at the upper and lower bounds of its support,
 i.e. $\rho_c(a)=0=\rho_c(b)$. We will see that
 this is valid for $1<s<s_1$ with $s_1=\frac{315}{256}$, which we will call regime {\bf I}.
 Another possibility is that $a=0$. In this case the density only needs to vanish at $b$, i.e. $\rho_c(b)=0$.
 This is actually valid for $s_1<s<s_2$ as we will see (with $s_2\sim\frac{5}{2}$ for $N\to\infty$)
 and we will call this regime regime {\bf II}.

Before coming to the explicit derivation of regime {\bf I} (section \ref{sec:regI}) andregime {\bf II}
 (section \ref{sec:regII}), let us present the general setting
 that we will use in both regimes to solve Eqn~\eqref{eq:saddle1}.

\subsection{Resolvent and effective energy}
\label{subsec:generalSolResEn}

In the next sections
(sections \ref{sec:regI} and \ref{sec:regII}), we will explain how to find the density $\rho_c$ solution of  Eqn~\eqref{eq:saddle1}
in regimes {\bf I} and {\bf II} i.e. for a density with support $[a,b]$ with $0\leq a <b$.
The method for the two regimes is quite different but uses in both cases complex analysis.
In both cases it is easier to compute  the resolvent (that is the Stieltjes transform of the density)
instead of the density. From the expression of the resolvent, one can then come back to the density.
However the minimal energy $E_s[\rho_c]$ (and thus the rate function $\Phi$)
can be computed directly from the resolvent (we do not really need the density), cf. Eqn~\eqref{eq:minE} below.

\subsubsection{Resolvent}
\label{subsubsec:resolvent}

 By definition the resolvent is the Stieltjes transform of the density:
\begin{equation}\label{eq:defRes}
\boxed{W(z)=\int_0^{\infty} \mathrm{d}x \; \frac{\rho_c(x)}{z-x}\;\;, \quad z\in \mathbb{C}\setminus [a,b]}
\end{equation}
where $[a,b]$ is the density support.

For a continuous density $\rho_c$ with support $[a,b]$
satisfying Eqn~\eqref{eq:saddle1} and the three constraints
$\int \rho_c=1$, $\int \mathrm{d}x\,x \rho_c=1$ and $\int \mathrm{d}x\,x^2 \rho_c=s$,
 the resolvent has the following properties:
\\
1/ $W(z)$ is analytic on $\mathbb{C}\setminus [a,b]$ with a cut on $[a,b]$.
\\
2/ The jump of $W(z)$ when passing through the cut is given by the density:
\begin{equation}
\rho_c(x)=\frac{1}{2 i \pi}\left(W(x-i 0^+)-W(x+i 0^+)\right)=\frac{-1}{\pi}{\rm Im} W(x+i 0^+)\;\;, \quad x\in[a,b]
\end{equation}
3/ The asymptotic behaviour of $W(z)$ for $|z|\to\infty$ is the following\footnote{We have indeed
$W(z)=\frac{1}{z}\int_0^{\infty} \mathrm{d}x \rho_c(x)+\frac{1}{z^2}\int_0^{\infty} \mathrm{d}x \, x \rho_c(x)
+\frac{1}{z^3}\int_0^{\infty} \mathrm{d}x \, x^2 \rho_c(x)+...$}:
\begin{equation}\label{eq:asymptW}
W(z)=\frac{1}{z}+\frac{1}{z^2}+\frac{s}{z^3}+O\left(\frac{1}{z^4}\right)\;\; \textrm{as $|z|\to\infty$}
\end{equation}
4/ $W(z)$ satisfies the following equation%\footnote{It comes from Eqn~\eqref{eq:saddle1}
%and because $W(x+i 0^+)+W(x-i 0^+)=2 {\rm Re} W(x+i 0^+)=
%2 \fint_0^{\infty}\mathrm{d}x' \frac{\rho_c(x')}{x-x'}$}:
\begin{equation}\label{eq:On1}
\boxed{W(x+i 0^+)+W(x-i 0^+)+ n W(-x)=t_2 x+t_1\;\;, \quad x\in[a,b]}
\end{equation}
\newline

Eqn~ \eqref{eq:On1} corresponds to the equation of the $O(n)$ model
for a potential $ V'(x)=t_2 x+t_1$. %It can be rewritten:
%\begin{equation}\label{eq:OnBis}
%{\rm Re}\, W(x+i 0^+)+ \frac{n}{2} W(-x)= \frac{V'(x)}{2}=\frac{t_2 x+t_1}{2}\;\;, \quad x\in[a,b]
%=t_2 x+t_1
%\end{equation}
The usual Bures case corresponds to $n=1$.

\subsubsection{Minimal energy}

We want to compute  the distribution of the purity
for large $N$ or equivalently the minimal energy $E_s[\rho_c]$ (cf. Eqn~\eqref{eq:pdfSigma2Norm}).
 $E_s[\rho_c]$ is given by Eqn~\eqref{eq:EffEnS}
computed for $\rho=\rho_c$. The optimal density $\rho_c$ satisfies the three constraints
$\int \rho_c=1$, $\int x \rho_c=1$, $\int x^2 \rho_c=s$, thus we have
$E_s[\rho_c]=\int_0^{\infty}\mathrm{d}x \int_0^{\infty}\mathrm{d}x' \rho_c(x)\rho_c(x')(-\ln|x-x'|+\frac{n}{2}\ln|x+x'|)$.
Using  the saddle point equation
\eqref{eq:saddle} and the constraints again, we get
\begin{equation}\label{eq:minE}
\boxed{E_s[\rho_c]=-\frac{t_0}{2}-\frac{t_1}{2}-\frac{t_2\, s}{4}}
\end{equation}
To obtain the expression of the Lagrange multipliers $t_{0,1,2}$ and thus the
energy $E_s[\rho_c]$, the steps are the following:
\\
$\bullet$ First we must find the solution $W(z)$ of Eqn~\eqref{eq:On1} for fixed
(but unknown)
$t_{1,2}$, $W(z)$ must be analytic
on $\mathbb{C}\setminus [a,b]$ with a cut on $[a,b]$.
In general, $W(z)$ will be of the form:
\begin{equation}
W(z)=h(z)+\overline{W}(z)
\end{equation}
where $h(z)$ is a particular solution of Eqn~\eqref{eq:On1} that we can choose as follows
\begin{equation}
\label{eq:solpart}
h(z)=\frac{2 V'(z)-n\, V'(-z)}{4-n^2}=\frac{t_1}{2+n}+\frac{t_2 z}{2-n}
\end{equation}
and where $\overline{W}(z)$ is a general solution (with one cut on $[a,b]$)
of the homogeneous equation:
\begin{equation}\label{EqHom}
 \overline{W}(x+i 0^+)+\overline{W}(x-i 0^+)+n \, \overline{W}(-x)=0\;\;{\rm for}\;\;
x\in [a,b]
\end{equation}
$\bullet$ Then we must impose the asymptotic behaviour of $W(z)$ (cf. Eqn~\eqref{eq:asymptW}). This is equivalent to impose
the constraints on the density $\rho_c$ ($\int\rho_c=1$, $\int x\rho_c =1$, $\int x^2\rho_c=s$).
We thus get the Lagrange multipliers $t_1$ and $t_2$ as well as the bounds of the density support
$a$ and $b$ as functions of the purity $s$.
\\
$\bullet$ We must also compute the last Lagrange multiplier $t_0$.
We show below that it can be expressed as a sum of integrals of $W(z)$, see Eqn~\eqref{eq:t0}.
\\
$\bullet$ Finally, knowing
$W(z)$, $t_1$, $t_2$ and also $t_0$ as functions of $s$,
 we get the expression of the energy $E_s[\rho_c]=-\frac{t_0}{2}-\frac{t_1}{2}-\frac{t_2\, s}{4}$.
 To recover the usual Bures case, we impose $n=1$.
\\
\\
\textbf{Computation of $t_0$:}

$t_0$ can be computed from Eqn~\eqref{eq:saddle} for $x=b$:
\begin{equation}
t_0=-
t_1 b
-t_2 \frac{b^2}{2}+
2\int_0^{\infty}\mathrm{d}x \rho_c(x)\left(\ln|b-x|-\frac{n}{2}\ln|b+x|\right)
\end{equation}
Moreover, for $z\notin [a,b]$, we have by definition:
 \begin{equation}
 W(z)=\int_{0}^{\infty} \mathrm{d}x \frac{\rho_c(x)}{z-x}=\frac{\mathrm{d}}{\mathrm{d}z}\int_{0}^{\infty} \mathrm{d}x \rho_c(x)\ln|z-x|
 \end{equation}
As $\rho_c$ has a support over $[a,b]$ we thus get $\int_{0}^{\infty} \mathrm{d}x \rho_c(x)\ln|z-x|=\int_{a}^{b} \mathrm{d}x \rho_c(x)\ln|z-x|$
and thus for $Z >b$:
\begin{equation}
\int_{a}^b \mathrm{d}x \rho_c(x)\ln|b-x|-\int_{a}^{b}\mathrm{d}x \rho_c(x)\ln|Z-x|=-\int_b^Z \mathrm{d}z W(z)
\end{equation}
or equivalently:
\begin{equation}
\int_{a}^b \mathrm{d}x \rho_c(x)\ln\left|\frac{b-x}{b}\right|
-\int_{a}^{b}\mathrm{d}x \rho_c(x)\ln\left|\frac{Z-x}{Z}\right|=-\int_b^Z \mathrm{d}z \left(W(z)-\frac{1}{z}\right)
\end{equation}
We can now take the limit $Z\to\infty$ above, we get:
\begin{equation}\label{eq:I1}
I_1\equiv\int_{a}^b \mathrm{d}x \rho_c(x)\ln\left|\frac{b-x}{b}\right|
=\int_b^{\infty} \mathrm{d}z \left(\frac{1}{z}-W(z)\right)
\end{equation}
Similarly one can show that:
\begin{equation}\label{eq:I2}
I_2\equiv\int_{a}^b \mathrm{d}x \rho_c(x)\ln\left|\frac{b+x}{b}\right|
=\int_{-\infty}^{-b} \mathrm{d}z \left(W(z)-\frac{1}{z}\right)
\end{equation}
therefore we get:
\begin{equation}\label{eq:t0}\boxed{\begin{aligned}
t_0&=-
t_1 b
-t_2 \frac{b^2}{2}+(2-n) \ln b+2
I_1-n I_2\\
\textrm{with}&\quad I_1
=\int_b^{\infty} \mathrm{d}z \left(\frac{1}{z}-W(z)\right)\;\;, \quad
I_2
=\int_{-\infty}^{-b} \mathrm{d}z \left(W(z)-\frac{1}{z}\right)
\end{aligned}}
\end{equation}

\section{Regime {\bf II}}
\label{sec:regII}

In this section, we assume that the density $\rho_c(x)$ solution of the saddle point equation
\eqref{eq:saddle1} has a finite support
of the form $[0,b]$, i.e. $a=0$.  The density is expected to be continuous, thus we must have $\rho_c(b)=0$.

As explained in section \ref{subsec:generalSolResEn}, we need to find
the resolvent $W(z)$ that must satisfy the conditions 1/-4/ of subsection  \ref{subsubsec:resolvent}.
The solution can be found using a method proposed by  Bueckner~\cite{Bueckner}
in the case of a polynomial potential $V(x)$
(here we have $V(x)=t_2 \frac{x^2}{2}+t_1 x+t_0$).
%To solve this equation, we apply Bueckner's method  (explained in \ref{subsec:buecknerPolyn})
%for the case of the explicit polynomial potential $g(x)=-V'(x)=-\frac{\mu_1}{2}-\mu_2\, x$
%and with $\gamma=1/2$, thus $\mu=2/3$.

\subsection{Resolvent: Bueckner's method~\cite{Bueckner}}
\label{subsec:buecknerPolyn}

%
%We consider the equation of the $O(n)$ model (cf. Eqn~\eqref{eq:OnBis})
%in the case of a density $\rho_c$ with support $[0,b]$:
%\begin{equation}\label{EqBuck}
%\fint_{0}^b dt \, \frac{\rho_c(t)}{x-t}
%-\frac{n}{2} \int_{0}^b dt \, \frac{\rho_c(t)}{x+t}=\frac{V'(x)}{2}
%\;\; {\rm for}\;\; x\in ]0,b[
%\end{equation}
%where $n \in ]-2,2[$
%and $V'(x)$ is a real polynomial of $x$.
%$\fint$ denotes the Cauchy principal value of an integral.
%Our physical case corresponds to
%$n=1$ and $V'(x)=t_1+t_2 x$ (polynomial of degree $1$).

We recall that the resolvent is defined as the Stieltjes transform of the density, cf. Eqn~\eqref{eq:defRes}.
%\begin{equation}
%W(z)=\int_0^{b} \mathrm{d}x \; \frac{\rho_c(x)}{z-x}\;\;, \quad z\in \mathbb{C}\setminus [0,b]
%\end{equation}
As we have seen in the previous section, the resolvent is of the form
$W(z)=h(z)+\overline{W}(z)$ where $h(z)$ is given in Eqn~\eqref{eq:solpart}
 and
where $\overline{W}(z)$ is a general one-cut solution (cut $[0,b]$) of
the homogeneous equation \eqref{EqHom}:
$ \overline{W}(x+i 0^+)+\overline{W}(x-i 0^+)+n \, \overline{W}(-x)=0$ for
$x \in [a,b]$.

The  solutions of the homogeneous equation Eqn~\eqref{EqHom}
form a module of dimension $2$ over the ring
%$R'$
of polynomials of $z^2$ with real coefficients.
In general, we have:
\begin{equation}
\overline{W}(z)=A(z^2)\, \phi_0(z)+B(z^2)\, \phi_1(z)
\end{equation}
where $A$ and $B$ are polynomials with real coefficients
and where $\{\phi_0,\phi_1\}$ is a basis of the module.
Let us make a change of variables by defining
$w=w(z)$ and $\mu$ (instead of $n$) as\footnote{In the computations we use very often the following relations
obtained from
$n=-2 \cos(\pi\mu)$:
\begin{align*} &\cos\left(\frac{\pi \mu}{2}\right)=\frac{\sqrt{2-n}}{2}\;,\qquad
\sin\left(\frac{\pi \mu}{2}\right)=\frac{\sqrt{2+n}}{2}\;,\qquad
\tan\left(\frac{\pi \mu}{2}\right)=\sqrt{\frac{2+n}{2-n}}
\end{align*}}:
\begin{equation}
\frac{z}{b}=\frac{1}{\sin w}\;\;
{\rm and}\;\; n=-2 \cos\left( \pi \mu \right)\;\; \mu \in ]0,1[
\end{equation}
The basis functions are then explicitly given by:
\begin{equation}\label{eq:defphi01}
\phi_0(z)=\frac{\cos\left[\mu\left(w+\frac{\pi}{2}\right)\right]}{
\cos\left[\frac{\mu \pi}{2}\right]}
\;\;
{\rm and}\;\;
\phi_1(z)=\frac{\sin\left[\mu\left(w+\frac{\pi}{2}\right)\right]}{
\tan w}
\end{equation}
When $|z|\to\infty$ we have $W(z)\sim\frac{1}{z}$, thus $W(z)$
tends to zero.
The polynomials $A$ and $B$ are  uniquely determined
by the condition $W(z)=\overline{W}(z)+h(z) \rightarrow 0$
as $|z|\rightarrow \infty$, where $h(z)$ is the polynomial
defined in Eqn~\eqref{eq:solpart}.
To determine $A$ and $B$ for a given polynomial potential $V'(x)$, we thus need to know
the asymptotic expansion of $\phi_0(z)$ and $\phi_1(z)$
at $\infty$. One can show that for $|z|\to\infty$:
\begin{equation}\label{eq:phi01asympt}
\begin{aligned}
\phi_0(z)&= 1-\mu \tan\left[\frac{\mu \pi }{2}\right]\frac{b}{z}
-\frac{\mu^2}{2}\frac{b^2}{z^2}+\frac{\mu (\mu^2-1)}{6}
\tan\left[\frac{\mu \pi }{2}\right]
\frac{b^3}{z^3}+O\left(\frac{1}{z^4}\right)\\
\phi_1(z)&= \sin\left[\frac{\mu \pi}{2}\right]\frac{z}{b}
+\mu \cos \left[\frac{\mu \pi}{2}\right]
-\left(\frac{\mu^2+1}{2}\right) \sin\left[\frac{\mu \pi}{2}\right] \frac{b}{z}
\\
&\;\; -\frac{\mu \left(\mu^2+2\right)}{6}
\cos\left[\frac{\mu \pi}{2}\right] \frac{b^2}{z^2}
+\frac{\left(\mu^4+2\mu^2-3 \right)}{24} \sin\left[\frac{\mu \pi}{2}\right]
\frac{b^3}{z^3}+O\left(\frac{1}{z^4}\right)
\end{aligned}
\end{equation}
\newline
$\bullet$ When $|z|\to \infty$, we must have $W(z)\to 0$.
Using the asymptotic expansion of $\phi_{0,1}$ in Eqn~\eqref{eq:phi01asympt},
we get $A$ and $B$:
\begin{equation}\label{eq:AB}
\begin{aligned}
&A=-\frac{t_1}{2+n}+\frac{b \mu t_2}{\sqrt{4-n^2}}\;\;\;\;,\;\;\;\;\;
&&B=-\frac{2 t_2 b}{(2-n)\sqrt{2+n}}\end{aligned}
\end{equation}
%%%%%%*********
We thus have an explicit expression for the resolvent $W(z)=h(z)+A \phi_0(z)+B \phi_1(z)$.
\\
$\bullet$ We now impose the constraints on $\rho_c$ or equivalently
we impose that $W(z)=\frac{1}{z}+\frac{1}{z^2}+\frac{s}{z^3}+...$.
We thus obtain three equations that determine $t_1$, $t_2$ and $b$. After some manipulations, we get:
\begin{equation}\label{eq:t12}
\begin{aligned}
&t_1 =\frac{2 \left(-6-3 n+2 b \sqrt{4-n^2} \mu \right)}{b^2 \mu ^2}\;\;\;\;,\;\;\;\;\;
t_2 =\frac{6\left(2 \sqrt{4-n^2}-b \mu  (2-n)\right)}{b^3 \mu  \left(1-\mu ^2\right)}
\\
& \qquad \qquad \qquad \qquad b^2\frac{\left(\mu ^2-1\right)}{12}+b\frac{\left(1-\mu ^2\right)}{2\mu }\sqrt{\frac{2+n}{2-n}}=s
\end{aligned}
\end{equation}
The upper bound of the density support is thus explicitly given by:
%(choice of -?):
\begin{equation}\label{eq:bsn}
\boxed{b=b(s)=
\frac{3}{\mu } \sqrt{\frac{2+n}{2-n}} \left(1-\sqrt{ 1 -\frac{4}{3} \left(\frac{2-n}{2+n}\right)  \frac{\mu ^2}{\left(1-\mu ^2\right)}\; s}\right)}
\end{equation}

Replacing in Eqn~\eqref{eq:AB}  $t_1$ and $t_2$ by their expression Eqn~\eqref{eq:t12}, we get:
\begin{equation}\label{eq:AB1}
\begin{aligned}
A&=\frac{6\left(1+\mu ^2\right)}{\mu ^2\left(1-\mu ^2\right)b^2}-\sqrt{\frac{2-n}{2+n}}\frac{2\left(2+\mu ^2\right)}{\mu  \left(1-\mu ^2\right)b}\\
B&=\frac{12}{\sqrt{2+n}\left(1-\mu ^2\right)b}-\frac{24 }{\sqrt{2-n}\,\mu  \left(1-\mu ^2\right)b^2}
\end{aligned}
\end{equation}
\\

 To summarize,
  the resolvent $W(z)$ is explicitly given in regime {\bf II} by:
 \begin{equation}\label{eq:Wreg2n1}
 \boxed{\begin{aligned}
  W(z)=&\frac{t_1}{2+n}+\frac{t_2 z}{2-n}+A \phi_0(z)+B \phi_1(z)\\
 & \textrm{with}\quad \phi_0(z)=\frac{\cos\left[\mu\left(w+\frac{\pi}{2}\right)\right]}{
\cos\left[\frac{\mu \pi}{2}\right]}
\;\;
{\rm and}\;\;
\phi_1(z)=\frac{\sin\left[\mu\left(w+\frac{\pi}{2}\right)\right]}{
\tan w}\\
&\qquad \textrm{where}\quad \frac{z}{b}=\frac{1}{\sin w}\;\;
{\rm and}\;\; n=-2 \cos(\pi\mu)
 \end{aligned}}
 \end{equation}
 The parameters $t_1$, $t_2$, $A$ and $B$ are functions of $b=b(s)$
 given in Eqn~\eqref{eq:t12} and \eqref{eq:AB1}, where
  the upper bound of the density support $b(s)$ is given by Eqn~\eqref{eq:bsn}. For the usual Bures case $n=1$, and thus
 $\mu=\frac{2}{3}$ we get:
 \begin{equation}\begin{aligned}
&A =\frac{351}{10 b^2}-\frac{22 \sqrt{3}}{5 b}\;\;,\quad B=-\frac{324}{5 b^2}+\frac{36 \sqrt{3}}{5 b}\\
& t_1=\frac{3 \left(-27+4 \sqrt{3} b\right)}{2 b^2}\;\; ,\quad t_2=\frac{54 \left(3 \sqrt{3}-b\right)}{5 b^3}
\end{aligned}
\end{equation}
and for the upper bound of the density support:
\begin{equation}
 b=b(s)=\frac{9}{2} \sqrt{3} \left(1-\sqrt{1-\frac{16 s}{45}}\right)
\end{equation}

\subsection{Optimal density}

The optimal density $\rho_c(x)$
is obtained by taking the imaginary part of $W$ (cf. subsection \ref{subsubsec:resolvent}):
\begin{equation}
\rho_c(x)=-\frac{1}{\pi} {\rm Im} \, W(x+i 0^+)\;\;
{\rm for}\;\; x\in [0,b]
\end{equation}
where $z=x +i0^+$ with $x\in [0,b]$ corresponds
to $w=\frac{\pi}{2}-i \eta$ with $\eta>0$.
More precisely, we have $\frac{b}{x}=\sin w=\cosh \eta$.
Therefore we get:
\begin{equation}
e^{\pm \eta}=\frac{b}{x}\pm \sqrt{\frac{b^2}{x^2}-1}
\;\; {\rm for }\;\; x=\frac{b}{\sin w}=\frac{b}{\cosh \eta}
\in [0,b]
\end{equation}

Using the explicit expresssion of $W(z)$ obtained in the previous subsection, we get:
\begin{equation}
\begin{aligned}
\rho_c(x)=\frac{\alpha}{\pi} \left(e^{-\mu \eta}-e^{\mu \eta}\right)+\frac{\beta}{\pi} \;  \left(
\frac{e^{-(\mu+1) \eta}
-e^{(\mu+1) \eta}}{\cosh\eta}\right)\;\;\;\;\textrm{where $\cosh\eta=\frac{b}{x}$}
\end{aligned}
\end{equation}
thus
\begin{equation}\label{eq:reg2optDens}
\begin{aligned}
\rho_c(x)=&\frac{\alpha}{\pi} \left\{\left(\frac{b}{x}-\sqrt{\frac{b^2}{x^2}-1}\right)^{\mu}-
\left(\frac{b}{x}+\sqrt{\frac{b^2}{x^2}-1}\right)^{\mu}\right\}\\
& +\frac{\beta}{\pi} \;\frac{x}{b}\;  \left\{
\left(\frac{b}{x}-\sqrt{\frac{b^2}{x^2}-1}\right)^{1+\mu}
-\left(\frac{b}{x}+\sqrt{\frac{b^2}{x^2}-1}\right)^{1+\mu}\right\}
\end{aligned}
\end{equation}

\begin{equation}
\begin{aligned}
\alpha &=
-\frac{ \sqrt{2-n}(2+\mu )}{(1-\mu ) \mu  b}+\frac{3 \sqrt{2+n} (1+\mu )}{(1-\mu ) \mu ^2 b^2}
\\
\beta &=\frac{3 \sqrt{2-n} }{\left(1-\mu ^2\right) b}-\frac{6\sqrt{2+n}}{\mu \left(1-\mu ^2\right) b^2}
\end{aligned}
\end{equation}
For the usual Bures case $\mu=2/3$, $n=1$:
\begin{equation}
\alpha%=\frac{135 \sqrt{3}}{4 b^2}-\frac{12}{b}
=\frac{3}{4 b^2}(45 \sqrt{3}-16 b)\;\; , \quad
 \beta %=-\frac{81 \sqrt{3}}{5 b^2}+\frac{27}{5 b}
 =\frac{27}{5 b^2}(b-3 \sqrt{3})
 \end{equation}
where $b=b(s)=\frac{9}{2} \sqrt{3} \left(1-\sqrt{1-\frac{16
      s}{45}}\right)$.

A plot of the density is shown in Fig. \ref{fig:dens} and \ref{fig:potdens}.

\subsection{Minimal energy}

To compute the minimal energy $E_s[\rho_c]=-\frac{t_0}{2}-\frac{t_1}{2}-\frac{t_2\, s}{4}$, cf. Eqn~\eqref{eq:minE},
we need to first determine $t_0$. We use the expression in
Eqn~\eqref{eq:t0}:
\begin{equation}\begin{aligned}
t_0 &=-t_1 b-\frac{t_2}{2} b^2+(2-n)\ln b +2 I_1- n I_2
%\\
%\textrm{with}\;\; 2 I_0-I_1 &=
%\frac{207 \sqrt{3}-405+b \left(60 \sqrt{3}-79-15 \ln 3-10 \ln 2\right)}{10 b}
\end{aligned}
\end{equation}
%%%%%%%%%%%%%%%%%%%%%%%%%%%%%%%%%%%%
Using the explicit expresion of $W(z)$ given in Eqn \eqref{eq:Wreg2n1}, we
can compute exactly the integrals $I_1$ and $I_2$ defined
in Eqn~\eqref{eq:t0}, we get:
\begin{equation}\begin{aligned}
2 I_1-n I_2 =&
\frac{3 \left(2 (2+n)\left(1-\mu ^2\right)+\sqrt{4-n^2} \mu \left(\mu ^2-3\right)\right)}{b\, \mu ^2 \left(-1+\mu ^2\right)}\\
&+(2-n)\gamma_E
 +(2-n)\psi\left(\frac{1-\mu}{2}\right)\\
&+ \frac{3 (n-2)}{1-\mu^2}+\sqrt{4-n^2}\left(\frac{4}{\mu }+\frac{\pi }{2}\right)+\left(1-\frac{n}{2}\right) (-5+\ln 4)
\end{aligned}
\end{equation}
where $\psi(x)=\frac{d}{\mathrm{d}x}\, \ln\Gamma(x)$ is the digamma function and $\gamma_E$ the Euler constant.
Thus we obtain the exact expression of the minimal energy:
\begin{equation}\begin{aligned}
E_s[\rho_c] = &
\left(\frac{n}{2}-1\right)\ln b+\frac{3 (2+n)}{2 b^2 \mu ^2}-\frac{5 \sqrt{4-n^2}}{2b \mu }\\
&+\left(\frac{n}{2}-1\right)\left\{\gamma_E+\psi\left(\frac{1-\mu}{2}\right)
+\ln 2\right\}+\frac{9}{8}(2-n)-\frac{\sqrt{4-n^2}}{4}\pi
\end{aligned}
\end{equation}
$E_s[\rho_c]$ is minimal for $b=\bar{b}=\frac{2}{\mu }\sqrt{\frac{2+n}{2-n}}$
which corresponds to $\bar{s}=\frac{2 (2+n) \left(1-\mu ^2\right)}{3 (2-n) \mu ^2}$.
As the distribution of the purity is highly peaked for large $N$, cf. Eqn~\eqref{eq:saddleprop},
the most probable value $\bar{s}$ of the rescaled purity is also its mean value.
Therefore the mean value of the purity for generalized Bures states 
is given by
\begin{equation}\label{eq:meanvaluegenBures}
\boxed{\langle \Sigma_2 \rangle \sim \frac{\bar{s}}{N} \;\;\; \textrm{as $N\to \infty$}\;\;\;
\textrm{with}\;\;\;\bar{s}=\frac{2 (2+n) \left(1-\mu ^2\right)}{3 (2-n) \mu ^2}}
\end{equation}
For the usual Bures case $n=1$, we recover $\bar{s}=5/2$.
\\

We can also compute the normalisation $\mathcal{Z}_N=e^{-N^2 E_{\bar{s}}[\rho_c]}$
(partition function), cf. Eqn~\eqref{eq:pdfSigma2Norm}:
\begin{equation}
\label{ctes}E_{\bar{s}}[\rho_c]=\left(\frac{n}{2}-1\right)\left[\gamma_E+\psi\left(\frac{1-\mu }{2}\right)-\frac{1}{2}\right]-\frac{1}{4} \sqrt{4-n^2}\,\pi +
\left(\frac{n}{2}-1\right) \ln\left(\frac{4}{\mu } \sqrt{\frac{2+n}{2-n}}\right)
\end{equation}
thus the rate function in regime {\bf II} 
$\Phi_{II}(s)=E_{s}[\rho_c]-E_{\bar{s}}[\rho_c]$ is explicitly given by:
\begin{equation}\label{eq:phiIIn}\boxed{\begin{aligned}
&\Phi_{II}(s)=\left(\frac{n}{2}-1\right)\ln b+\frac{3 (2+n)}{2 b^2 \mu ^2}-\frac{5 \sqrt{4-n^2}}{2 b \mu }+
\frac{(2-n)}{8} \left[7+2 \ln\left(\frac{4 (2+n)}{\mu ^2(2-n)}\right)\right]\\
& \textrm{where}\;\;\;\;\;b=b(s)=
\frac{3}{\mu } \sqrt{\frac{2+n}{2-n}} \left(1-\sqrt{ 1 -\frac{4}{3} \left(\frac{2-n}{2+n}\right)  \frac{\mu ^2}{\left(1-\mu ^2\right)}\; s}\right)
\end{aligned}}
\end{equation}
%where $b=b(s)$, the upper bound of the density support is given in Eqn~\eqref{eq:bsn}.
For $n=1$, Eqn~\eqref{eq:phiIIn} specializes to
%$\Phi_{II}(s)=E_s\left[\rho_c\right]-E_{\bar{s}}\left[\rho_c\right]$
%with $E_{\bar{s}}\left[\rho_c\right]=\frac{1}{4}+\frac{\ln 2}{2}$:
\begin{equation}\label{PhiII}
\boxed{
\begin{aligned}
\Phi_{II}(s)=-\frac{\ln b}{2}+\frac{81}{8 b^2}-\frac{15 \sqrt{3}}{4 b}
+\frac{7}{8}+\frac{1}{2}\ln\left(3 \sqrt{3}\right)\\
\textrm{with}\quad b=b(s)=\frac{9}{2} \sqrt{3} \left(1-\sqrt{1-\frac{16 s}{45}}\right)
\end{aligned}}
\end{equation}
Its minimum is reached at $b=\bar{b}=3 \sqrt{3}$,
i.e. at $s=\bar{s}=\frac{5}{2}$.
\\

From Eqn \eqref{eq:bsn}, we see that $b(s)$ is real (and positive) only for
$s\leq  s_0=\frac{3}{4} \left(\frac{2+n}{2-n}\right)  \frac{1-\mu ^2}{\mu ^2}$
(i.e.$s\leq \frac{45}{16}$ for $n=1$).
Moreover $\rho_c$ becomes negative when $\alpha+2\beta >0$, i.e.when $b<b_1=\frac{3}{\mu}\sqrt{\frac{2+n}{2-n}}\frac{(1-\mu)}{2-\mu}$
(for $n=1$, $b_1=9\sqrt{3}/8$), which is non-physical.
Thus the regime can be valid only for $b>b_1$, i.e.for $s>s_1$ where 
\begin{equation}
\boxed{s_1=\frac{3}{4}\left(\frac{2+n}{2-n}\right)\frac{(1-\mu )^2(1+\mu )(3-\mu )}{\mu ^2(2-\mu )^2}}
\end{equation}
thus $s_1=\frac{315}{256}$ for $n=1$.
The range $1<s<s_1$ on the left of regime {\bf II} is described by regime {\bf I},
with a density $\rho_c$ with finite support $[a,b]$ with $0<a<b$ as we will see.

It is useful to compute what happens close to $s=s_1^+$ in order to compare later with regime {\bf I}
(for $s=s_1^-$). In particular, we get:
\begin{equation}\begin{aligned}
E_{s_1}[\rho_c]=&\left(\frac{n}{2}-1\right)\Bigg\{\gamma_E+\psi\left(\frac{1-\mu}{2}\right)+
 \ln\left(6 \sqrt{\frac{2+n}{2-n}} \frac{(1-\mu )}{\mu (2-\mu )}\right)\\
&\qquad\qquad\qquad +\frac{ \left(-3+10\mu -11 \mu ^2\right)}{12 (1-\mu )^2}
 \Bigg\}
-\sqrt{4-n^2}\,\frac{\pi}{4} 
\end{aligned}
\end{equation}
We  may also compute the expansion of $\Phi_{II}(s)$ when $s \rightarrow s_1^+$:
\begin{equation}
\label{exphi2}\begin{aligned}
&\Phi_{II}(s)  = \frac{2 - n}{2}\left\{\frac{-9 + 14\mu - \mu^2}{24(1 - \mu)^2} +
 \ln\left(\frac{2}{3}\,\frac{2 - \mu}{1 - \mu}\right)\right\} \\
 &+ \frac{(2 - n)(3 - \mu)(1 + \mu)}{12(1 - \mu)^2}\Big(\frac{s_1 - s}{s_1}\Big) + \frac{(2 - n)(3 - \mu)^2}{8(1 - \mu)^2}\,\Big(\frac{s_1 - s}{s_1}\Big)^2 + O\big((s_1 - s)^3\big)
\end{aligned}
\end{equation}
We will see also that, although regime {\bf II} exists for $s$ in the range $[s_1,s_0]$ that contains in particular
the mean $\bar{s}$, there exists also another solution for $s>\bar{s}$ 
 that becomes more stable when $s>s_2$ for some $s_2\in [\bar{s},s_0[$
(its energy is lower energy than the present one). We will show that $s_2\sim \bar{s}$ as $N\to\infty$.
This second solution is characterized by the fact that one eigenvalue is much larger than the other,
the density is then described by one single eigenvalue plus a continuous part, this is regime {\bf III}.
%\newline

%thus we get:
%\begin{equation}
%t_0=\ln b+\frac{9 \sqrt{3}}{2 b}-\frac{5}{2}-\frac{3 \ln 3}{2}-\ln 2
%\end{equation}
%The minimal energy is explicitly given as a function of $b=b(s)$ and thus of the purity $s$:
%\begin{equation}\label{eq:EsrhocReg2}
%\boxed{E_s[\rho_c]=
%-\frac{\ln b}{2}+\frac{81}{8 b^2}-\frac{15 \sqrt{3}}{4 b}+\frac{9}{8}+3 \frac{\ln 3}{4}+\frac{\ln 2}{2}}
%\end{equation}
%where $b=b(s)=\frac{9}{2} \sqrt{3} \left(1-\sqrt{1-\frac{16 s}{45}}\right)$.
%\\

%
%Regime {\bf II}
% is well defined  only for $\frac{315}{256}<s<\frac{45}{16}$.
%Indeed $b(s)$ is real (and positive) only for $s<\frac{45}{16}$,
%and
%the density $\rho_c$ becomes negative for $s<\frac{315}{256}$, which is non-physical.
%
%For $s<s_1$ with $s_1=\frac{315}{256}$, the density has actually a support
%over $[a,b]$ with $a>0$, this is regime {\bf I}.

%
%At the transition point
%$s_1=\frac{315}{256}$, we have $b(s_1)=\frac{9 \sqrt{3}}{8}$, the energy becomes
%$E_{s_1}[\rho_c]=\frac{11}{24}+2 \ln 2-\frac{\ln 3}{2}$ and
%$\phi_{II}(s_1)=\frac{5}{24}+\frac{3}{2}\ln 2-\frac{\ln 3}{2}$.
%\newline

\section{Regime {\bf III}}
\label{sec:regIII}

For large $s$, there is no continuous density $\rho_c$ solution of Eqn~\eqref{eq:saddle1}.
In the limit $s\to N$, i.e. $S\to 1$, the quantum state becomes pure, which means that only one eigenvalue
of the density matrix $\sigma$ is nonzero.
For $s$ large enough, the maximal eigenvalue must detach from the sea of the other eigenvalues
to become larger and larger, thus it has to be taken into account separately
from the other eigenvalues.
We describe it by $\lambda_{\rm max}=T$ and the other eigenvalues
by a continuous density $\rho_{N-1}(\lambda)=\frac{1}{N-1}\sum_{i\neq {\rm max}}
\delta\left(\lambda-\lambda_i \right)$ with support $]0,\zeta]$
(similarly to regime {\bf II}), cf Fig. \ref{fig:potdens}.
%This assumption will actually be valid for $s>s_2$, with $s_2\sim 5/2$ when $N\to\infty$,
%this is regime {\bf III}.

Exactly as for the third regime in the distribution of purity for random Hilbert-Schmidt states~\cite{BElong}, 
one can actually show that all the eigenvalues apart from the maximal one remain of order $1/N$ as in regimes {\bf I} and {\bf II},
whereas the maximal eigenvalue 
$\lambda_{\rm max}$ is in regime {\bf III} of order $1/\sqrt{N}$ -thus much larger than the other eigenvalues for large $N$.
Here, for simplicity, we will not show that it is the case (but the proof can be done exactly as in the Hilbert-Schmidt
case \cite{BElong}), we will just assume this behaviour and see that it is coherent.
Assuming these scalings, the density of the $(N-1)$ smallest eigenvalues can be replaced for large $N$
by the rescaled density $\rho(x)=\frac{1}{N}\, \rho_{N-1}\left(\frac{x}{N}\right)$ which is expected
to have a continuous and finite limit when $N\to\infty$
(with support $[0,b]$ as in regime {\bf II}), whereas $\lambda_{\rm max}\asymp \frac{1}{\sqrt{N}}$.

This change of behaviour of the maximal eigenvalue between regime {\bf II} and {\bf III} from $1/N$ to $1/\sqrt{N}$
 can be understood as a Bose-Einstein type  (condensation) transition. 
The purity $\Sigma_2=\sum_i\lambda_i^2$ is fixed to a value $S=s/N$. When the eigenvalues are very close to each other, they can be described by a continuous density $\rho(x)$, which corresponds actually to regime {\bf II}, but this regime is valid only for $s$ smaller than a certain value $s_0$ (cf. previous section), which means that the purity $\Sigma_2$, given in the continuous case by
$\frac{1}{N}\int \mathrm{d}x \,x^2\,\rho(x)$, can not exceed the value $s_0/N$. For larger $s$, there must be (at least)\footnote{One can actually
show that a case where two or more eigenvalues detach is less favorable, cf. \cite{BElong}.}
 one eigenvalue that detaches from the other eigenvalues and becomes larger.
 One can then write the purity as $\frac{s}{N}=\Sigma_2=\frac{1}{N}\int \mathrm{d}x \rho(x) x^2+\lambda_{\rm max}^2$ to leading order for large $N$. Thus
$\lambda_{\rm max}^2\leq \frac{s}{N}\leq \frac{s_0}{N}
 +\lambda_{\rm max}^2$, which implies that $\lambda_{\rm max}^2$ has to become of same order
 as the purity, thus $\lambda_{\rm max}$ has to be of order $1/\sqrt{N}$.

\subsection{New saddle point}

We assume here as explained above that the eigenvalues are described by an isolated eigenvalue $\lambda_{\rm max}=T\asymp 1/\sqrt{N}$
plus a continuous density $\rho_{N-1}(\lambda)=\frac{1}{N-1}\sum_{i\neq {\rm max}}
\delta\left(\lambda-\lambda_i \right)$, which can be written for large $N$ as 
$\rho_{N-1}(\lambda )=N\rho\left(\lambda N\right)$  where 
$\rho$ is expected to have a finite continuous limit as $N\to\infty$, with a finite support $[0,b]$.
Because of the presence of an isolated eigenvalue, the saddle point has to be modified.
The distribution of the purity $\Sigma_2=\sum_i\lambda_i^2$ can be written as:
\begin{equation}\label{eq:newsadlle}
\mathcal{P}\left(\Sigma_2 = \frac{s}{N},N\right)=\frac{1}{\mathcal{Z}_N} \int \mathrm{d}t \int \mathcal{D}\rho \; e^{-N^2\mathcal{E}_s[\rho,T]}
\end{equation}
where the effective energy is now a function of $T$ and $\rho$
and is given for large $N$ by:
\begin{equation}\label{eq:energybrutreg3}
\begin{aligned}
\mathcal{E}_s\left[\rho,T\right]&=
\int_0^{b}\int_{0}^b \mathrm{d}x \mathrm{d}x' \rho(x) \rho(x') \,
\left(-\ln|x-x'|+\frac{ n}{2}
\ln|x+x'|\right)\\
&+t_0 \left[\int_{0}^{b} \mathrm{d}x \rho(x) -1 \right]
+t_1 \left[ \int_{0}^{b} \mathrm{d}x\,x \rho(x) +T-1 \right]\\
&
+\frac{t_2}{2} \left[  \int_0^{b} \mathrm{d}x\,x^2 \rho(x) +T^2 N-s \right]
+O\left(\frac{\ln N}{N}\right)
\end{aligned}
\end{equation}
where $t_0$, $t_1$ and $t_2$ are Lagrange multipliers (as in regime {\bf II}) enforcing the three constraints resp.
$\int \rho=1$ (normalisation of the density of eigenvalues), $\int x \rho +T=1$ (i.e.$\sum_i\lambda_i=1$) and
$\int x^2 \rho+T^2 N=s$ (i.e.fixed purity $\Sigma_2=s/N$).

Exactly as in regime {\bf II}, the functional integral in Eqn \eqref{eq:newsadlle} can be computed for large $N$ by a saddle point method -the minimal energy dominates the integral. With the correct normalisation, we get:
\begin{equation}\label{eq:newsadlle1}
\mathcal{P}\left(\Sigma_2 = \frac{s}{N},N\right)\approx  e^{-N^2\left(\mathcal{E}_s[\rho_c,T_c]-\mathcal{E}_{\bar{s}}[\rho_c,T_c]\right)}
\end{equation}
where $(\rho_c,T_c)$ minimizes $\mathcal{E}_s[\rho,T]$, thus
$\left.\frac{\delta E_s}{\delta \rho}\right|_{\rho_c,T_c}=0=\left.\frac{\partial E_s}{\partial T}\right|_{\rho_c,T_c}$.

The saddle point equation $0=\left.\frac{\partial E_s}{\partial T}\right|_{\rho_c,T_c}$
reads:
\begin{equation}\label{eq:saddle2reg3}
t_1+t_2 N T_c=0
\end{equation}
The saddle point equation $0=\left.\frac{\delta E_s}{\delta \rho}\right|_{\rho_c,T_c}$
is an integral equation on $\rho$:
\begin{equation}\label{eq:saddle1reg3}
2 \int_0^{b} \mathrm{d}x'  \rho_c(x') \,
\left(\ln|x-x'|-\frac{ n}{2}
\ln|x+x'|\right)=
t_0 
+t_1 x
+\frac{t_2}{2} x^2 \;\;\;,\; x\in[0,b]
\end{equation}
After differentiating with respect to $x$, we find:
\begin{equation}\label{eq:saddle1reg3bis}
\int_0^{b} \mathrm{d}x'  \frac{\rho_c(x')}{x-x'} \,
-\frac{  n }{2}\int_0^{b} \mathrm{d}x'  \frac{\rho_c(x')}{x+x'} \,
=
\frac{t_1 +t_2 x}{2}\;\;\;,\; x\in[0,b]
\end{equation}
Eqn~\eqref{eq:saddle1reg3bis} is exactly the same as in regime {\bf II},
except for the fact that the constraints on $\rho_c$ are slightly modified:
here we have $\int \rho_c=1$, $\int \rho_c x=1-T_c$, $\int \rho_c x^2=s-T_c^2 N$
whereas the $T_c$ were negligible in regime {\bf III}.

\subsection{Solution for regime {\bf III}}

We can use the same Bueckner's method as in regime {\bf II} to compute the resolvent $W(z)=\int \mathrm{d}x \frac{\rho_c(x)}{z-x}$
where $\rho_c$ is solution of Eqn~\eqref{eq:saddle1reg3bis}, with the constraints
$\int \rho_c=1$, $\int \rho_c x=1-T_c$, $\int \rho_c x^2=s-T_c^2 N$.
$W(z)$ must be solution of  
\begin{equation}\label{eq:reg3reseq}\begin{aligned}
 W(x+i 0^+)+W(x-i 0^+)+n \,W(-x)=t_2 x+t_1\;\;\; \textrm{for $x \in [0,b]$}\\
 \textrm{with}\;\;\;\; W(z)=\frac{1}{z}+\frac{1-T}{z^2}+\frac{s-T_c^2 N}{z^3}+...\;\;\;\textrm{as $|z|\to\infty$}
 \end{aligned}
 \end{equation}
The asymptotic condition is equivalent to the three conditions on the density $\rho_c$ (cf. above).
The solution of Eqn~\eqref{eq:reg3reseq} is as in regime {\bf II} of the form:
\begin{equation}
W(z)=h(z)+A\, \phi_0(z)+B\, \phi_1(z)
\end{equation}
where $h(z)=\frac{t_1}{2+n}+\frac{t_2 z}{2-n}$ is a particular solution, cf. Eqn~\eqref{eq:solpart},
and where $\phi_0$ and $\phi_1$ form a basis of solutions of the homogeneous equation, cf. Eqn~\eqref{eq:Wreg2n1}:
\begin{equation}\label{eq:Wreg2n1bis}
\begin{aligned}
  & \phi_0(z)=\frac{\cos\left[\mu\left(w+\frac{\pi}{2}\right)\right]}{
\cos\left[\frac{\mu \pi}{2}\right]}
\;\;
{\rm and}\;\;
\phi_1(z)=\frac{\sin\left[\mu\left(w+\frac{\pi}{2}\right)\right]}{
\tan w}\\
&\qquad \textrm{where}\quad \frac{z}{b}=\frac{1}{\sin w}\;\;
{\rm and}\;\; n=-2 \cos(\pi\mu)
 \end{aligned}
 \end{equation}
When $|z|\to \infty$, we must have $W(z)\to 0$, thus exactly as regime {\bf II}, cf. Eqn~\eqref{eq:AB}, we get:
\begin{equation}\label{eq:ABbis}
\begin{aligned}
&A=-\frac{t_1}{2+n}+\frac{b \mu t_2}{\sqrt{4-n^2}}\;\;\;\;,\;\;\;\;\;
&&B=-\frac{2 t_2 b}{(2-n)\sqrt{2+n}}\end{aligned}
\end{equation}
We now impose the constraints on $\rho_c$ or equivalently
we impose that $W(z)=\frac{1}{z}+\frac{1}{z^2}+\frac{s}{z^3}+...$ when $|z|\to \infty$.
This step differs from regime {\bf II}.
We get:
\begin{align}
t_1&=\frac{2 (2+n) \left(-3+3 T_c+2 b \sqrt{\frac{2-n}{2+n}} \mu \right)}{b^2 \mu ^2}\label{eq:t1reg33}\\
t_2&=\frac{6 (2-n) \left(2 (1-T_c)\sqrt{\frac{2+n}{2-n}}-b \mu \right)}{b^3 \mu  \left(1-\mu ^2\right)}\label{eq:t2reg33}\\
12 \left(s-N T_c^2\right) \mu &= 6 b \sqrt{\frac{2+n}{2-n}} (1-T_c) \left(1-\mu ^2\right)-b^2 \mu  \left(1-\mu ^2\right)\label{3bbis}
\end{align}
Eqn~\eqref{3bbis} together with the other saddle point equation
 $ \frac{\partial E_s}{\partial T}=0$, ie
$t_1+t_2 N T_c=0$, give in the large $N$ limit, for $T_c\gg 1/N$:
\begin{equation}\begin{aligned}
T_c &=\sqrt{\frac{s-\bar{s}}{N}}+
\frac{\bar{s}}{N}
\left(\frac{s-\frac{9}{8}\bar{s}}{s-\bar{s}}\right)+
O\left(\frac{1}{N^{3/2}}\right)\;\;\;\;\\
 b &=\bar{b}+\frac{\bar{b}}{\sqrt{N}}
%\frac{2}{\mu }\sqrt{\frac{2+n}{2-n}}
\frac{
 \left(\frac{3}{2}\bar{s}-s\right)}{\sqrt{s-\bar{s}}}
+O\left(\frac{1}{N}\right)\\
&\textrm{with}\;\;\;\; \bar{b}=\frac{2}{\mu }\sqrt{\frac{2+n}{2-n}}
\end{aligned}
\end{equation}
where $\bar{s}=\frac{2}{3}\left(\frac{2+n}{2-n}\right)\frac{\left(1-\mu ^2\right)}{\mu ^2}$ is the mean value of the 
rescaled purity, cf. Eqn~\eqref{eq:meanvaluegenBures} (for $n=1$, i.e.the usual Bures case, we have $\bar{s}=5/2$)
and where $\bar{b}$ is the value of the upper bound of the density support
at $s=\bar{s}$ in regime {\bf II}.
Then, using Eqn~\eqref{eq:t1reg33} and \eqref{eq:t2reg33} we find the expression of the Lagrange multipliers for large $N$:
\begin{align}
t_1 &=1-\frac{n}{2}+\frac{1}{\sqrt{N}}\frac{(2-n)s}{2 \sqrt{s-\bar{s}}}+O\left(\frac{1}{N}\right)\\
t_2 &=\frac{-1}{\sqrt{N}}\frac{(2-n) }{2 \sqrt{s-\bar{s}}}+O\left(\frac{1}{N}\right)
\end{align}

The computation of the minimal energy $\mathcal{E}_s[\rho_c,T_c]$ is done along the same lines as for regime {\bf II}.
It is given by (cf. Eqn~\eqref{eq:energybrutreg3} and \eqref{eq:saddle1reg3}):
\begin{equation}\begin{aligned}
\mathcal{E}_s[\rho_c,T_c]&=-\frac{t_0}{2}-\frac{t_1}{2}(1-T_c)-\frac{t_2}{4}(s-T_c^2 N)\\
&\textrm{where}\;\;\;
t_0 =-t_1 b-\frac{t_2}{2} b^2+(2-n)\ln b +2 I_1- n I_2
\end{aligned}
\end{equation}
with
$ I_1
=\int_b^{\infty} \mathrm{d}z \left(\frac{1}{z}-W(z)\right)$ and
$I_2 =\int_{-\infty}^{-b} \mathrm{d}z \left(W(z)-\frac{1}{z}\right)$.
The integrals $I_1$ and  $I_2$ can be computed after the change of variable $z\to w$ where
$\frac{z}{b}=\frac{1}{\sin w}$.

After some simplifications, we get when $N\to \infty$:
%cf BuresReg3OxNew1.nb
\begin{equation}
\mathcal{E}_s[\rho_c,T_c]=E_{\bar{s}}[\rho_c]+\frac{(2-n) \sqrt{s-\bar{s}}}{2 \sqrt{N }}+...
\end{equation}
where $E_{\bar{s}}[\rho_c]$ is the value of the energy at $s=\bar{s}$ in regime {\bf II}, cf. Eqn~\eqref{ctes}:
\begin{equation}
E_{\bar{s}}[\rho_c]=\left(\frac{n}{2}-1\right)\left[\gamma_E+\psi\left(\frac{1-\mu }{2}\right)-\frac{1}{2}\right]-\sqrt{4-n^2}\,\frac{\pi}{4} +
\left(\frac{n}{2}-1\right) \ln\left(\frac{4}{\mu } \sqrt{\frac{2+n}{2-n}}\right)
\end{equation}
Thus  $\mathcal{E}_{\bar{s}}[\rho_c,T_c]=E_{\bar{s}}[\rho_c]$ and we get for regime {\bf III} as $N\to \infty$:
\begin{equation}\label{phi3nqcq}
\boxed{\mathcal{P}\left(\Sigma_2=\frac{s}{N},N\right)\approx e^{-N^{\frac{3}{2}} \Phi_{III}(s)}\;\;\;\;\;\textrm{with}
\;\;\; \Phi_{III}(s)=\frac{(2-n) }{2 }\, \sqrt{s-\bar{s}}}
\end{equation}

The solution of regime {\bf III} exists for $s\geq \bar{s}$, cf. Eqn~\eqref{phi3nqcq}.
For $\bar{s}<s<s_0$, the two regimes {\bf II} and {\bf III} coexist.
The regime that is valid for a given $s$ is the one that has the
minimal energy, cf Fig. \ref{fig:plotphi}.
The transition between {\bf II} and {\bf III} thus happens at $s=s_2$ defined by
$E_{s_2}[\rho_c]=\mathcal{E}_{s_2}[\rho_c,T_c]$, i.e. $N^2\Phi_{II}(s_2)=N^{3/2}\, \Phi_{III}(s_2)$.
We find explicitly:
\begin{equation}\label{eq:s2gen}
\boxed{s_2=\bar{s}+\frac{\bar{s}^{4/3}}{N^{1/3}}+...\;\;\;\;\;\;\textrm{with \;\;$\bar{s}=\frac{2}{3}\left(\frac{2+n}{2-n}\right)\frac{\left(1-\mu ^2\right)}{\mu ^2}$}}
\end{equation}
In the limit $N\to\infty$, $s_2$ tends to the mean value $\bar{s}$.

We define the rate function $\Phi(s)$ as
\begin{equation}
\Phi(s)\sim -\frac{1}{N^2} \ln\mathcal{P}\left(\Sigma_2=\frac{s}{N},N\right)=\left\{\begin{array}{ll}
\Phi_{II}(s) & \textrm{for $s_1<s<s_2$}\\
\Phi_{III}(s)/\sqrt{N} & \textrm{for $s>s_2$}
\end{array}\right.
\end{equation}
At the transition point, the rate function $\Phi(s)$ is continuous but
its first derivative is discontinuous, cf Fig. \ref{fig:plotphi}:
\begin{equation}\begin{aligned}
\left.\frac{d\Phi(s)}{ds}\right|_{s_2^+}&=\frac{1}{\sqrt{N}}\left.\frac{d\Phi_{III}(s)}{ds}\right|_{s_2}
\sim \frac{(2-n)}{4 N^{\frac{1}{3}}\, \bar{s}^{\frac{2}{3}}}
\\ 
\left.\frac{d\Phi(s)}{ds}\right|_{s_2^-}&=\left.\frac{d\Phi_{II}(s)}{ds}\right|_{s_2}
\sim \frac{(2-n)}{ N^{\frac{1}{3}}\, \bar{s}^{\frac{2}{3}}}
\end{aligned}
\end{equation}
The transition between regime {\bf II} and regime {\bf III} is thus a first order transition
(exactly as in the Hilbert-Schmidt case \cite{BElong}).

%%%%%%%%%%%%%%%%%%%%%%%%%%%%%

\section{Regime {\bf I}}
\label{sec:regI}

In this section, we assume that the density $\rho_c(x)$ solution of the saddle point equation
\eqref{eq:saddle1} has a finite support of the form $[a,b]$ with $0<a<b$. By continuity, the density must vanish at $a$ and $b$, thus $\rho_c(b)=0=\rho_c(a)$.

\subsection{Resolvent}

As in regime {\bf II}, $W(z)$ can always be written:
\begin{equation}
W(z)=h(z) + A(z^2)\,\chi_0(z) + B(z^2)\, \chi_1(z)
\end{equation}
where $h(z)=\frac{t_1}{2+n}+\frac{t_2 z}{2-n}$
 is the particular solution given in Eqn~\eqref{eq:solpart}, and where $\chi_0$ and $\chi_1$ form a basis of the solutions of the homogeneous equation
$\overline{W}(x+i 0^+)+\overline{W}(x-i 0^+)+ n \overline{W}(-x)= 0$ for $x\in [a,b]$. Now $[a,b]$ is disjoint from $[-b,-a]$, thus guessing an expression for $\chi_0$ and $\chi_1$ is more difficult.
The techniques to find a basis $(\chi_0,\chi_1)$ were basically developed in \cite{EK95}, and reformulated in terms of algebraic geometry in \cite{GBthese}. We first need to introduce elliptic functions and theta functions. For a review on those kind of special functions, see the book \cite{guo}.

\subsubsection{Elliptic parametrization and special functions}

Let us first introduce some useful parameters.
 We define the modulus $k\in [0,1]$ and the complementary modulus $k'$ as follows:
 \begin{equation}
\boxed{k=\frac{a}{b}\;,\qquad k'=\sqrt{1-k^2}}
\end{equation}
The complete elliptic integral $K$ and the complete elliptic integral for the complementary modulus $K'$
are defined by:
\begin{equation}
K\equiv \int_0^1 \frac{\mathrm{d}t}{\sqrt{(1-t^2)(1-k^2 t^2)}}\;,\qquad
K'\equiv \int_0^1 \frac{\mathrm{d}t}{\sqrt{(1-t^2)(1-k'^2 t^2)}}
\end{equation}
For $k\in[0,1]$, we have $K\in\mathbb{R}$ and $K'\in\mathbb{R}$.
We also define $(\tau,\q)$ and $(\tau',\qp)$ such that:
\begin{equation}
\boxed{\begin{aligned}
\tau &=\frac{i K}{K'}\;, \qquad & \tau' &=-\frac{1}{\tau}=\frac{i K'}{K}
\end{aligned}}
\end{equation}
 \begin{equation}
\boxed{\begin{aligned}
\q &=e^{i \pi \tau}\;, \qquad & \qp &=e^{i\pi\tau'}
\end{aligned}}
\end{equation}
For $k\in[0,1]$, we have $\qp\in\mathbb{R}$ and $\q\in\mathbb{R}$.
When $k\to 0$ then $\qp\to 0$. When $k\to 1$ then $\q\to 0$.
\newline

We recall $n = - 2\cos\pi\mu$, $\mu \in ]0,1[$. Let us introduce the elliptic parametrization
as in \cite{GaetanB}:
\begin{equation}\label{eq:ellparam}
\boxed{u(\x) = \frac{ib}{2K'}\int_{-a}^{\x} \frac{\mathrm{d}z'}{\sqrt{(z'^2 - a^2)(z'^2 - b^2)}} \;\;, \quad
\x\in\mathbb{C}}
\end{equation}
where the integral runs from $-a$ to $\x$ in the complex plane following a path that has to be specified ($u(\x)$
depends on the path). The determination of the squareroot $\kappa(z) = \sqrt{(z^2 - a^2)(z^2 - b^2)}$ 
is chosen\footnote{This can be obtained by writing
$\kappa(z)=-\sqrt{z-a}\sqrt{z+a}\sqrt{z-b}\sqrt{z+b}$ where
the four squareroots are chosen with usual determination, i.e. cut on $\mathbb{R}_-$.}
 so that it is positive along the direct path from $-a$ to points $z \in ]-a,a[$. Then, $\kappa$ is an analytic function on $\mathbb{C}\setminus[-b,-a]\cup[a,b]$, and takes a minus sign across the two cuts $[-b,-a]\cup[a,b]$.

For a path included in $\mathbb{H}=\{\x\in\mathbb{C},\;\;{\rm Im}\,\x>0\}$ (for example
a path
from $-a$ to $+\infty$ or to $-\infty$ of the form $z'=y+i0^+$ with $y\in\mathbb{R}$), we have the remarkable values:
%(see appendix \ref{app:inteell}):
\begin{equation}\begin{aligned}
u(-a)&=0, \qquad & u(0) &\equiv u^0 = \frac{\tau}{2}, \qquad &u(a)&= u^{a} = \tau \\
 u(b) \equiv  u^b &=  -\frac{1}{2} + \tau,\qquad &  u(\infty)& \equiv u^{\infty} = \frac{-1 + \tau}{2},\qquad
 &u(-b)&=-\frac{1}{2}
\end{aligned}
 \end{equation}
However, for  a path included in $(-\mathbb{H})$, for example from $-a$ to $+\infty$
 (or $-a$ to $-\infty$) of the form $z'=y-i0^+$ with $y\in\mathbb{R}$, we have other values:
\begin{equation*}\begin{aligned}
u(-a)&=0, \qquad & u(0) &  = \frac{\tau}{2}, \qquad &u(a)&=\tau \\
 u(b)  &=  \frac{1}{2} + \tau,\qquad &  u(\infty)&  = \frac{1 + \tau}{2},\qquad
 &u(-b)&=\frac{1}{2}
\end{aligned}
 \end{equation*}
 With a path from $-a$ to $-a$ encircling once the segment $[-b,-a]$ clockwise, we get
 $u(-a)=\frac{1}{2}+\frac{1}{2}=1$.

We may construct a Riemann surface $\mathcal{C}$ by gluing two copies of $\mathbb{C}$ along the segments $[-b,-a]$ and $[a,b]$, such that the squareroot $\kappa(z)$ defined an analytic function on $\mathcal{C}$. The squareroot determination on the second copy is chosen opposite to the determination on the first copy, so that $\kappa$ has no discontinuity on the $[-b,-a]\cup[a,b] \subseteq \mathcal{C}$.  Moreover we have $u(+\infty)=u(-\infty)$ and actually all the points $|z|\to\infty$ can be identified to one point called $\infty$. Each sheet is thus topologically a sphere with two cuts.
 Therefore the surface obtained by gluing the sheets together is topologically equivalent to a torus. This torus has two distinguished uncontractible paths. One of them
is a path encircling $[a,b]$ with positive orientation (or equivalently $[-b,-a]$). Each time the
path from $-a$ to $\x$ encircles $[a,b ]$ (or $[-b,-a]$), it adds $+1$ to the value of $u(\x)$. Another uncontractible path is a path going from $-a$ to $+a$ in one sheet and then from $+a$ to $-a$
in the other sheet. This loop adds $\tau+\tau=2\tau$ to the value of $u(\x)$. Those two paths form a basis of the homology group of the torus.

 By inverting $u(\x)$, we get a function $\x(u)$ that is thus doubly periodic with periods $1$ and $2\tau$
 (with $\tau \in i\mathbb{R}_+^*$ here),
 i.e. $\x(u+1)=\x(u)$ and $\x(u+2\tau)=\x(u)$.
More precisely, one can show that the parametrization has the following properties:\footnote{
The property $\x(-u)=\x(u)$ comes from the fact that the two sheets represent the complex plane but
with two opposite
determinations of the squareroot. The starting point of the path, $-a$, is the same point
on both sheets but the end point $\x\in\mathbb{C}$ can be chosen on one sheet or the other.
Thus, if for a given path from $-a$ to $\x$ we get $u=u(\x)$ then the path obtained by inverting the two sheets
gives $u(\x)=-u$.
}
\begin{equation}\label{eq:periodx(u)}
\boxed{\x(u+1)=\x(u)\;,\quad \x(-u)=\x(u)\;, \quad \x(u+\tau)=-\x(u)}
\end{equation}
and thus
\begin{equation}
\begin{aligned}
  \x(\overline{u}) &= \x(u)\;,\quad & \x(\tau - u) &= -\x(u)
%\kappa(u + 1) &= \kappa(u), & \kappa(-u) &= -\kappa(u), & \kappa(\overline{u}) &= -\kappa(u),& \kappa(\tau - u) &= \kappa(u)
\end{aligned}
\end{equation}
where we have defined $\overline{u}$ as $\overline{u} = 2\tau - u$.

If we restrict the function $u(\x)$ to one upper half plane,
one can show that $u:\mathbb{H} \to \mathbb{C}$ is a conformal map from $\mathbb{H}$ to the interior
of a rectangle with vertices $-\frac{1}{2} $, $0$, $\tau$ and $-\frac{1}{2} + \tau$.
The inverse function $\x(u)$ is thus an analytic function from the interior of this rectangle to $\mathbb{C}$. It can be analytically continued on $\mathbb{C}$ using the symmetries and periodicities of Eqn~\eqref{eq:periodx(u)}. We thus get a double-periodic analytic function $u\to \x(u)$ from $\mathbb{C}$ to $\mathbb{C}$,
namely an ``elliptic'' function.
\\

 The function $\x(u)$ thus belongs to the class of elliptic functions.
From the definition of $u(\x)$ Eqn~\eqref{eq:ellparam},
one can show that $\x(u)$ is actually closely related to the Jacobi sine function $\mathrm{sn}$. Indeed:
\begin{equation}
\boxed{\x(u)=a \, {\rm sn}_k(\phi)\;\;\; \textrm{where}\;\;\;
\phi=\left(u-\frac{\tau}{2}\right)\frac{2 K'}{i}}
\end{equation}
$\x(u)$ has a pole at $u^{\infty}=\frac{-1 + \tau}{2}$. From its definition in Eqn~\ref{eq:ellparam}, the expansion of $\x(u)$ when $u \rightarrow u^{\infty}$ reads:
\begin{equation}
\label{eq:sxp} \x(u) = \frac{x_{-1}}{u - u^\infty} + x_1(u - u^\infty) + O\big((u - u^\infty)^3\big)
\end{equation}
with:
\begin{equation}
\label{eq:xone} x_{-1} = \frac{ib}{2K'} = \frac{b\tau}{2K},\qquad x_{1} = \frac{a^2 + b^2}{6x_{-1}}
\end{equation}

\subsubsection{Homogeneous equation}

Using the elliptic parametrization defined in the previous section, we can now solve
the homogeneous equation
\begin{equation}\label{eq:hom}
\overline{W}(x+i 0^+)+\overline{W}(x-i 0^+)+ n \overline{W}(-x)= 0\;\;, \qquad x\in[a,b]
\end{equation}
For $x\in[a,b]$, we have $u=u(x+i0^+)\in \tau+\left[-\frac{1}{2},0\right]$ for a direct path from $-a$ to
$x+i0^+$ (included in $\mathbb{H}$), whereas $u(x - i0^-) = 2\tau - u - 1$. Besides, $u(-x - i0^+) = \tau -  u$ and $u(-x + i0^+) = u - \tau$.
By abuse of notations, we shall write $\overline{W}(u)$ for $\overline{W}(\x(u))$. If $\overline{W}(z)$ is a $1$-cut solution (cut on $[a,b]$)
of Eqn~\eqref{eq:hom}, then $\overline{W}(u)+\overline{W}(2\tau -u - 1) + n\overline{W}(\tau -u)  = 0$ for $u\in\tau+\left[-\frac{1}{2},0\right]$.
%(cf. Eqn~\eqref{eq:transfuz}). 
Moreover, $\overline{W}(u + 1) = \overline{W}(u)$ and $\overline{W}(-u)=\overline{W}(u)$
as $\x(u+1)=\x(u)$ and $\x(-u)=\x(u)$\footnote{
However, as $\overline{W}(z)$ has a cut on $[a,b]$ we have
$\overline{W}(x+i0^+)\neq \overline{W}(x-i0^+)$ thus $\overline{W}(2\tau-u)\neq\overline{W}(u)$.}. $\overline{W}(u)$ is initially defined in the interior of the rectangle of vertices $-1/2$, $1/2$, $\tau + 1/2,\tau - 1/2$. We have just seen that its values on the boundary of the rectangle are related, and we can use those relations to define $\overline{W}(u)$ as an analytic function for $u \in \mathbb{C}$. Then, the relations are equalities between analytic functions, so must hold on the whole complex plane:
\begin{equation}\label{eq:OmegaHom}
\boxed{\begin{aligned}
%\forall u \in \mathbb{C}\quad \omega(u + 1) = \omega(u),\quad \omega(u) = -\omega(-u),\quad \omega(u - 2\tau) - n\omega(u - \tau) + %\omega(u) = 0
\forall u \in \mathbb{C}\,, \quad &\Omega(u + 1) = \Omega(u),\quad  \Omega(-u) = \Omega(u),\\
&\Omega(2\tau -u) + n\Omega(\tau -u) + \Omega(u) = 0
\end{aligned}}
\end{equation}
To summarize, to any $1$-cut solution $\overline{W}(z)$ of Eqn~\ref{eq:hom}, we have associated a meromorphic function $u \mapsto \overline{W}(u)$ defined on the whole complex plane.

If we focus on the system of equations
\begin{equation}
\label{system}\Omega(u + 1) = \Omega(u),\qquad \Omega(u+2\tau) + n\Omega(u+\tau) + \Omega(u) = 0,
\end{equation}
the space of solutions is of dimension $2$ on the field of meromorphic functions invariant under $1$- and $2\tau$-translations.
We recall that $n=-2 \cos(\pi\mu)$. Then, if we introduce $\mathbf{T}$ the operator of $\tau$-translation, one can remark that
$\ker\left(\mathbf{T}^2+n\mathbf{T}+{\rm id}\right)=\ker\left(\mathbf{T}-e^{i\pi\mu}{\rm id}\right)
\oplus \ker\left(\mathbf{T}-e^{-i\pi\mu}{\rm id}\right)$.
 A basis of solutions is thus given by functions which take a phase $e^{\pm i\pi\mu}$ under $\tau$-translation
 and are invariant under $1$-translation. We may call "pseudo-elliptic" this kind of functions.

 One can show, cf. \cite{EK95}, that the unique solution $g(u)=\Omega(u)$ of Eqn~\eqref{eq:OmegaHom} satisfying
the
boundary conditions
$g(u(\x))\sim\frac{1}{\x}$ when $\x\to\infty$
and  $g(u(\x))\propto\frac{1}{\kappa(\x)}$ when $\x\to a,b,-a,-b$
where $\kappa(\x)=\sqrt{(\x^2-a^2)(\x^2-b^2)}$, is given by a ratio of Jacobi theta functions:
\begin{equation}\label{eq:defgmu}
\boxed{g(u(\x))=\frac{\x}{\kappa(\x)}\; \frac{\vartheta_1\left(u(\x)-\frac{\tau}{2}+\frac{\mu}{2}|\tau\right)\,
\vartheta_1\left(-\frac{1}{2}|\tau\right)}{\vartheta_1\left(u(\x)-\frac{\tau}{2}|\tau\right)\,
\vartheta_1\left(\frac{\mu-1}{2}|\tau\right)}}
\end{equation}
The Jacobi theta function here is by definition (cf. \cite{guo})
\begin{equation}\label{eq:defvartheta}
\vartheta_1\left(v|\tau\right)=i\sum_{p \in \mathbb{Z}} (-1)^p\,e^{i\pi\tau(p + 1/2)^2 + (2p - 1)\pi v}
\end{equation}
It satisfies
$\vartheta_1\left(v+1|\tau\right)=-\vartheta_1\left(v|\tau\right)$
and $\vartheta_1\left(v+\tau|\tau\right)$ $=-e^{-i\pi(\tau + 2v)}\,\vartheta_1\left(v|\tau\right)$.
 $g(u(\x))$ and $g(u(-\x))$ form a basis of solutions of Eqn~\ref{system}. By abuse of notations, we shall write $g(\x)$ for $g(u(\x))$. Notice that $g(\x)$ has a zero when $\x = c$, with $c = a\,\mathrm{sn}_k(i\mu K')$.

$g(\x)$ does have a discontinuity on $[-b,-a]$. Indeed, we have not taken into account yet the equation $\overline{W}(u) = \overline{W}(-u)$ which mean that $\overline{W}(z)$ has no discontinuity on $[-b,-a]$. We can now enforce parity to construct a basis $(\chi_0,\chi_1)$ of $1$-cut solution of Eqn~\ref{eq:hom}:
 \begin{equation}\label{eq:chi01}
\boxed{ \begin{aligned}
 \chi_0(\x)&=\frac{\kappa^2(\x)}{{\x}^2 - c^2}\,\frac{f(\x)}{2 - n}\;\; \textrm{and}\;\;
\chi_1(\x)=\frac{\x^2}{\x^2 - c^2}
\,\frac{\widehat{f}(\x)}{2 + n}\\
&\qquad \textrm{with}\;\; \kappa(\x)=\sqrt{(\x^2-a^2)(\x^2-b^2)}
\end{aligned}}
\end{equation}
where $f$ and $\widehat{f}$ are defined by:
\begin{equation}
\boxed{f(\x)  =  \frac{g(\x) + e^{i\pi\mu}g(-\x)}{1 - e^{i\pi\mu}} \;\;\textrm{and}\;\;
\widehat{f}(\x)  =  \frac{\kappa(\x)}{\x}\,\frac{g(\x) - e^{i\pi\mu}g(-\x)}{1 + e^{i\pi\mu}} }
\end{equation}
where $g$ is the pseudo-elliptic function
defined in Eqn~\eqref{eq:defgmu}.

\subsubsection{Resolvent}

As we have seen above, the resolvent $W(z)$ solution of
%Eqn~\eqref{eq:On}
$W(x+i 0^+)+W(x-i 0^+)+ n W(-x)= V'(x)=t_2 x+t_1$
is given by:
\begin{equation}
\label{eq:esu} W(z)=h(z) + A(z^2)\,\chi_0(z) + B(z^2)\, \chi_1(z)
\end{equation}
where $h(z)$ is the  particular solution
$h(z)=\frac{2 V'(z)-n\, V'(-z)}{4-n^2}=\frac{t_1}{2+n}+\frac{t_2 z}{2-n}$
and where  $\chi_0$ and $\chi_1$ are given by Eqn~\eqref{eq:chi01} above.

$A(z^2)$ and $B(z^2)$ are polynomials of $z^2$ that are determined using the asymptotic condition
$W(z)\to 0$ as $|z|\to\infty$.

Then we impose the three conditions on the density $\rho_c$ or equivalently
we impose the asymptotic expansion of $W(z)$ when $|z|\to\infty$,
i.e. $W(z)=\frac{1}{z}+\frac{1}{z^2}+\frac{s}{z^3}+...$.
There is also an additional condition coming from the fact
that $W(z)$ must be analytic at $z=\pm c$ (as $c\notin [a,b]$).
These four conditions fix the value of $t_1$, $t_2$,
$a$ and $b$ (as functions of $s$).
\\

In order to write down these conditions, we need the asymptotic expansion of $g(z)$ when $z \rightarrow \infty$. It can be computed with a minimal number of new notations by first establishing that \cite{GaetanB}:
\begin{equation}
\frac{1}{g(z)}\,\frac{\mathrm{d}g(z)}{\mathrm{d}z} = \frac{-i\gamma + \frac{z\kappa(z) + c\kappa(c)}{z^2 - c^2}}{\kappa(z)} - \frac{1}{\kappa(z)}\,\frac{\mathrm{d}\kappa(z)}{\mathrm{d}z}
\end{equation}
where:
\begin{equation}
\label{eq:clambdagamma}\gamma = \frac{b}{2K'}\,(\ln\vartheta_1)'\big(\frac{1 - \mu}{2}\big|\tau\big),\qquad c = a\,\mathrm{sn}_k(i\mu K')
\end{equation}
It is also convenient to introduce $\lambda = \frac{\kappa(c)}{ic} \in \mathbb{R}_+^*$. Then, the expansion of $g(z)$ takes the form:
\begin{equation}\label{eq:gds}
\begin{aligned}
g(z)=&
\frac{1}{z} + \frac{-i\gamma}{z^2} + \frac{1}{2}(a^2 + b^2 - c^2 - \gamma^2)\,\frac{1}{z^3}  \\
& +  \frac{i}{6}\,\left\{\gamma^3 + \gamma\left(3c^2 - 4(a^2 + b^2)\right) - 2c^2\lambda\right\})\,\frac{1}{z^4}   \\
 &  +
\frac{1}{24}\Big\{
9 a^4+9 b^4-3 c^4+6 a^2 b^2-6 a^2 c^2-6 b^2 c^2-10 \gamma^2 (a^2+b^2)\\
&\qquad\qquad +6 c^2\gamma^2+\gamma^4-8 c^2 \gamma \lambda
\Big\}\frac{1}{z^5}
 +O\left(\frac{1}{z^6}\right)
\end{aligned}
\end{equation}

\noindent $\bullet$ The resolvent $W(\x)$ must be analytic at $\x=\pm c$, it imposes:
\begin{equation}\label{eq:analytc}
\lambda\sqrt{\frac{2 + n}{2 - n}}\,A(c^2) + B(c^2) = 0
\end{equation}
\\
$\bullet $ When $|z|\to\infty$, $W(\x)$ tends to zero. Therefore we get:
\begin{equation}
A \equiv -t_2,\qquad B \equiv -t_1 + t_2\,\gamma\,\sqrt{\frac{2 + n}{2 - n}}
\end{equation}
\\
$\bullet$ When $|z|\to\infty$, $W(z)$ behaves more precisely as
$W(z)=\frac{1}{z}+\frac{1}{z^2}+\frac{s}{z^3}+..$ (cf. subsection  \ref{subsubsec:resolvent}).
Taking also account of Eqn~\eqref{eq:analytc}, we get the following system of four equations
that fix the value of the Lagrange multipliers $t_{1,2}$ and the bounds $a,b$
(for a given purity $s$):
\begin{equation}\label{eq:systreg1}\begin{aligned}
0&=\frac{t_2}{2 - n}\left(\lambda - \gamma\right)+ \frac{t_1}{\sqrt{4 - n^2}}  \\
1&=\frac{t_2}{2 - n}\frac{1}{2}\left(a^2 + b^2 - c^2 - \gamma^2\right) + \frac{t_1}{\sqrt{4 - n^2}}\,\gamma  \\
 \sqrt{\frac{2 + n}{2 - n}}&=\frac{t_2}{2 - n}\,\frac{1}{3}\left(-\lambda c^2 + \gamma(a^2 + b^2) - \gamma^3\right)
+ \frac{t_1}{\sqrt{4 - n^2}}\frac{1}{2}\big(\gamma^2 - c^2\big)  \\
s&= \frac{t_2}{2 - n}\,\frac{1}{8}\big(a^4 + b^4 + \gamma^4 - 3c^4 -2a^2b^2 + 2c^2(a^2 + b^2)
 - 2\gamma^2(a^2 + b^2 + c^2)\big) \\
&\quad + \frac{t_1}{\sqrt{4 - n^2}}\,\frac{1}{6}\big(2\lambda c^2 + \gamma(a^2 + b^2 + 3c^2) - \gamma^3\big)
\end{aligned}
\end{equation}

\subsubsection{Parametric solution}

We can solve the system of Eqn~\eqref{eq:systreg1} parametrically with the variable $k = a/b$, or equivalently with the variable $\tau$.
\begin{equation}\label{eq:t12bs}\begin{aligned}
t_1 & =  - (2 - \n)\,\alpha_1\, \frac{\delta_3}{\delta_2^2} \;, \qquad &
t_2 & =  \frac{(2 - \n)^2}{2 + \n}\,\frac{\delta_3^2}{\delta_2^3}  \\
b & =  \sqrt{\frac{2 + \n}{2 - \n}}\,\frac{\delta_2}{\delta_3} \;, \qquad &
s & =  \frac{2 + \n}{2 - \n}\,\frac{\delta_4\delta_2}{\delta_3^2}
\end{aligned}
\end{equation}
where
\begin{equation}\label{eq:alphadelta}
\begin{aligned}
\alpha_1 & =  \widetilde{\lambda} - \widetilde{\gamma}  \\
\delta_2 & =  \frac{1}{2}\big(1 + k^2 - \widetilde{c}^2 +
 \widetilde{\gamma}^2\big) - \widetilde{\gamma}\widetilde{\lambda} \\
\delta_3 & =  \frac{1}{6}\big(\widetilde{\gamma}^3 + \widetilde{\lambda}\widetilde{c}^2\big)
 + \widetilde{\gamma}\Big(\frac{1 + k^2}{3} - \frac{\widetilde{\gamma}\widetilde{\lambda}}{2}
  - \frac{\widetilde{c}^2}{2}\Big)  \\
\delta_4 & =  -\frac{\widetilde{c}^4}{24} -
\frac{\widetilde{c}^2}{12}(1 + k^2) + \frac{k^2}{12} + \frac{k^4 + 1}{8}
- \frac{1}{6}\widetilde{\lambda}\widetilde{\gamma}\big(1 + k^2 + \widetilde{c}^2 - \widetilde{\gamma}^2\big) \\
&  - \frac{\widetilde{\gamma}^4}{24} - \frac{\widetilde{\gamma}^2}{12}(1 + k^2)
+ \frac{\widetilde{\gamma}^2\widetilde{c}^2}{4}
\end{aligned}
\end{equation}
Those are expressed in terms of the basic parameters, which are function of $k$ (or $\tau$) only:
\begin{equation}\label{eq:ctltgt}\begin{aligned}
\widetilde{c} & =  c/b = k\,\mathrm{sn}_k(i\mu K')  \;,\qquad
\widetilde{\lambda} =  \lambda/b = \Big|\frac{\mathrm{dn}_k(i\mu K')\mathrm{cn}_k(i\mu K')}{i\mathrm{sn}_k(i\mu K')}\Big|  \\
\widetilde{\gamma} & =  \gamma/b = \frac{1}{2K'}\,(\ln\vartheta_1)'\big(\frac{1 - \mu}{2}\big|\tau\big)
\end{aligned}
\end{equation}
We also record the relations \cite{guo}:
\begin{equation}
\tau = \frac{iK'}{K},\qquad k = \Big(\frac{\vartheta_4(0|\tau)}{\vartheta_3(0|\tau)}\Big)^2,\qquad K' = \frac{\pi}{2}\vartheta_3^2(0|\tau)
\end{equation}
Eqn~\eqref{eq:t12bs} determines the purity $s$ as a function of the modulus $k=a/b$. This relation can in principle be inverted
to get $k$ as a function of $s$, but it seems too difficult to be done in closed form. Finally all the parameters $t_1$, $t_2$, $b$ and $a=k b$ are functions of $k$ and thus implicitly of $s$.

\subsection{Distribution of purity}

From Eqn~\eqref{eq:minE}, we know that the minimal energy is given by
\begin{equation}\begin{aligned}
& E_s[\rho_c]=-\frac{t_0}{2}-\frac{t_1}{2}-\frac{t_2\, s}{4}\\
&\qquad \textrm{with}\;\;\; t_0=-
t_1 b
-t_2 \frac{b^2}{2}+(n-2)\ln b+2
I_1-n I_2
\end{aligned}
\end{equation}
where
$ I_1=\int_b^{\infty} \mathrm{d}z \left(\frac{1}{z}-W(z)\right)$and
$I_2
=\int_{-\infty}^{-b} \mathrm{d}z \left(W(z)-\frac{1}{z}\right)$, cf. Eqn~ \eqref{eq:t0}. Then the distribution of purity is given by:
\begin{equation}
\Phi_I(s) = E_s[\rho_c] - E_{\bar{s}}[\rho_c]
\end{equation}
where $E_{\bar{s}}[\rho_c]$ is the constant introduced in Eqn~\eqref{ctes}.

We thus need to integrate the resolvent with respect to $\dd z$, and the solution in the form of Eqn~\eqref{eq:esu} is not convenient for this task. It is however a good idea to use again the elliptic parametrization, and define $\overline{\omega}(u) = z'(u)\overline{W}(u)$. The new function $\overline{\omega}$ satisfies for all $u \in \mathbb{C}$
\begin{equation}
\label{eq:bubu1}\begin{aligned}
\overline{\omega}(u)  = - \overline{\omega}(-u),\qquad \overline{\omega}(u + 1) = \overline{\omega}(u) & \\
\overline{\omega}(u) - n\overline{\omega}(u - \tau) + \overline{\omega}(u - 2\tau) = 0 &
\end{aligned}
\end{equation}
We justify in Appendix~\ref{energy} that, after some algebra, the minimal energy can be written:
\begin{equation}
\label{enene}E_s[\rho_c] = -\frac{t_1}{2} - \frac{t_2}{4}\Big(s - \frac{a^2 + b^2}{3}\Big) + "J_0"
\end{equation}
where $"J_0"$ is the constant term (i.e. independent of $\epsilon$)
in the asymptotic expansion of $J_{\epsilon}$ when $\epsilon \rightarrow 0$, the latter being defined by:
\begin{equation}
J_{\epsilon} = \int_{u^{\infty}+i \epsilon}^{u^b}
 \dd u\big(-\overline{\omega}(u) - \frac{n}{2}\,\overline{\omega}(\tau - u)\big) - \frac{2 - n}{2}\,\ln\Big(\frac{x_{-1}}{i\epsilon}\Big)
\end{equation}
where $x_{-1}$ is given Eqn~\eqref{eq:xone}. Another approach to compute the resolvent, providing a residue formula for $\overline{\omega}(u)$, was developed in \cite{GBthese}. It allows to perform in a systematic way the integration with respect to $\dd u$ on some paths of the complex plane, thus to carry further the computation of Eqn~\eqref{enene}. The resulting expression for $\overline{\omega}(u)$ is given in Eqn~\eqref{eq:omegabarcalc}.

\subsubsection{Residue formula for the resolvent}

From now on, we set $\nu = 1 - \mu$, so that $n = 2\cos\pi\nu$.
It was claimed in \cite{GBthese} that\footnote{Eqn~\eqref{eq:bubu} can be justified in two steps. First, one notices that $\mathcal{G}(u_0,u)$ is constructed to be a solution of Eqn~\eqref{eq:bubu1} with respect to the variable $u_0$, which implies that the residue expression is also a solution of Eqn~\eqref{eq:bubu1}. Second, the behaviour at the poles of the left and right hand side of Eqn~\eqref{eq:bubu} match, so that the difference is holomorphic on the whole complex plane. But when $n \in ]-2,2[$, the functional relation implies that such functions are bounded, and cannot be constant unless $0$. Thus, as an application of Liouville theorem, this difference vanishes.}
\begin{equation}
\begin{aligned}
\label{eq:bubu}\overline{\omega}(u_0) & = 2 \Res_{u \rightarrow u^{\infty}} \mathcal{G}(u_0,u)\,\overline{\omega}(u) \\
& = 2 \Res_{u \rightarrow u^{\infty}} \mathcal{G}(u_0,u)\,x'(u)\Big(-\frac{2V'(x(u)) - nV'(-x(u))}{4 - n^2} + \frac{1}{x(u)}\Big)
\end{aligned}
\end{equation}
where $\Res_{u \rightarrow u^{\infty}} $ means the residue when $u\to u^{\infty}$.
This looks like a Cauchy residue formula, and the "Cauchy kernel" is given by:
\begin{equation}
\label{eq:juju}
\begin{aligned}
\mathcal{G}(u_0,u) & = \frac{1}{2}\,\frac{1}{e^{i\pi\nu} - e^{-i\pi\nu}}\Big\{-e^{-i\pi\nu}\zeta_{\nu}(u_0 + u) + e^{i\pi\nu}\zeta_{\nu}(u_0 - u) \\
& \phantom{= \frac{1}{2}\,\frac{1}{e^{i\pi\nu} - e^{-i\pi\nu}}} + e^{-i\pi\nu}\zeta_{\nu}(-u_0 + u) - e^{i\pi\nu}\zeta_{\nu}(-u_0 - u)\Big\}
\end{aligned}
\end{equation}
where $\zeta_{\nu}(u) = -z'(u + u^{\infty})g(u + u^{\infty})$ can be written\footnote{We prefer to write $\zeta_{\nu}$ instead of $\zeta$, to avoid confusion with the usual notation for the Weierstra{\ss} zeta function.}:
\begin{equation}
\zeta_{\nu}(u) = \sum_{p \in \mathbb{Z}} e^{-i\pi p \nu}\,\pi\,\mathrm{cotan}\pi(u + m\tau) = \frac{\vartheta_1(u - \nu/2|\tau)}{\vartheta_1(u|\tau)}\,\frac{\vartheta_1'(0|\tau)}{\vartheta_1(-\nu/2|\tau)}
\end{equation}
It has the following important properties, merely a rewriting of those of $g$:
\begin{equation}
\zeta_{\nu}(u + 1) = \zeta_{\nu}(u),\qquad \zeta_{\nu}(u + \tau) = e^{i\pi\nu}\zeta_{\nu}(u),\qquad \zeta_{\nu}(u) \mathop{\sim}_{u \rightarrow 0} \frac{1}{u}
\end{equation}
Notice the simple pole of $\zeta_{\nu}(u)$ when $u \rightarrow 0$, with residue $1$.

Let us do now the residue computation for $V'(x) = t_2 x + t_1$. Notice that the term $\frac{x'(u)}{x(u)}$ contributes for a simple pole with residue $-1$. Moreover, the term involving $V$ is a total derivative, thus has no simple pole in its partial fraction expansion at infinity. We thus need to evaluate it up to order $O((u - u^{\infty})^{-2})$. The result is:
\begin{equation}\label{eq:omegabarcalc}
\begin{aligned}
\overline{\omega}(u_0) & = 2\,\Res_{u \rightarrow u^{\infty}} \mathcal{G}(u_0,u) \\
& = 2\Res_{u \rightarrow u^{\infty}} \mathcal{G}(u_0,u)\,\Big(\frac{v_2}{(u - u^{\infty})^3} + \frac{v_1}{(u - u^{\infty})^2} -\frac{1}{(u - u^{\infty})} + O(1)\Big) \\
& = -2\mathcal{G}(u_0,u^{\infty}) + 2v_1 \partial_2\mathcal{G}(u_0,u^{\infty}) + v_2\partial_2^2 \mathcal{G}(u_0,u^{\infty})
\end{aligned}
\end{equation}
where $\partial_2\mathcal{G}(u_0,u) $ is the partial derivative of
$\mathcal{G}$ with respect to its second argument (here $u$), and with:
\begin{equation}
v_2 = \frac{t_2x_{-1}^2}{2 - n},\qquad v_1 = \frac{t_1x_{-1}}{2 + n}
\end{equation}
where $x_{-1}$ is defined in Eqn~\eqref{eq:xone} .
Then, we want the combination:
\begin{equation}
\begin{aligned}
 -\overline{\omega}(u_0) - \frac{n}{2}\overline{\omega}(\tau - u_0) & = - \overline{\omega}(u_0) + \frac{n}{2}\overline{\omega}(u_0 - \tau) \\
& = -2\widetilde{\mathcal{G}}(u_0,u^{\infty}) + 2v_1\partial_2\widetilde{\mathcal{G}}(u_0,u^{\infty}) + v_2\partial_2^2\widetilde{\mathcal{G}}(u_0,u^{\infty})
\end{aligned}
\end{equation}
where $\widetilde{\mathcal{G}}(u_0,u)  = -\mathcal{G}(u_0,u) +
\frac{n}{2}\mathcal{G}(u_0 - \tau,u) $.
From Eqn~\eqref{eq:juju}, and using the relations
$\zeta_{\nu}(u+\tau)=e^{i \pi\nu}\: \zeta_{\nu}(u)$ and $n=2
\cos(\pi\nu)$, we find:
\begin{equation}
\begin{aligned}
\widetilde{\mathcal{G}}(u_0,u) & = -\mathcal{G}(u_0,u) + \frac{n}{2}\mathcal{G}(u_0 - \tau,u) \\
& = \frac{1}{4}\Big(e^{-2i\pi\nu}\zeta_{\nu}(u_0 + u) - \zeta_{\nu}(u_0 - u) + \zeta_{\nu}(-u_0 + u) - e^{2i\pi\nu}\zeta_{\nu}(-u_0 - u)\Big)
\end{aligned}
\end{equation}
Moreover $u^{\infty}=\frac{\tau-1}{2}$ thus $2 u^{\infty}=\tau\; {\rm
  mod}\;\mathbb{Z}$ and thus:
\begin{equation}
\label{fofo}\begin{aligned}
\widetilde{\mathcal{G}}(u_0,u^{\infty}) = & \frac{e^{i\pi\nu/2} - e^{-i\pi\nu/2}}{4}\Big(-e^{-i\pi\nu/2}\zeta_{\nu}(u_0 - u^{\infty}) - e^{i\pi\nu/2}\zeta_{\nu}(u^{\infty} - u_0)\Big) \\
\partial_2\widetilde{\mathcal{G}}(u_0,u^{\infty}) & =  \frac{e^{i\pi\nu/2} + e^{-i\pi\nu/2}}{4}\Big(e^{-i\pi\nu/2}\zeta'_{\nu}(u_0 - u^{\infty}) + e^{i\pi\nu/2}\zeta'_{\nu}(u^{\infty} - u_0)\Big) \\
\partial_2^2\widetilde{\mathcal{G}}(u_0,u^{\infty}) & = \frac{e^{i\pi\nu/2} - e^{-i\pi\nu/2}}{4}\Big(-e^{-i\pi\nu/2}\zeta''_{\nu}(u_0 - u^{\infty}) - e^{i\pi\nu/2}\zeta''_{\nu}(u^{\infty} - u_0)\Big)
\end{aligned}
\end{equation}

\subsubsection{Consequences}

Now, we can perform the integral $J_{\epsilon}$ and find $"J_0"$. The terms $\partial_2^k\widetilde{\mathcal{G}}(u_0,u^{\infty})$ for $k = 1,2$ can easily be integrated given the formulas \eqref{fofo}. We need the asymptotic expansion of $\zeta_{\nu}(u)$ when $u \rightarrow 0$, which can be derived from its expression in terms of theta functions, or from that of $g(z)$ when $z \rightarrow \infty$:
\begin{equation}
\zeta_{\nu}(u) = \frac{1}{u} + C_0 + C_1 u + o(u)\qquad \textrm{as
  $u\to 0$}
\end{equation}
where:
\begin{equation}
\label{esy}C_0 =  - \frac{i\gamma}{x_{-1}},\qquad C_1 = \frac{1}{2x_{-1}^2}\Big(\frac{a^2 + b^2}{3} - c^2 - \gamma^2\Big)
\end{equation}
We thus find:
\begin{equation}
"J_0" = \frac{t_1}{2}\,\sqrt{\frac{2- n}{2 + n}}\,(ix_{-1}C_0) + \frac{t_2}{4}\,(x_{-1}^2 C_1) + "\widetilde{J}_0"
\end{equation}
where remains a piece $"\widetilde{J}_0"$ for which we could not find a more explicit form in general
\begin{equation}
\begin{aligned}
"\widetilde{J}_0" & = \lim_{\epsilon \rightarrow 0} \Big\{\frac{e^{i\pi\nu/2} - e^{-i\pi\nu/2}}{2}\Big(e^{-i\pi\nu/2}\int_{i\epsilon}^{\tau/2} \dd u\,\zeta_{\nu}(u) - e^{i\pi\nu/2}\int_{-i\epsilon}^{-\tau/2} \dd u\,\zeta_{\nu}(u)\Big) \\
& - \frac{2 - n}{2}\,\ln\Big(\frac{x_{-1}}{i\epsilon}\Big)\Big\}
\end{aligned}
\end{equation}

Actually, when $\nu$ is rational, $"\widetilde{J}_0"$ can be written in closed form, but this expression depends on the denominator of $\nu$. All the same, $g$ and $\zeta_{\nu}$ can be reduced to algebraic functions when $\nu$ is rational. Those expressions are quite complicated to work with, and we have prefered to state results with elliptic functions and theta functions, whose origin is more transparent. However, since one is particularly interested in $n = 1$ (corresponding to $\nu = 1/3$, $\mu = 2/3$), i.e. the original Bures measure problem, let us state in that case the result for $"\widetilde{J}_0"$. Let us define implicitly $\widehat{k}$, $\widehat{K}$ and $\widehat{K}'$, such that:
\begin{equation}
\tau = \frac{iK}{K'} = \frac{i\widehat{K}}{3\widehat{K}'},\qquad \widehat{k} = \Big(\frac{\vartheta_4(0|3\tau)}{\vartheta_3(0|3\tau)}\Big)^2,\qquad \widehat{K} = \frac{\pi}{2}\,\vartheta_3^2(0|3\tau)
\end{equation}
With such definitions, $\widehat{K}$ and $\widehat{K'}$ are the complete elliptic integrals associated to the modulus $\widehat{k}$. Then, we have:
\begin{equation}
\zeta_{1/3}(u) = 2i\widehat{K}'\Big(\frac{1}{\mathrm{sn}_{\widehat{k}}(2i\widehat{K}'\,u)} + \frac{\sqrt{3} - i}{2}\,\frac{1}{\mathrm{sn}_{\widehat{k}}(2i\widehat{K}'\,(u + \tau))} + \frac{\sqrt{3} - i}{2}\,\frac{1}{\mathrm{sn}_{\widehat{k}}(2i\widehat{K'}\,(u + \tau))}\Big)
\end{equation}
The primitive of $\frac{1}{\mathrm{sn}_{\widehat{k}}}$ is \cite{guo}:
\begin{equation}
\int^v \frac{\dd v}{\mathrm{sn}_{\widehat{k}}(v)} = \frac{1}{2}\ln\Big[\frac{\mathrm{cn}_{\widetilde{k}}(v) - \mathrm{dn}_{\widehat{k}}(v)}{\mathrm{cn}_{\widehat{k}}(v) + \mathrm{dn}_{\widehat{k}}(v)}\Big]
\end{equation}
We introduce the notation $R_l = \frac{\mathrm{cn}_{\widehat{k}}(2l\widehat{K}'/3) - \mathrm{dn}_{\widehat{k}}(2l\widehat{K}'/3)}{\mathrm{cn}_{\widehat{k}}(2l\widehat{K}'/3) + \mathrm{dn}_{\widehat{k}}(2l\widehat{K}'/3)}$ and after some algebra using the properties of Jacobi elliptic functions:
\begin{equation}
"\widetilde{J}_0"\Big|_{n = 1} = \frac{1}{2}\ln\Big(\frac{K'}{b\sqrt{1 - \widehat{k}^2}}\Big) + \frac{1}{4}\Big[\ln R_1 + \frac{\sqrt{3}}{2}\ln R_2 - \sqrt{3} \ln R_3\Big]
\end{equation}
However, we will not use this expression in the remaining of this article.

\subsection{Limit $s\to 1$: completely mixed states}

The limit $s\to 1$ (i.e. $S\to 1/N$) corresponds to completely mixed states (minimal purity).
This limit is reached when all eigenvalues of the density matrix $\sigma$
 are equal. In other words, we have in the limit $a\to b$, i.e. $k\to 1$ or equivalently $\q=e^{i \pi \tau}\to 0$.
In this regime, the fast converging series in $q$ defining $\vartheta_1$ are convenient to find the asymptotic expansion of all interesting quantities.

\subsubsection{Asymptotics of $t_1,t_2,a,b$ and $s$}

The basic parameters to compute are:
\begin{equation}
\begin{aligned}
\widetilde{c} & =  k\,\mathrm{sn}_k(i\mu K') =  i k\,\frac{\mathrm{sn}_{k'}(\mu K')}{\mathrm{sn}_{k'}(\mu K')}  = i \frac{\vartheta_4(0|\tau)}{\vartheta_3(0|\tau)}\,\frac{\vartheta_1(\mu/2|\tau)}{\vartheta_2(\mu/2|\tau)} \\
& = -i\sqrt{\frac{2 + n}{2 - n}}\Big\{1 - 4\q + 2(4 + n)\q^2  - 8(2 + n)\q^3 + 2(16 + 9n + n^2)\q^4 \\
& \phantom{=} -8(7 + 5n + n^2)\q^5 + 2(2 + n)(24 + 8n + n^2)\q^6 - 8(2 + n)(10 + 4n + n^2)\q^7 \\
& \phantom{=} + 2(128 + 129n + 49n^2 10n^3 + n^4)\q^8  + o(\q^8)\Big\}
\end{aligned}
\end{equation}
\begin{equation}
\begin{aligned}
\widetilde{\lambda} & = \Big|\frac{\mathrm{dn}_{k}(i\mu K')\,\mathrm{cn}_k(i\mu K')}{i\mathrm{sn}_k(i\mu K')}\Big| = \Big|\frac{\mathrm{dn}_{k'}(\mu K')}{\mathrm{cn}_{k'}(\mu K')\,\mathrm{sn}_{k'}(\mu K')}\Big| \\
& = \Big(\frac{\vartheta_2(0|\tau)}{\vartheta_3(0|\tau)}\Big)^2\,\frac{\vartheta_3(\mu/2|\tau)\,\vartheta_4(\mu/2|\tau)}{\vartheta_1(\mu/2|\tau)\,\vartheta_2(\mu/2|\tau)}  \\
& = \frac{4}{\sqrt{4 - n^2}}\Big\{1 - 4\q + \q^2(16 - n^2) - 4(12 - n^2)\q^3 + (128 - 13n^2)\q^4  \\
& \phantom{=} - 12(26 - 3n^2)\q^5 + (704 - 84n^2 - n^4)\q^6 - 4(376 - 46n^2 - n^4)\q^7  \\
& \phantom{=} + 3(1024 - 127n^2 - 4n^4)\q^8 + o(\q^8)\Big\}
\end{aligned}
\end{equation}
\begin{equation}
\begin{aligned}
k & = \Big(\frac{\vartheta_2(0|\tau')}{\vartheta_3(0|\tau')}\Big)^2 = \Big(\frac{\vartheta_4(0|\tau)}{\vartheta_3(0|\tau)}\Big)^2  \\
& = 1 - 8\q + 32\q^2 - 96\q^3 + 256\q^4 - 624\q^5 + 1408\q^6 \\
 & \phantom{=} - 3008\q^7 + 6144\q^8 + o(\q^8) \\
K' & = \frac{\pi}{2}\,\vartheta_3^2(0|\tau) = \frac{\pi}{2}\Big\{1 + 4\q + 4\q^2 + 4\q^4 + 8\q^5 + 4\q^8 + o(\q^8)\Big\}
\end{aligned}
\end{equation}
\begin{equation}
\begin{aligned}
\widetilde{\gamma} & = \frac{(\ln \vartheta_1)'(\nu/2)}{2K'}  \\
& = \sqrt{\frac{2 + n}{2 - n}}\Big\{1 - 4\q + 2(8 - n)\q^2 - 8(6 - n)\q^3 + 2(64 - 11n - n^2)\q^4 \\
& \phantom{=} - 8(39 - 7n - n^2)\q^5 + 2(352 - 64n - 10n^2 - n^3)\q^6 - 8(188 - 34n - 6n^2 - n^3)\q^7  \\
 & \phantom{=} + 2(1536 - 279n - 51n^2 - 10n^3 - n^4)\q^8 + o(\q^8)\Big\}
\end{aligned}
\end{equation}
Now, the expansions of the parameters defined in Eqn~\eqref{eq:alphadelta} read:
\begin{equation}
\begin{aligned}
\alpha_1 & = \sqrt{\frac{2 - n}{2 + n}}\Big\{1 - 4\q + 2(8 + n)\q^2 - 8(6 + n)\q^3 + 2(64 + 11n - n^2)\q^4  \\
& \phantom{=} - 8(39 + 7n - n^2)\q^5 + (352 + 64n -10n^2 + n^3)\q^6 - 8(188 + 34n - 6n^2 + n^3)\q^7 \\
& \phantom{=} + 2(1536 + 279n - 51n^2 + 10n^3 - n^4)\q^8 + o(\q^8)\Big\} \\
\delta_2 & = 4(2 - n)\Big\{\q^2 - 8\q^3 + (40 - n)\q^4 - 8(20 - n)\q^5 + (556 - 38n + n^2)\q^6 \\
& \phantom{=}- 8(218 - 18n + n^2)\q^7 + (5056 - 474n - 38n^2 + n^3)\q^8 + o(\q^8)\Big\} \\
\delta_3 & = 4\sqrt{4 - n^2}\Big\{\q^2 - 12\q^3 + (88 - n)\q^4 - 4(124 - 3n)\q^5 + (2348 - 84n + n^2)\q^6  \\
& \phantom{=} - 4(2446 - 112n + 3n^2)\q^7 + (36928 - 2006n - 84n^2 + n^3)\q^8 + o(\q^8)\Big\}  \\
\delta_4 & =  4(2 - n)\Big\{\q^2 - 16\q^3 +(156 - n)\q^4 - 16(72 - n)\q^5 + (7020 - 150n + n^2)\q^6 \\
& \phantom{=} - 16(2314 - 66n + n^2)\q^7 + (174240 - 6126n - 150n^2 + n^3)\q^8 + o(\q^8)\Big\}
\end{aligned}
\end{equation}
We deduce the expansion of the physical parameters:
\begin{equation}
\begin{aligned}
t_1 & = -\frac{1}{4\q^2} - \frac{8 + 3n}{4} - \frac{10 + 7n}{2}\q^2  + \frac{n(n - 1)}{2}\q^4 + o(\q^4) \\
t_2 & = \frac{1}{4\q^2} + \frac{n + 8}{4} + \frac{5(n + 2)}{2}\,\q^2 + \frac{n(n + 5)}{2}\,\q^4 + o(\q^4) \\
b & = 1 + 4\q - 16\q^3 - 2n\q^4 + 8(7 - n)\q^5 + 12n\q^6 + o(\q^6) \\
a & = 1 - 4\q  + 16\q^3 - 2n\q^4 - 8(7 - n)\q^5 + 12n\q^6 + o(\q^6) \\
s & = 1 + 4\q^2 - 32\q^4 + 4(44 - 5n)\q^6 + o(\q^6)
\end{aligned}
\end{equation}
Notice that the position of $a < 1 < b$ can be deduced from each other by changing the sign of all odd powers in $\q$. This can now be converted to an expansion in the variable $(s - 1) \rightarrow 0$. We first compute:
\begin{equation}
\q = \frac{(s - 1)^{1/2}}{2} + \frac{(s - 1)^{3/2}}{2} + \frac{68 + 5n}{64}(s - 1)^{5/2} + o\big((s - 1)^{5/2}\big)
\end{equation}
from which we obtain:
\begin{equation}
\begin{aligned}
t_1 & = -\frac{1}{s - 1} - \frac{3n}{4} - \frac{9n}{16}\,(s - 1) + o(s - 1) \\
t_2 & = \frac{1}{s - 1} + \frac{n}{4} + \frac{5}{16}\,(s - 1) + o(s - 1) \\
b & = 1 + 2\sqrt{s - 1} - \frac{n}{8}\,(s - 1)^2 + o\big((s - 1)^2\big) \\
a & = 1 - 2\sqrt{s - 1} - \frac{n}{8}\,(s - 1)^2 + o\big((s - 1)^2\big)
\end{aligned}
\end{equation}

\subsubsection{Asymptotics of the purity distribution}

We need the asymptotic expansion of $\zeta_{\nu}(w)$ for $w$ pure imaginary between $-\tau/2$ and $\tau/2$:
\begin{equation}
\label{expz}\begin{aligned}
\zeta_{\nu}(u) & = \frac{\vartheta_1(u - \nu/2|\tau)}{\vartheta_1(u|\tau)}\,\frac{\vartheta_1'(0|\tau)}{\vartheta_1(-\nu/2|\tau)} \\
& = \pi\,\mathrm{cotan}(\pi u) - \pi\,\mathrm{cotan}(\pi\nu/2) + 4\pi\sin\big(\pi(2u - \nu)\big)\q^2  \\
& \phantom{=} + 4\pi\big\{\sin\big(\pi(2u - 2\nu)\big) + \sin\big(\pi(4u - \nu)\big)\big\}\q^4 \\
& \phantom{=} + 4\pi\big\{\sin\big(\pi(2u - 3\nu)\big) + \sin\big(\pi(6u - \nu)\big)\}\q^6 \\
& \phantom{=} + 4\pi\big\{\sin\big(\pi(2u - 4\nu)\big) + \sin\big(\pi(4u - 2\nu)\big) + \sin\big(\pi(8u - \nu)\big)\big\}\q^8 \\
& \phantom{=} + O(\q^5)
\end{aligned}
\end{equation}
The $O(\q^5)$ is indicated after taking into account the fact that
$u \in [-\tau/2,\tau/2]$ and $|\tau| \rightarrow \infty$. If $u$
remains bounded, it is in fact a $O(\q^{9})$. Those expressions
could be used to obtain an asymptotic expansion of the spectral
density. Here, we will focus on the purity distribution $\Phi_I(s)$,
for which we need the expansion of $"\widetilde{J}_0"$. Since the
$O(\q^5)$ to integrate 
in fact reaches the order of magnitude $O(\q^5)$ from a smaller order of magnitude, and exponentially in the variable $u$, it remains a $O(\q^5)$ after integration over $u$.
\begin{equation}
\label{exju} \begin{aligned}
"\widetilde{J}_0" & = -\ln \q + \frac{2 - n}{2}\Big\{\ln\Big[\frac{K'}{\pi b}\Big] + n\q^2 + (-2 + \frac{n}{2} + n^2)\q^4  + o(\q^4)\Big\}  \\
& = -\ln[2^{1 - n/2}\q] + \frac{(2 - n)(4 + n)}{2}\Big\{\q^2 + \frac{2n - 3}{2}\,\q^4\Big\} + o(\q^4)
\end{aligned}
\end{equation}

Notice also that, by expanding Eqn~\eqref{expz} around $u \rightarrow 0$, we can identify the asymptotic expansion of the constants $C_0$ and $C_1$ (although it could also be obtained from the previous expansions and Eqn~\eqref{esy}):
\begin{equation}
\begin{aligned}
C_0 & = \pi\sqrt{\frac{2 + n}{2 - n}}\Big\{ -1 - 2(2 - n)\q^2 - 2(2 - n)(1 + n)\q^4 - 2(2 - n)n^2 \q^6 + o(\q^6)\Big\} \\
C_1 & = \frac{\pi^2}{3}\Big\{- 1 + 12n\,\q^2 - 12(2 - 2n - n^2)\q^4 + 12n^3 \q^6 + o(\q^6)\Big\}
\end{aligned}
\end{equation}
Finally, the purity distribution $\Phi_I(s) = E_s[\rho_c] - E_{\overline{s}}[\rho_c]$ has a small $\q$ expansion:
\begin{equation}
\begin{aligned}
\Phi_I(s) & = -\ln \q + \frac{1}{4} +\frac{2 -
  n}{2}\Big\{\ln\Big(\frac{2}{\mu}\sqrt{\frac{2 + n}{2 - n}}\Big) +
\psi\Big(\frac{1 - \mu}{2}\Big) - \psi(1) - \frac{1}{2}\Big\} \\
& \phantom{=} + \frac{\pi}{4}\sqrt{4 - n^2}  
 + \frac{8 - n}{2}\q^2 - 3\frac{8 - 7n}{4}\q^4 + o(\q^4)
\end{aligned}
\end{equation}
and in terms of the variable $s$:
\begin{equation}\label{unun}
\begin{aligned}
\Phi_I(s)& = -\frac{1}{2}\ln\Big(\frac{s - 1}{2}\Big) +  \frac{1}{4} +\frac{2 - n}{2}\Big\{\ln\Big(\frac{2}{\mu}\sqrt{\frac{2 + n}{2 - n}}\Big) + \psi\Big(\frac{1 - \mu}{2}\Big) - \psi(1) - \frac{1}{2}\Big\} \\
& \phantom{=} + \frac{\pi}{4}\sqrt{4 - n^2}  + \frac{8 - n}{8}(s - 1) + \frac{104 + 5n}{64}(s - 1)^2 + o\big((s - 1)^2\big)
\end{aligned}
\end{equation}
The rate function $\Phi_I(s)$ has thus a logarithmic
(integrable) divergence when $s \rightarrow 1$, with the same leading
coefficient $-1/2$ for all values of $n$. The distribution of the
purity thus vanishes rapidly when $s\to 1$:
\begin{equation}
\mathcal{P}\left(\Sigma_2=\frac{s}{N},N\right)\propto
(s-1)^{\frac{N^2}{2}}\qquad \textrm{as $N\to \infty$ and $s\to 1$}
\end{equation}
The probability that the Bures random state is a highly mixed state, ie
$s$ close to $1$ (it corresponds to the case where the state is
highly entangled with its environment) is thus
very small (it was the same for random Hilbert-Schmidt states \cite{BElong}).

\subsubsection{Remark on separability}

Let us assume that $\mathcal{H}$ is the Hilbert space of a bipartite
system, i.e. $\mathcal{H}=\mathcal{H}_A\otimes\mathcal{H}_B$ with $N =
N_AN_B$
(with $N$ the dimension of  $\mathcal{H}$, $N_A$ the one of
$\mathcal{H}_A$ and $N_B$ of  $\mathcal{H}_B$ ). We recall that a
state is said to be separable if it is a convex sum of tensor product
states, ie if the density matrix is of the form 
\begin{equation}
\sigma=\sum_i p_i \sigma_i^A\otimes \sigma_i^B\qquad \textrm{ with $p_i\geq 0$ and
$\sum_ip_i=1$}
\end{equation}
and where $\sigma_i^A\in\mathcal{L}(\mathcal{H}_A)$,
$\sigma_i^B\in\mathcal{L}(\mathcal{H}_B)$.
otherwise it is entangled.
%$\sigma = t\,\sigma_{1}\otimes\sigma_2 + (1 -
%t)\,\sigma_1'\otimes\sigma_2'$, with $t \in [0,1]$,
%$\sigma_{i},\sigma_i' \in  \mathcal{L}(\mathbb{C}^{N_i})$. 
Gurvits and Barnum \cite{Gurv} showed that states with purity
$\Sigma_2$ 
smaller than $\frac{1}{N - 1}$ are necessarily separable (independently of the bipartition $A,B$). In our notations, this corresponds to states having $s \leq s' = \frac{N}{N - 1}$, which is close to $1$ for large $N$. We can deduce from Eqn~\eqref{unun} a logarithmic equivalent for the probability that a random Bures state has purity less that $\frac{1}{N - 1}$:
\begin{equation}
\mathcal{P}\Big(\frac{s}{N} \leq \frac{1}{N - 1}\Big) \approx N^{-\frac{N^2}{2}}
\end{equation}
Since the probability that a random Bures states is separable is of order $N^{-N^2/4}$ (Corollary 2 in \cite{Ye}), we have:
\begin{equation}
\mathcal{R}_N = \ln \frac{\mathcal{P}\Big(\frac{s}{N} \leq \frac{1}{N - 1}\Big)}{\mathcal{P}(\mathrm{separable})} \simeq N{-N^2/2}
\end{equation}
This means that the criterion of separability based on purity is superexponentially loose when $N$ is large.

\subsection{Transition {\bf I}-{\bf II}}

The transition between regime {\bf I} and regime {\bf II} happens
at $s=s_1$ when $a\to 0$, i.e. $\tau \rightarrow 0$ or equivalently $\q \rightarrow 1$, $\qp=e^{i\pi\tau'} = e^{-i\pi/\tau} \rightarrow 0$. It is now convenient to perform a modular transformation on the theta functions, to write them as fast converging series of $\qp$ instead of $\q$:
\begin{equation}
\vartheta_1(v|\tau) = (-i\tau)^{-1/2}\,e^{-i\pi v^2/\tau}\,\vartheta_1(v/\tau|\tau'),\qquad \tau' = 1/\tau
\end{equation}
It turns out that an expansion up to $O((\qp)^{1 + \mu})$ of all parameters is necessary to find the order of the transition between {\bf I} and {\bf II}.

\subsubsection{Asymptotics of $t_1,t_2,a,b$ and $s$}

$(\qp)^{\mu}$ and $Q = (\qp)^{1 - \mu}$ are both involved in the asymptotics when $\qp \rightarrow 0$. For non-rational $\mu$, they are algebraic independent, so all the terms of the form $(\qp)^{\mu p}Q^{j}$ are of distinct order of magnitude. For rational $\mu$, one will have to several collect of these terms to find the asymptotic expansion at a given order. Since we assume $1/2 < \mu < 1$, and we want expressions up to $o((\qp)^{1 + \mu} = (\qp)^{2\mu}Q)$, we can use $(\qp)^{4\mu} \in o((\qp)^{1 + \mu})$ to expand all expressions up to order $3$ in the variable $(\qp)^{\mu}$, and then find the general coefficient of the expansion in $Q$. We can also simplify expressions since $(\qp)^{3\mu}Q \in o((\qp)^{3\mu})$ and $(\qp)^{2\mu}Q^j \in o((\qp)^{1 + \mu})$ whenever $j \geq 2$.
\begin{equation}
\label{eq:K}\begin{aligned}
\widetilde{c}^2 & = k^2\,\mathrm{sn}_{k}(i\mu K') = \Big(\frac{\vartheta_2(0|\tau')}{\vartheta_3(0|\tau')}\,\frac{\vartheta_1(\mu\tau'/2|\tau')}{\vartheta_4(\mu\tau'/2|\tau')}\Big)^2  \\
 &= - \sum_{j \geq 0} 4jQ^j + 8(\qp)^{\mu}\big[Q + \sum_{j \geq 1} 4(j - 1)Q^{j}\big] -4(\qp)^{2\mu}\big[Q + 12Q^2\big] + o((\qp)^{1 + \mu}) \\
\widetilde{\lambda} & = \Big|\frac{\mathrm{dn}_k(i\mu K')\,\mathrm{cn}_k(i\mu K')}{\mathrm{sn}_k(i\mu K')}\Big| = \Big|\Big(\frac{\vartheta_4(0|\tau')}{\vartheta_3(0|\tau')}\Big)^2\,\frac{\vartheta_3(\mu\tau'/2|\tau')\vartheta_2(\mu\tau'/2|\tau')}{\vartheta_1(\mu\tau'/2|\tau')\vartheta_4(\mu\tau'/2|\tau')}\Big| \\
& = 1 + \sum_{j \geq 1} 2jQ^j + 2(\qp)^{\mu}\big[1 - 2Q - 5Q^2 - \sum_{j \geq 3} 8Q^j\big] + 2(\qp)^{2\mu}\big[1 - 5Q\big] + 2(\qp)^{3\mu} + o((\qp)^{1 + \mu}) \\
\widetilde{\gamma} & = \frac{1}{2}(\ln \vartheta_1)'\Big(\frac{1 - \mu}{2}\Big|\tau\Big) = \frac{\pi}{2K}\Big[\mu - 1 + \frac{i}{\pi}(\ln\vartheta_1)'\Big(\frac{(1 - \mu)\tau'}{2} \Big| \tau'\Big)\Big] \\
& = \mu + \sum_{j \geq 1} 2jQ^j -4(\qp)^{\mu}\big[\mu Q+ \sum_{j \geq 2} 2jQ^j\big] - 2(\qp)^{2\mu}Q + o((\qp)^{1 + \mu}) \\
k^2 & = \Big(\frac{\vartheta_2(0|\tau')}{\vartheta_3(0|\tau')}\Big)^4 =  16(\qp)^{\mu}Q + o((\qp)^{1 + \mu}) \\
K & = \frac{\pi}{2}\,\vartheta_3^2(0|\tau') = \frac{\pi}{2}\Big(1 + 4(\qp)^{\mu}Q + o((\qp)^{1 + \mu})\Big)
\end{aligned}
\end{equation}
Then, we deduce the expansion for the parameters defined in Eqn~\eqref{eq:alphadelta}:
\begin{equation}
\begin{aligned}
\alpha_1 & = 1 - \mu + 2(\qp)^{\mu}\big[1 - 2(1 - \mu)Q - Q^2\big] + 2(\qp)^{2\mu}\big[1 - 4Q\big] + 2(\qp)^{3\mu} + o((\qp)^{1 + \mu})  \\
\delta_2 & = \frac{(1 - \mu)^2}{2} + 2(\qp)^{\mu}\big[-\mu +2\mu(2 - \mu)Q - (2 - \mu)Q^2\big] + 2\mu (\qp)^{2\mu}\big[- 1 + 8Q\big] - 2\mu (\qp)^{3\mu} + o((\qp)^{1 + \mu}) \\
\delta_3 & = \frac{\mu(2 - \mu)(1 - \mu)}{6} + (\qp)^{\mu}\big[-\mu^2 - 2\mu(2 - \mu)(1 - \mu)Q + (2 - \mu)^2Q^2\big] \\
& \phantom{=} \mu^2(\qp)^{2\mu}\big[-1 + 12Q\big] - \mu^2(\qp)^{3\mu} + o((\qp)^{1 + \mu}) \\
\delta_4 & = \frac{(1 - \mu)^2(1 + \mu)(3 - \mu)}{24} + \frac{(\qp)^{\mu}(1 - \mu)}{3}\big[-\mu(1 + \mu) - 2\mu(1 - \mu)(2 - \mu)Q + (2 - \mu)(3 - \mu)Q^2\big] \\
& \phantom{=} - \frac{\mu (\qp)^{2\mu}}{3}\big[(1 - \mu)(1 + \mu) + 8(2 + \mu^2)Q\big] - \frac{\mu(1 - \mu)(1 + \mu)}{3}(\qp)^{3\mu} + o((\qp)^{1 + \mu})
\end{aligned}
\end{equation}
We then find that $s$ reaches from below the limit:
\begin{equation}
s^* = s_1 = \frac{2 + n}{2 - n}\,\frac{3}{4}\,\frac{(1 - \mu)^2(1 + \mu)(3 - \mu)}{\mu^2(2 - \mu)^2}
\end{equation}
which gives the value of $s_1$. The limit of the other physical parameters read:
\begin{equation}
\begin{aligned}
t_1^* & = -(2 - n)\,\frac{2}{3}\,\frac{\mu(2 - \mu)}{(1 - \mu)^2} \\
t_2^* & = \frac{(2 - n)^2}{2 + n}\,\frac{2}{9}\,\frac{\mu^2(2 - \mu)^2}{(1 - \mu)^4} \\
b^* & = \sqrt{\frac{2+n}{2 - n}}\,3\,\frac{1 - \mu}{\mu(2 - \mu)} \\
a^* & = 0
\end{aligned}
\end{equation}
and their asymptotic expansion is:
\begin{equation}
\label{eq:B}\begin{aligned}
\Big(\frac{t_1 - t_1^*}{t_1^*}\Big) & = \frac{4\,(\qp)^{\mu}}{(2 - \mu)(1 - \mu)^2\mu}\big[\mu(1 + \mu) - 4\mu(2 - \mu)Q + (2 - \mu)(3 - \mu)Q^2\big] \\
& \phantom{=} + \frac{4\,(\qp)^{2\mu}}{(1 - \mu)^4(2 - \mu)}\big[(1 + \mu)^2(1 + 2\mu) + 16(-1 - 3\mu + \mu^2)Q\big] \\
 & \phantom{=} +\frac{4(1 + \mu)^4(1 + 3\mu)}{(1 - \mu)^6(2 - \mu)}\,(\qp)^{3\mu} + o((\qp)^{1 + \mu}) \\
\Big(\frac{t_2 - t_2^*}{t_2^*}\Big) & = \frac{12\,(\qp)^{\mu}}{(2 - \mu)(1 - \mu)^2\mu}\big[\mu^2 - 2\mu(2 - \mu)Q + (2 - \mu)^2Q^2\big] \\
& \phantom{=} + \frac{24\mu\,(\qp)^{2\mu}}{(2 - \mu)^2(1 - \mu)^4}\big[(1 + \mu)(1 + 2\mu - \mu^2) - 4\mu^3(2 - \mu)(5 - \mu)Q\big] \\
& \phantom{=} + \frac{4\mu(1 + \mu)^2(6 + 27\mu + 16\mu^2 - 9\mu^3)}{(2 - \mu)^2(1 - \mu)^6}\,(\qp)^{3\mu} + o((\qp)^{1 + \mu})\\
\big(\frac{b - b^*}{b^*}\Big) & = \frac{2\,(\qp)^{\mu}}{\mu(2 - \mu)(1 - \mu)^2}\big[-\mu^2(1 + \mu) + 2\mu(2 - \mu)(3 - 2\mu + \mu^2)Q - (3 - \mu)(2 - \mu)^2Q^2\big] \\
& \phantom{=} + \frac{4\mu\,(\qp)^{2\mu}}{(2 - \mu)^2(1 - \mu)^3}\big[-(1 + \mu)^2(2 + \mu) + 4(2 - \mu)(5 + \mu^2)Q\big]  \\
& \phantom{=} - \frac{4\mu(1 + \mu)^3(2 + \mu)^2}{(2 - \mu)^3(1 - \mu)^4}\,(\qp)^{3\mu} + o((\qp)^{1 + \mu}) \\
\Big(\frac{s - s^*}{s^*}\Big) & = \frac{4\,(\qp)^{\mu}}{(3 - \mu)(2 - \mu)(1 - \mu)^2\mu(1 + \mu)}\big[-\mu^2(1 + \mu)^2 + 2\mu(2 - \mu)(9 - 10\mu + 5\mu^2)Q \\
& \phantom{= \frac{4\,(\qp)^{\mu}}{(3 - \mu)(2 - \mu)(1 - \mu)^2\mu(1 + \mu)} } - (2 - \mu)^2(3 - \mu)^2Q^2\big] \\
& \phantom{=} - \frac{8\mu\,(\qp)^{2\mu}}{(3 - \mu)(2 - \mu)^2(1 - \mu)^3}\big[(1 + \mu)^3 + 4(2 - \mu)(1 - 5\mu)Q\big] \\
& \phantom{=} - \frac{4\mu(1 + \mu)^4(2 + \mu)(2 + 3\mu)}{(3 - \mu)(2 - \mu)^3(1 - \mu)^4(1 + \mu)}\,(\qp)^{3\mu} + o((\qp)^{1 + \mu})
\end{aligned}
\end{equation}
This could in turn be converted to expansions in the variable $(s^* - s) \rightarrow 0$. We just notice here that at leading order:
\begin{equation}
\qp \sim \Big(3\,\frac{2 - n}{2 + n}\,\frac{(2 - \mu)^3\mu}{(1 + \mu)^2}\,(s^* - s)\Big)^{1/\mu}
\end{equation}
and since $\qp > 0$, it is clear that the regime {\bf I} must correspond to $s < s^*$.

\subsubsection{Asymptotics of the purity distribution}

We need the asymptotic expansion of $\zeta_{\nu}(\tau w)$ for $w$ real valued between $-1/2$ and $1/2$:
\begin{equation}
\label{eq:trans}\zeta_{\nu}(w\tau) = \frac{2i\pi}{\tau}\Big(\frac{e^{i\pi(1 - \mu)w}}{e^{2i\pi w} - 1} - \frac{(\qp)^{(1 - \mu)}}{1 - (\qp)^{1 - \mu}}\,e^{i\pi(1 - \mu)w} + (\qp)^{1 + \mu}e^{-(1 + \mu)i\pi w}  + o((\qp)^{1 + \mu})\Big)
\end{equation}
By expanding around $w \rightarrow 0$, we also identify the asymptotic expansion of the coefficients $C_0$ and $C_1$:
\begin{equation}
\begin{aligned}
\frac{C_0}{i\pi/\tau} & = -\mu - \frac{2(\qp)^{1 - \mu}}{1 - (\qp)^{1 - \mu}} + 2(\qp)^{1 + \mu} + o((\qp)^{1 + \mu})  \\
\frac{C_1}{\pi^2/\tau^2} & = \frac{1}{2}\Big(\frac{1}{3} - \mu^2\Big) + 2(1 - \mu)\,\frac{(\qp)^{1 - \mu}}{1 - (\qp)^{1 - \mu}} + 2(1 + \mu)\,(\qp)^{1 + \mu} + o((\qp)^{1 + \mu})
\end{aligned}
\end{equation}
We also deduce the asymptotic expansion of $"\widetilde{J}_0"$ (the derivation is presented in detail in Appendix~\ref{appjo}):
\begin{equation}
\begin{aligned}
"\widetilde{J}_0" & = - \frac{2 - n}{2}\ln\Big(\frac{\pi b}{K}\Big) - \sqrt{4 - n^2}\,\frac{\pi}{4} - \frac{2 - n}{2}\Big[\psi(1) - \psi\Big(\frac{1 - \mu}{2}\Big)\Big] \\
& \phantom{=} + \frac{(\qp)^{1 - \mu}}{1 - (\qp)^{1 - \mu}}\,\frac{2 - n}{1 - \mu} + \frac{2 - n}{1 + \mu}\,(\qp)^{1 + \mu} + o((\qp)^{1 + \mu})
\end{aligned}
\end{equation}
where $\psi = (\ln \Gamma)'$ is the digamma function, and $b$ and $K$ must also be expanded according to Eqns~\eqref{eq:K} and \eqref{eq:B}. Notice that the terms of $"\widetilde{J}_0"$ involving $\pi$ and $\psi$ are also present in $E_{\overline{s}}[\rho_c]$ given in Eqn~\eqref{ctes}. So, they cancel each other in $\Phi_I(s) = E_s[\rho_c] - E_{\overline{s}}[\rho_c]$. We thus find the small $\qp$ expansion:
\begin{equation}
\begin{aligned}
\Phi_I(s) & = \frac{2 - n}{2}\Big\{\frac{-9 + 14\mu - \mu^2}{24(1 - \mu)^2} + \ln\Big(\frac{2}{3}\,\frac{2 - \mu}{1 - \mu}\Big)\Big\} \\
& \phantom{=} + \frac{(2 - n)\,(\qp)^{\mu}}{3(1 - \mu)^4\mu}\big[\frac{\mu^2(1 + \mu)^2}{2 - \mu} - 2\mu(9 - 10\mu + 5\mu^2)Q + (3 - \mu)^2(2 - \mu)Q^2\big] \\
& \phantom{=} + \frac{2(2 - n)\mu\,(\qp)^{2\mu}}{3(2 - \mu)^2(1 - \mu)^6(1 + \mu)}\big[(1 + \mu)^3(-1 - 4\mu + \mu^2 + \mu^3) \\
& \phantom{=} + 4(2 - \mu)^2(-13 + 5\mu - 11\mu^2 - 5\mu^3)Q\big]  \\
& \phantom{=} + \frac{(2 - n)\mu(1 + \mu)^3(4 + 28\mu + 7\mu^2 - 31\mu^3 + 5\mu^4 + 3\mu^5)}{3(2 - \mu)^3(1 - \mu)^8}\,(\qp)^{3\mu} \nonumber \\
& \phantom{=} + o((\qp)^{1 + \mu})
\end{aligned}
\end{equation}
If we turn this expression into an expansion in the variable $(s^* - s)$, we notice that up to a global factor, the coefficient of $(\qp)^{\mu}$ and of $(\qp)^{\mu}Q$ are precisely those found in the expansion of $s^* - s$. This means that $\Phi_I(s)$ contains a linear term in $(s^* - s)$, and no $(s^* - s)^{1/\mu}$ as we would have naively expected. The next correction comes at order $(\qp)^{2\mu} \sim (s^* - s)^2$, and going one step further, we find a singular term of order $(\qp)^{2\mu}Q = (\qp)^{1 + \mu} \sim (s^* - s)^{1 + 1/\mu}$. More precisely:
\begin{equation}
\begin{aligned}
\Phi_I(s) & = \frac{2 - n}{2}\Big\{\frac{-9 + 14\mu - \mu^2}{24(1 - \mu)^2} + \ln\Big(\frac{2}{3}\,\frac{2 - \mu}{1 - \mu}\Big)\Big\} \\
& \phantom{=} + \frac{(2 - n)(3 - \mu)(1 + \mu)}{12(1 - \mu)^2}\Big(\frac{s^* - s}{s^*}\Big) + \frac{(2 - n)(3 - \mu)^2}{8(1 - \mu)^2}\,\Big(\frac{s^* - s}{s^*}\Big)^2  \\
& \phantom{=} + \frac{32(2 - n)\mu^2}{(1 - \mu)^5(1 + \mu)}\,\Big(\frac{(3 - \mu)(2 - \mu)(1 - \mu)^2}{4\mu(1 + \mu)}\,\frac{s^* - s}{s^*}\Big)^{1 + 1/\mu} \\
& \phantom{=} + o\big((s^* - s)^{1 + 1/\mu}\big)
\end{aligned}
\end{equation}
In fact, this expansion matches up to $O((s^* - s)^2)$ that of $\Phi_{II}(s)$, see Eqn~\eqref{exphi2}, and the transition between regime {\bf I} and {\bf II} is characterized by:
\begin{equation}
\Phi_{I}(s) - \Phi_{II}(s) \sim \frac{32(2 - n)\mu^2}{(1 - \mu)^5(1 + \mu)}\,\Big(\frac{(3 - \mu)(2 - \mu)(1 - \mu)^2}{4\mu(1 + \mu)}\,\frac{s^* - s}{s^*}\Big)^{1 + 1/\mu}
\end{equation}
So, the purity distribution and its first and second derivative are continuous when $s \rightarrow s_1^*$, but its third left derivative blows up. The transition {\bf I}-{\bf II} is therefore of order $1 + \frac{\pi}{\mathrm{arccos}(n/2)}$ for all $n \in ]0,2[$.

\subsection*{Acknowledgments}

Part of this work was conducted during the session "Random matrices and vicious walkers" at l'\'Ecole des Houches, to which we are grateful both for atmosphere and funding. G.B. thanks Guillaume Aubrun for discussion and references, and benefited from the support of the ANR project "Grandes Matrices Al\'eatoires" ANR-08-BLAN-0311-01.
C.N. thanks Satya N. Majumdar, Gr\'egory Schehr and Pierpaolo Vivo for
useful discussions and references.

\appendix

\section{General expression for the minimal energy}
\label{energy}
We start from the expression:
\begin{equation}
E_s[\rho_c] = -\frac{t_1}{2} - \frac{t_2 s}{4} + \Big(-I_1 + \frac{n}{2} I_2 - \frac{2 - n}{2} \ln b + \frac{t_1 b}{2} +\frac{t_2 b^2}{4}\Big)
\end{equation}
where:
\begin{equation}
\begin{aligned}
& -I_1 + \frac{n}{2}I_2  \\
& =  \int_{b}^{\infty} \dd z\Big(W(z) - \frac{1}{z}\Big) + \frac{n}{2}\int_{-\infty}^{-b} \dd z\Big(W(z) - \frac{1}{z}\Big) \\
& =  \int_{\infty}^{b} \dd z\Big\{-\overline{W}(z) - \frac{2V'(z) - nV'(-z)}{4 - n^2} + \frac{1}{z} + \frac{n}{2}\Big(-\overline{W}(-z) - \frac{2V'(-z) - nV'(z)}{4 - n^2} - \frac{1}{z}\Big)\Big\} \\
& =  \lim_{M \rightarrow + \infty} \int_{M}^{b} \dd z \Big(-\overline{W}(z) - \frac{n}{2}\overline{W}(-z) + \frac{2 - n}{2z} - \frac{V'(z)}{2}\Big) \\
& =  \lim_{M \rightarrow +\infty} \Big\{\int_{u(M)}^{u_b} \dd u\Big(-\overline{\omega}(u) - \frac{n}{2}\overline{\omega}(\tau - u) \Big)- \frac{(2 - n)\ln (M/b)}{2} + \frac{\widetilde{V}(M)}{2}\Big\} - \frac{\widetilde{V}(b)}{2} \\
& = \lim_{\epsilon \rightarrow 0^+} \Big\{\int_{u^{\infty} + i\epsilon}^{u_b} \dd u\Big(-\overline{\omega}(u) - \frac{n}{2}\overline{\omega}(\tau - u) \Big) - \frac{(2 - n)\ln [x(u^{\infty} + i\epsilon)/b]}{2} + \frac{\widetilde{V}(x(u^{\infty} + i\epsilon))}{2}\Big\} \\
& - \frac{t_2 b^2}{4} - \frac{t_1 b}{2}
\end{aligned}
\end{equation}
We have written $\widetilde{V}(x) = \frac{t_2x^2}{2} + t_1 x$. Notice that $\widetilde{V}(x(u^{\infty} + i\epsilon))$ admits an expansion in the limit $\epsilon \rightarrow 0$ with a constant term due to the $x^2$, and equal to $t_2x_1x_{-1}$, where:
\begin{equation}
\label{AA1} x(u^{\infty} + \delta) \mathop{=}_{\delta \rightarrow 0} \frac{x_{-1}}{\delta} + x_1\delta + O(\delta^3)
\end{equation}
is the asymptotic expansion of $x$ near infinity. Moreover, we know $x_1$ from Eqn~\eqref{eq:xone}.
If we introduce
\begin{equation}
\label{Jeps}J_{\epsilon} = \int_{u^{\infty}+i \epsilon}^{u_b} \dd u\Big(-\overline{\omega}(u) - \frac{n}{2}\overline{\omega}(\tau - u)\Big) - \frac{2 - n}{2}\ln\Big(\frac{x_{-1}}{i\epsilon}\Big)
\end{equation}
and $"J_0"$ the constant term in the asymptotic expansion of $J_{\epsilon}$ when $\epsilon \rightarrow 0$, we can write:
\begin{equation}
E_s[\rho_c] = -\frac{t_1}{2} - \frac{t_2}{4}\Big(s - \frac{1}{3}(a^2 + b^2)\Big) + "J_0"
\end{equation}

\section{Transition I-II: expansion of $\tilde{J}_0$}
\label{appjo}

We want to compute in an asymptotic expansion when $\qp \rightarrow 0$:
 \begin{equation}
 \begin{aligned}
 \widetilde{J}_{\epsilon} & = \frac{e^{i\pi\nu/2} - e^{-i\pi\nu/2}}{2}\Big\{e^{-i\pi\nu/2}\int_{i\epsilon/\tau}^{1/2} \zeta_{\nu}(\tau w)\dd(\tau w) - e^{i\pi\nu/2}\int_{-i\epsilon/\tau}^{-1/2} \zeta_{\nu}(\tau w)\Big\} \\
 & - \frac{2 - n}{2}\ln\Big(\frac{x_{-1}}{i\epsilon}\Big)
 \end{aligned}
 \end{equation}
and in particular the constant term $"\widetilde{J}_0"$ in its asymptotic expansion when $\epsilon \rightarrow 0$. We have seen in Eqn~\eqref{eq:trans} that at the transition between regime {\bf I} and {\bf II},
\begin{equation}
\lim_{s \rightarrow s^*} \tau \zeta_{\nu}(\tau w) = (2i\pi)\,\frac{e^{i\pi(1 - \mu)w}}{e^{2i\pi w} - 1}
\end{equation}
and all the other corrections to $\zeta_{\nu}(\tau w)$ are regular when $w \rightarrow 0$. So, the computation of the subleading terms of $"\widetilde{J}_0"$ (written in Eqn~\eqref{exju}) is straightforward, and we shall concentrate on the leading term, i.e. $\lim_{s \rightarrow s^*} "\widetilde{J}_0"$.

We perform the change of variables $y = e^{2i\pi w}$ and introduce contours as in the picture (the quarter of circles $\mathcal{C}_{\pm}$ are both defined with anticlockwise orientation). We now have an integral on a portion of the unit circle in the $y$-plane. Notice that the function $\zeta_{\nu}(u(y))$ has a cut on the negative real axis, and going from above the cut to below the cut along the unit circle correspond to send $u$ to $u + \tau$, thus for any $y \in \mathbb{R}_-$:
\begin{equation}
\label{eq:1}\zeta_{\nu}(u(y - i0^+)) = e^{i\pi\nu}\zeta_{\nu}(u(y + i0^+))
\end{equation}
\begin{figure}[h!]
\begin{center}
\includegraphics[width=0.65\textwidth]{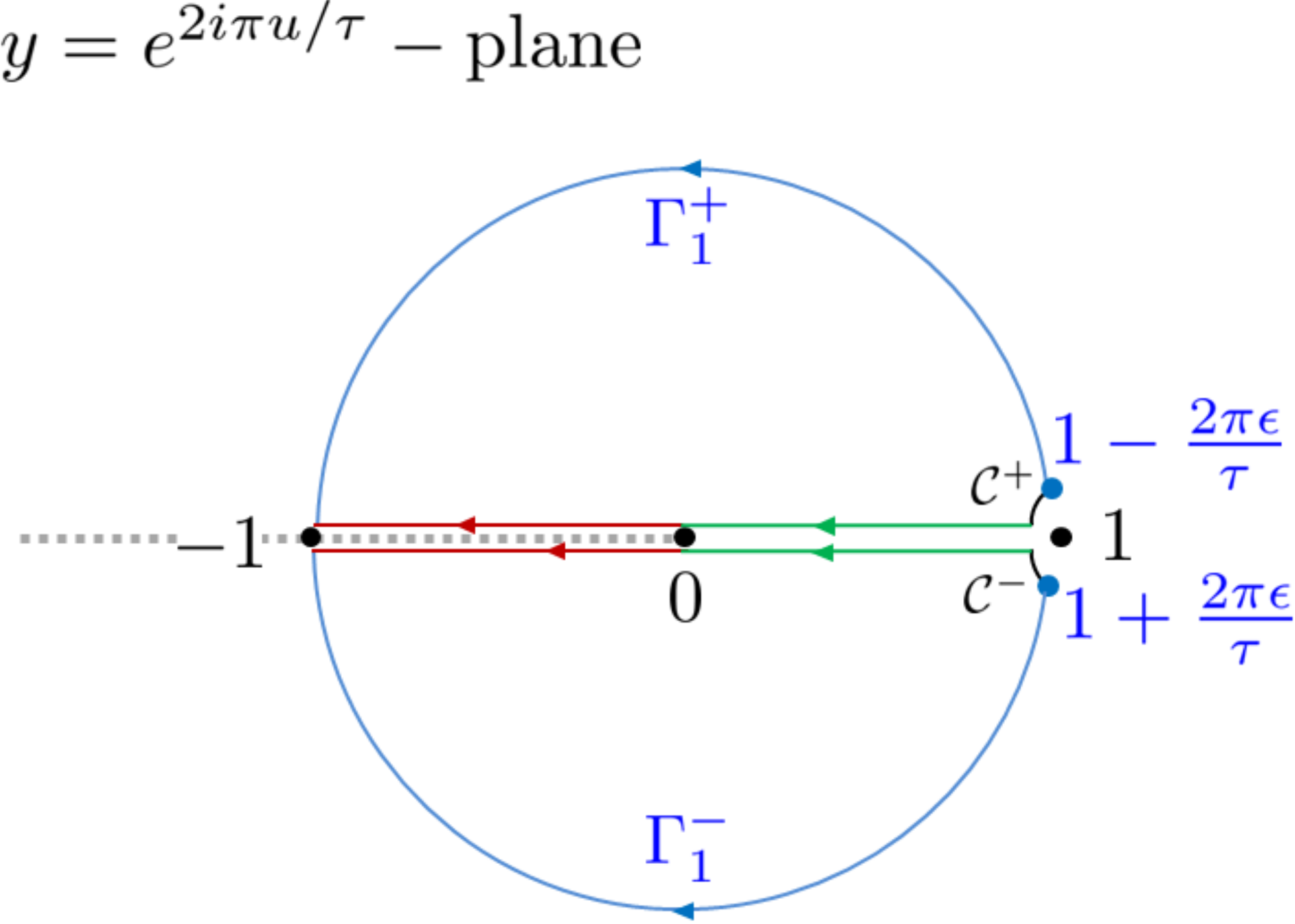}
\end{center}
\end{figure}
We have:
\begin{equation}
\begin{aligned}
\widetilde{J}_{\epsilon}\Big|_{s = s^*} & = \frac{e^{i\pi\nu/2} - e^{-i\pi\nu/2}}{2}\,\Big\{e^{-i\pi\nu/2}\int_{\Gamma_1^+} \frac{\dd y\,y^{\nu/2 - 1}}{y - 1} -  e^{i\pi\nu/2}\int_{\Gamma_1^-} \frac{\dd y\,y^{\nu/2 - 1}}{y - 1}\Big\} \\
&  \phantom{=} - \frac{2 - n}{2}\ln\Big(\frac{b\tau}{2i\epsilon K'}\Big) \\
& = \frac{e^{i\pi\nu/2} - e^{-i\pi\nu/2}}{2}\Big\{e^{-i\pi\nu/2}\Big(\int_{\mathcal{C}^+} \frac{\dd y\,y^{\nu/2 - 1}}{y - 1} - \int_{0}^{1 - \big|\frac{2\pi\epsilon}{\tau}\big|} \frac{\dd y\,y^{\nu/2 - 1}}{y - 1} - \int_{-1}^{0} \frac{\dd y\,(-e^{i\pi\nu/2})\,|y|^{\nu/2 - 1}}{y - 1}\Big) \\
& \phantom{=} - e^{i\pi\nu/2}\Big(- \int_{\mathcal{C}^-} \frac{\dd y\,y^{\nu/2 - 1}}{y - 1} - \int_{0}^{1 - \big|\frac{2\pi\epsilon}{\tau}\big|} \frac{\dd y\,y^{\nu/2 - 1}}{y - 1} - \int_{-1}^{0} \frac{\dd y\,(-e^{-i\pi\nu/2})\,|y|^{\nu/2 - 1}}{y - 1}\Big)\Big\} \\
& \phantom{=} - \frac{2 - n}{2}\ln\Big(\frac{b\tau}{2i\epsilon K'}\Big) \\
& = \frac{e^{i\pi\nu/2} - e^{-i\pi\nu/2}}{2}\Big\{\frac{i\pi}{2}(e^{-i\pi\nu/2} + e^{i\pi\nu/2}) + (e^{i\pi\nu/2} - e^{-i\pi\nu/2})\int_{0}^{1 - \big|\frac{2\pi\epsilon}{\tau}\big|} \frac{\dd y\,y^{\nu/2 - 1}}{y - 1}\Big\} \\
& \phantom{=} - \frac{2 - n}{2}\ln\Big(\frac{b\tau}{2i\epsilon K'}\Big) \\
& = -\frac{\pi}{4}\sqrt{4 - n^2} - \frac{2 - n}{2}\Big\{\int_{0}^{1 - \big|\frac{2\pi\epsilon}{\tau}\big|} \frac{\dd y\,y^{\nu/2 - 1}}{y - 1} + \ln\Big(\frac{b\tau}{2i\epsilon K'}\Big)\Big\} + o(1) \\
\end{aligned}
\end{equation}
where the $o(1)$ stands for a quantity going to $0$ when $\epsilon \rightarrow 0$. Thus:
\begin{equation}
"\widetilde{J}_0"\Big|_{s = s^*}  =   -\sqrt{4 - n^2}\,\frac{\pi}{4} - \frac{2 - n}{2}\Big\{\ln(2) + \int_{0}^{1} \dd y\,\frac{y^{\nu/2 - 1} - 1}{y - 1}\Big\}
\end{equation}
The last integral is elementary:
\begin{equation}
\begin{aligned}
\int_{0}^{1} \dd y\,\frac{y^{\nu/2 - 1} - 1}{y - 1} & = -\lim_{\alpha \rightarrow 1^-} \Big(\int_{0}^{1} \dd y\,y^{\nu/2 - 1}(1 - y)^{-\alpha} - \int_{0}^{1} \dd y\,(1 - y)^{-\alpha}\Big) \\
& =  -\lim_{\alpha \rightarrow 1^{-}} \Big(\frac{\Gamma(\nu/2)\Gamma(1 - \alpha)}{\Gamma(\nu/2 + 1 - \alpha)} - \frac{1}{1 - \alpha}\Big)
\end{aligned}
\end{equation}
Since $\Gamma(1 - \alpha) = \frac{1}{1 - \alpha} + \gamma_E + O(1 - \alpha)$ and $\Gamma'(1) = -\gamma_E$ while $\Gamma(1) = 1$, we obtain:
\begin{equation}
\int_{0}^{1} \dd y\,\frac{y^{\nu/2 - 1} - 1}{y - 1} = - \gamma_E - \frac{\Gamma'(\nu/2)}{\Gamma(\nu/2)} = \psi(1) - \psi(\nu/2)
\end{equation}
where $\psi = (\ln \Gamma)'$ is the digamma function.

\end{document}